\documentclass[manuscript]{aastex631}
\usepackage{amsmath}
\allowdisplaybreaks
\usepackage{xcolor}
\usepackage{graphicx}	
\usepackage{amsmath}	
\usepackage{bbm}

\shorttitle{Machine Learning Methods for Exoplanet Atmospheric Parameter Retrievals}
\shortauthors{Forestano et al.}
\graphicspath{{./}{figures/}}

\begin{document}

\title{Supervised Machine Learning Methods with Uncertainty Quantification for Exoplanet Atmospheric Retrievals from Transmission Spectroscopy}

\correspondingauthor{Roy T.~Forestano}
\email{roy.forestano@ufl.edu}

\author[0000-0002-0355-2076]{Roy T.~Forestano}
\altaffiliation{Equal contribution author.}
\affiliation{Physics Department, University of Florida, Gainesville, FL 32611, USA}

\author[0000-0003-4182-9096]{Konstantin T.~Matchev}
\altaffiliation{Equal contribution author.}
\affiliation{Department of Physics and Astronomy, University of Alabama, Tuscaloosa, AL 35487, USA}

\author[0000-0003-3074-998X]{Katia Matcheva}
\altaffiliation{Equal contribution author.}
\affiliation{Department of Physics and Astronomy, University of Alabama, Tuscaloosa, AL 35487, USA}

\author[0000-0002-6683-6463]{Eyup B.~Unlu}
\altaffiliation{Equal contribution author.}
\affiliation{Physics Department, University of Florida, Gainesville, FL 32611, USA}



\begin{abstract}
Standard Bayesian retrievals for exoplanet atmospheric parameters from transmission spectroscopy, while well understood and widely used, are generally computationally expensive. In the era of the JWST and other upcoming observatories, machine learning approaches have emerged as viable alternatives that are both efficient and robust. In this paper we present a systematic study of several existing machine learning regression techniques and compare their performance for retrieving exoplanet atmospheric parameters from transmission spectra. We benchmark the performance of the different algorithms on the accuracy, precision, and speed. The regression methods tested here include partial least squares (PLS), support vector machines (SVM), $k$ nearest neighbors (KNN), decision trees (DT), random forests (RF), voting (VOTE), stacking (STACK), and extreme gradient boosting (XGB). We also investigate the impact of different preprocessing methods of the training data on the model performance. We quantify the model uncertainties across the entire dynamical range of planetary parameters. The best performing combination of ML model and preprocessing scheme is validated on a the case study of JWST observation of WASP-39b.
\end{abstract}






\section{Introduction}
\label{sec:introduction}

Over the last three decades, the study of extrasolar system planets has shifted from discovery to inference with particular interest in the characterization of their chemical compositions and temperature profiles. The chemical inventory of an exoplanet atmosphere is impacted by the planet formation processes, evolutionary modifications, and its interactions with the local space environment, thus allowing us to place constraints on the existing evolutionary models from the retrieved atmospheric composition.  Transit spectroscopy is currently the most widely used observational technique to study the chemical composition of transiting exoplanets \citep{Schneider1994,Charbonneau2000}. During transit, the planet atmosphere is observed in transmitted light when a planet passes in front of its host star, i.e., the primary eclipse, and in emitted and/or reflected light when a planet travels behind its host star, referred to as the secondary eclipse. Both transmission and emission spectra can be obtained during a transit event. In this work we focus on the analysis of transmission spectra from transiting gas giant exoplanets.

As the stellar light passes through the atmosphere of the transiting planet, it is subjected to absorption and scattering from gas molecules and other atmospheric particulates including cloud particles, haze particles, etc. These particulates leave a characteristic spectroscopic signature in the observed spectrum. Thus, the spectral analysis of the transmitted stellar light allows us to determine key characteristics of the exoplanet's atmosphere, including its temperature profile, chemical abundances, cloud opacity, and further, have the ability to uncover signs of life.

The mathematical procedure for deriving the atmospheric properties from the observed spectrum is known as inversion or atmospheric retrieval. 
Atmospheric retrieval models are complex computational models which aim to determine the thermal structure and chemical composition of the planet atmosphere from the observed spectrum by exploring the high-dimensional parameter space to best fit the observed data \citep{Madhusudhan2018}. The primary component of a retrieval model is a  radiative transfer model (RTM) \citep{Waldmann_2015,Kitzmann2020,Harrington2021,Cubillos2021,Blecic2021,Welbanks2021}. RTMs take in information about the geometry of the planet-star system, including masses, radii, orbital parameters, etc., and specify the atmospheric properties, such as chemical abundances, cloud coverage, and pressure-temperature profile. Some models have the ability to incorporate more advanced aspects, such as large-scale dynamical effects \citep{2023RemS...15..635P}, day/night asymmetries \citep{2022A&A...658A..42P,2022ApJ...929...20M,2022ApJ...933...79W} and latitudinal/longitudinal variations \citep{2022A&A...658A..41F}. More realistic atmospheric models are inherently more complex, which inevitably leads to a larger number of unconstrained parameters and increased computational cost. 

Traditional atmospheric retrieval methods are based on statistical inference through the use of sampling approaches, such as Markov chain Monte Carlo (MCMC) \citep{Madhusudhan_2009,Cubillos_2013,Line_2013} or Nested Sampling \citep{Benneke_2012,Waldmann_2015,Oreshenko_2017,Gandhi_2018}, to perform Bayesian parameter estimation \citep{Nixon2020}. The complexity of Bayesian retrievals typically scales with the number of model parameters, therefore, by increasing the number of chemical constituents or adding geometrical dimensions such as latitudinal or longitudinal variability, run-times can increase dramatically. Furthermore, due to possible degeneracies in the atmospheric parameters one needs to fully explore the parameter space and find all viable solutions, which adds to the computational challenge. 

With the James Webb Space Telescope (JWST) \citep{2016ApJ...817...17G} online and with its successor, the Roman Telescope, due to launch in 2027, as well as a number of exoplanet-specific prospective launches, including the Twinkle Space Telescope \citep{2019ExA....47...29E} and the European Space Agency (ESA) Ariel mission \citep{2021arXiv210404824T}, the amount of spectral data observed from transiting exoplanets is expected to increase rapidly. The large number of observations and the improved spectral resolution pose further computational demands on the existing atmospheric retrieval methods and call for the development of alternative, more efficient atmospheric retrieval techniques.

Recently, supervised machine learning (ML) techniques have been employed as alternatives to the standard atmospheric retrieval methods. A number of supervised ML approaches have been tried to obtain atmospheric parameters from the observed spectra:  
deep belief networks \citep{Waldmann2016ApJ},
deep neural networks \citep{Yip_2021}
random forests \citep{Marquez2018,Oreshenko2020,Fisher2020,Nixon2020,Guzman2020},
generative adversarial networks (GANs) \citep{Zingales2018},
convolutional neural networks (CNNs) \citep{Soboczenski2018,Yip_2021,Ardevol2022},
recurrent neural networks (like LSTMs) \citep{Yip_2021},
an ensemble of Bayesian Neural Networks \citep{Cobb2019},
transformer-nspired deep learning architectures \citep{unlu2023reproducing}, 
conditional invertible neural networks (cINNs) \citep{Haldemann2022},
normalizing flows \citep{2023arXiv230909337A,2024ApJ...961...30Y}, sequential neural posterior estimation \citep{2024A&A...681L..14A} and
simulation-based inference \citep{2025ApJ...984L..32L}.
While this plethora of applicable approaches is a virtue, it also introduces the issue of epistemic uncertainty, which reflects how well the model has learned the underlying patterns in the data. Different ML architectures in principle give slightly different answers, thus contributing to the overall uncertainty of the retrievals \citep{Yip_2021}. 

At the same time, unsupervised machine learning techniques has also been shown to be useful, e.g., for deriving an informed prior for the retrieval \citep{Hayes2020}, for optimal preprocessing of the data or quick planetary categorization \citep{2022PSJ.....3..205M}, or for identifying unusual spectra \citep{Forestano_anomaly_2023}. An important component of any ML analysis is the pre-processing of the data into a form which makes it easy for the ML model to learn, i.e., to extract the physically meaningful patterns in the data. This usually implies some sort of dimensionality reduction \citep{2022PSJ.....3..205M} or transforming the original features by shifting, rescaling, etc. --- a process commonly referred to as feature engineering. 

In this paper, we test and compare the performance of several standard interpretable ML regression algorithms for predicting the temperature and chemical abundances from transmission spectra. The regression methods considered here include partial least squares (PLS), support vector machines (SVM), $k$ nearest neighbors (KNN), decision trees (DT), random forests (RF), voting, stacking, and extreme gradient boosting (XGB). Each model is trained on the Ariel Big Challenge (ABC) database, and then comprehensively tested on a diverse set of planets. Traditional evaluation metrics for ML regression include parameters like mean absolute error (MAE), root mean squared error (RMSE), mean absolute percentage error (MAPE), coefficient of determination (R-squared) and others. However, the performance of any given model can vary quite a bit throughout the parameter space and cannot be captured with a single parameter. Therefore, we shall examine and discuss the uncertainties in the model predictions by analyzing and visualizing the deviation from the true values over the different parts of parameter space. We shall also investigate the effect of different methods of pre-processing the data on the performance of the ML algorithms. These pre-processing methods are described below in Sections~\ref{sec:standardization} and \ref{sec:normalization}

The paper is organized as follows. Section~\ref{sec:database} describes the database used in our study. Section~\ref{sec:feature_engineering} introduces the two different approaches for preprocessing the data which will be tried in the analysis to follow. Section~\ref{sec:methods} contains an overview of the different machine learning methods used in the paper. Section~\ref{sec:analysis} presents the methodology and results from our numerical analysis. Section~\ref{sec:WASP39b} presents an application of the top performing ML regressor to a real data example of WASP-39b. Section~\ref{sec:summary} concludes with a summary of the results and an outlook for the future. To avoid interrupting the flow of the paper, many large figures are collected in an appendix for easy reference.

\section{Database Description}
\label{sec:database}

The database used in this paper was modeled after the Ariel Big Challenge (ABC) Database, which was introduced for public use in the 2022 Ariel Machine Learning Data Challenge as paert of a NeurIPS 2022 competition \citep{Taurex3, Yip_competition}. This challenge aimed to encourage the development of efficient and robust supervised ML models for exoplanetary atmospheric retrievals. As described by \cite{Changeat_2022}, the ABC dataset is a synthetic spectroscopic dataset created through the TauREx forward RTM \citep{Taurex3} and the official instrument simulator for the ESA Ariel mission \citep{Mugnai_2020}. It consists of 5900 actual unique planets from the Ariel preliminary Target List selected from the official TESS candidate list \citep{2019AJ....157..242E, 2022AJ....164...15E}.

\begin{figure}[t]
\begin{center}
\includegraphics[width=0.5\columnwidth]{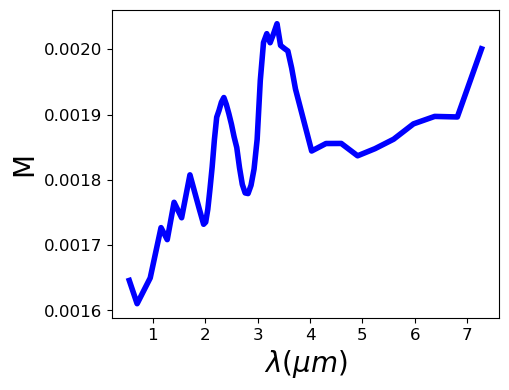}
\end{center}
    \caption{Sample transmission spectrum from the database used in this work. $M$ is the observed modulation of the stellar flux as defined in eq.~(\ref{eq:M}).}
    \label{fig:spectrum}
\end{figure}

While each planet retains its unique stellar and planetary parameters, e.g. radius, mass, distance, etc., the planet atmospheric chemical composition was randomly generated and the temperature was chosen to be the planet equilibrium temperature. This set up allows for very realistic physics scenarios with the caveat that the planet/star parameters are not uniformly sampled throughout the training/testing database. We will revisit the implications of this choice in Section~\ref{sec:summary}.  In our realization of the database, we used the same forward model parameters and assumptions as in the original ABC dataset but without the instrument-specific noise effects, i.e. the spectra have no noise added to them. An example transit spectrum is shown in Figure $\ref{fig:spectrum}$.

For the TauREx simulation we use an isothermal atmospheric setup with 100 layers equally spaced in log-pressure coordinates from a minimum of 1 Pa ($10^{-5}$ bar) to a maximum of $10^{6}$ Pa ($10$ bar). The planets are considered hot Jupiters with hydrogen-helium dominated atmospheres. The $He/H_2$  ratio is fixed at 0.17 assuming solar abundances, while each planet has a range of different trace amounts of gasses. The database includes five different trace absorbers with randomly sampled concentrations $X_i$ for $i \in \{ H_2O, CH_4, CO_2, CO, NH_3 \}$. With the volume mixing ratio of all trace gasses being less than $10^{-3}$, the atmospheric mean molecular mass of all planets is kept constant at $2.29$ amu. The radiative transfer calculations include line absorption by the trace gases, collision-induced absorption (CIA) due to $H_2$-$H_2$ and $H_2$-$He$ interactions, and Rayleigh scattering by the main atmospheric gases, while cloud and/or haze effects were excluded. Each star was modeled as a black body source at a specified temperature. For each planet, the final full resolution transmission spectrum $M(\lambda)$, consisting of $76,744$ wavelength bins, was binned down to $52$ bins, reflecting the spectral binning of the Ariel instruments.

Other planet-star system parameters included in the database are the star-Earth distance, star temperature, star radius, star mass, star-planet distance, planet radius, planet mass, planet surface gravity, and the planet orbital period.
\begin{figure}[t]
\begin{center}
\includegraphics[width=0.9\columnwidth]{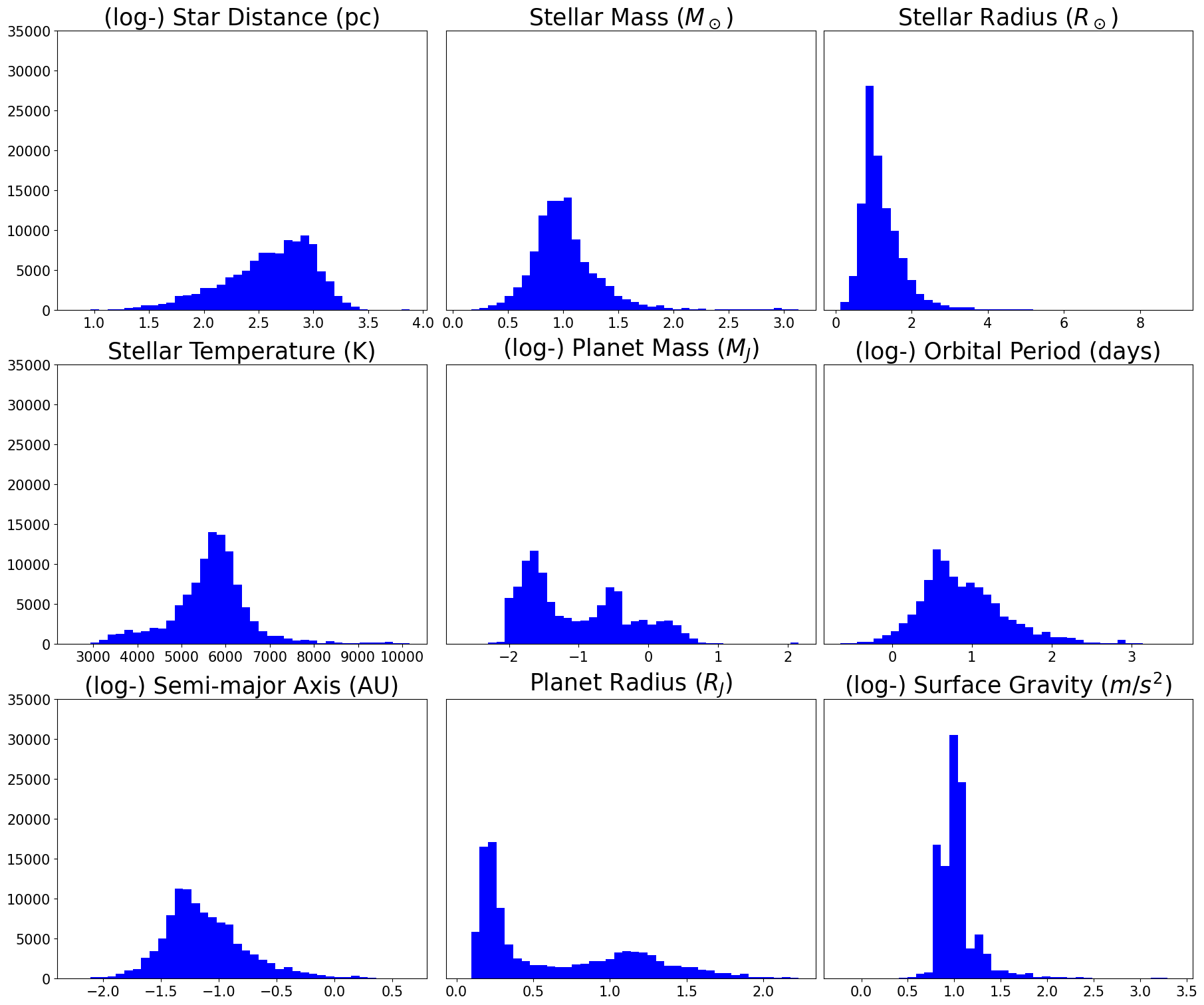}
\end{center}
    \caption{Distributions of selected stellar and planetary physical parameters over the ABC dataset. As labeled, these include the star distance, star mass, star radius, star temperature, planet mass, planet orbital period, planet distance, planet radius, and planet surface gravity.}
    \label{fig:auxiliary}
\end{figure}
Histograms of the distributions of all stellar and planet parameters included in the ABC database are presented in Figure~\ref{fig:auxiliary} displaying the auxiliary parameters and Figure~\ref{fig:FM} displaying the target parameters. Note that while some of the auxiliary parameters shown in Figure \ref{fig:auxiliary} are extraneous and are not used in the analysis; we include them here for completeness of the database description. Figure~\ref{fig:auxiliary} and the left panel in Figure~\ref{fig:FM} show that the parameter distributions in the database are not uniform which can lead to training biases in the machine learning models and can affect the performance of the models in the underrepresented regions of the parameter space (see Sec. \ref{sec:summary}).

\begin{figure}[t]
\begin{center}
\includegraphics[width=0.95\columnwidth]{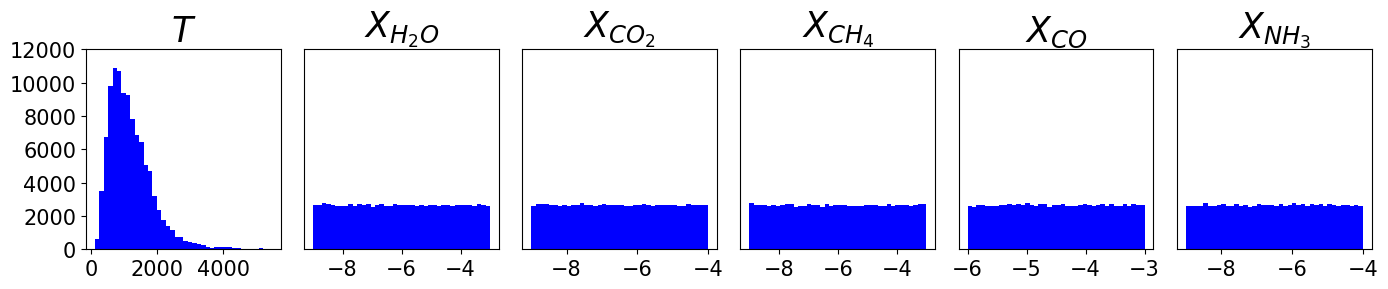}
\end{center}
\caption{Distributions of the target parameters, including the temperature $T$ and log-mixing ratios, $\log(X_i)$, for $i \in \{ H_2 O, CO_2, CH_4, CO, NH_3\}$.}
\label{fig:FM}
\end{figure}

To generate $105,887$ planet instances, i.e. individual spectra, from $5,900$ unique target planets, different planet atmospheres were sampled. While the primary contribution to each atmosphere is $H_2$ and $He$, trace amounts of $H_2O$, $CO_2$, $CH_4$, $CO$ and $NH_3$ were also included. The log-mixing ratios log$(X_i)$, i.e. the log-concentration of each species in the ABC database, were sampled from a uniform distribution, as shown in Figure $\ref{fig:FM}$. The sampling ranges were: 
$X_{H_2O} \in (10^{-9}, 10^{-3})$, 
$X_{CO_2} \in (10^{-9}, 10^{-4})$, 
$X_{CH_4} \in (10^{-9}, 10^{-3})$, 
$X_{CO} \in (10^{-6}, 10^{-3})$,
$X_{NH_3} \in (10^{-9}, 10^{-4})$ \citep{2020AJ....160...80C}. The sampling among constituents was done independently, assuming no particular chemical model.

\section{Data Preprocessing}
\label{sec:feature_engineering}

The transit spectra are generated with the {\tt TauREx3} forward RTM \citep{Taurex3}. 
During transit, the fractional change in the observed flux is defined as
\begin{align}
    M(\lambda) &= \frac{\Phi_0(\lambda) - \Phi_T(\lambda)}{\Phi_0(\lambda)}\, ,
    \label{eq:M}
\end{align}
where $\Phi_0(\lambda)$ is the original stellar flux, while $\Phi_T(\lambda)$ is the minimum flux observed during transit at a given wavelength $\lambda$. As mentioned previously, each spectrum $M(\lambda)$ (see Fig.~\ref{fig:spectrum}) contains unique information about the radiative transfer processes occurring within a planet's atmosphere. The spectrum can then be used to infer the atmospheric structure of an exoplanet. 

Using the {\tt TauREx3} framework, we generate $105,887$ synthetic exoplanetary transit spectra $M_i(\lambda_j)$ at $52$ different wavelengths $\lambda_j, (j=1,2,,\ldots, 52)$ between $0.55 \, {\mu}m$ and $7.275\, {\mu}m$. The purpose of this work is to test the performance of several standard supervised machine learning regression algorithms under various data preprocessing techniques described in Sections~\ref{sec:standardization} and \ref{sec:normalization}, and to estimate the corresponding uncertainties. 

\subsection{Notation and setup}
\label{sec:notation}

The typical structure of a dataset for a supervised machine learning task takes the form of an $s \times (f+t)$ matrix, i.e. dataframe, for $s$ samples, $f$ features, and $t$ targets, i.e.
\begin{equation}
\begin{array}{cccccccc}
    x_1^{(1)}, &  x_1^{(2)}, & \ldots , & x_1^{(f)}; & y_1^{(1)},& y_1^{(2)},& \ldots , & y_1^{(t)}\\
    x_2^{(1)}, &  x_2^{(2)}, & \ldots , & x_2^{(f)}; & y_2^{(1)},& y_2^{(2)},& \ldots , & y_2^{(t)}\\
    \vdots     & \vdots      & \vdots   & \vdots     & \vdots  & \vdots & \vdots & \vdots \\
    x_s^{(1)}, &  x_s^{(2)}, & \ldots , & x_s^{(f)}; & y_s^{(1)},& y_s^{(2)},& \ldots , & y_s^{(t)}.\\
\end{array}
\label{eq:dataset}
\end{equation}
Here the $f$-dimensional vector $\mathbf{x} \equiv (x^{(1)}, x^{(2)}, \ldots, x^{(f)})$ represents the independent feature variables and the $t$-dimensional vector $\mathbf{y} \equiv (y^{(1)}, y^{(2)}, \ldots , y^{(t)})$ corresponds to the dependent target variables. The database contains $s$ samples, i.e., $s$ instantiations of the vectors $\mathbf{x}$ and $\mathbf{y}$. From now on we shall label the samples by a subscript $i$, $(i=1,2,\ldots, s)$. For the ABC database, $s=105,887$ planets, $f=52$ wavelengths, and $t = 6$ targets including the equilibrium planet temperature and the concentrations of five chemicals.

In the problem of atmospheric retrievals, the feature variables represent the binned spectra at different wavelengths:
\begin{equation}
x_i^{(j)} = M_i(\lambda_j), 
\end{equation}
while the target variables are the atmospheric parameters used to generate the spectrum, in our case
\begin{equation}
y_i^{(k)} = \left\{ T, X_{H_2 O}, X_{CO_2}, X_{CH_4}, X_{CO}, X_{NH_3}\right\}_i.
\label{eq:target_variables}
\end{equation}

\subsection{Standardization of the spectra}
\label{sec:standardization}

A common preprocessing technique used in machine learning is the 
\textit{standardization} of the data, which consists of centering, i.e., subtracting the mean 
\begin{equation}
\bar{x}^{(j)}= \mathbb{E}_i [x_i^{(j)}] = \frac{1}{s}\sum_{i=1}^s x_i^{{(j)}}
\label{eq:meanx_stand}
\end{equation} 
and scaling, i.e., dividing by the standard deviation
\begin{equation}
\sigma_x^{(j)} = \sqrt{ \mathbb{E}_i \left[ \left(x_i^{(j)} - \bar{x}^{(j)}\right)^2 \right] } 
= \sqrt{\frac{1}{s}\sum_{i=1}^s  \left(x_i^{(j)} - \bar{x}^{(j)}\right)^2 }
\label{eq:sigmax_stand}
\end{equation}
for each individual feature in the dataset. Specifically, the standardized dataset is obtained as
\begin{equation}
x_{Si}^{(j)} \equiv  S[x_i^{(j)}] \equiv 
\frac{x_i^{(j)}-\bar{x}^{(j)}}{\sigma_x^{(j)}},
\label{eq:xS_def}
\end{equation}
where $S[\cdot]$ denotes the standardization operator with respect to some input $\cdot$, and $\mathbb{E}_i[\cdot]$ represents the expectation value of the input $\cdot$ over the distribution given by the index $i$. We can similarly standardize the target variables as 
\begin{align}
y_{Si}^{(k)} \equiv S[y_i^{(k)}] = \frac{y_i^{(k)}-\bar{y}^{(k)}}{\sigma_y^{(k)}},
\label{eq:yS_def}
\end{align}
where 
\begin{equation}
\bar{y}^{(k)} \equiv \mathbb{E}_i [y_i^{(k)}] = \frac{1}{s}\sum_{i=s}^s y_i^{{(k)}}
\label{eq:meany_stand}
\end{equation}
and
\begin{equation}
\sigma_y^{(k)} = \sqrt{\frac{1}{s}\sum_{i=1}^s  \left(y_i^{(k)} - \bar{y}^{(k)}\right)^2 }.
\label{eq:sigmay_stand}
\end{equation}
Note that the averaging in equations (\ref{eq:meanx_stand}), (\ref{eq:sigmax_stand}), (\ref{eq:meany_stand}) and (\ref{eq:sigmay_stand}) is over the sample index $i$. Standardization is commonly used in machine learning tasks to allow the model to treat the different feature variables as well as the different target variables on equal footing \citep{Bishop_ML}. 

\begin{figure}[t]
\begin{center}
\includegraphics[width=0.48\columnwidth]{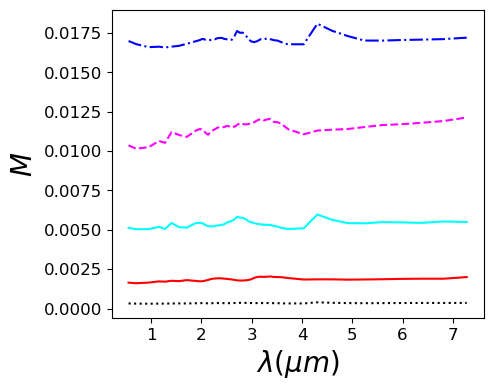}
\includegraphics[width=0.475\columnwidth]{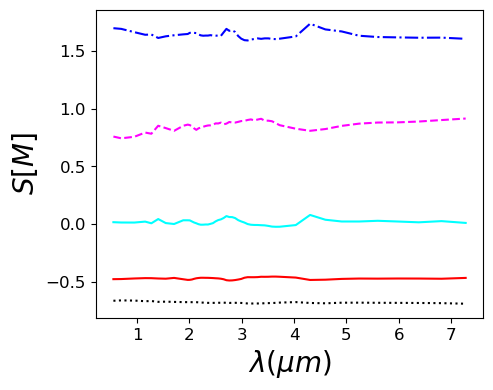}
\includegraphics[width=0.5\columnwidth]{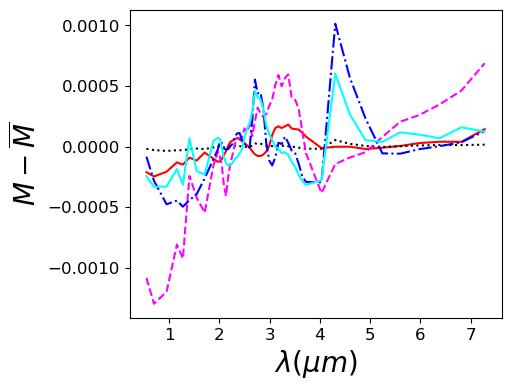}
\includegraphics[width=0.47\columnwidth]{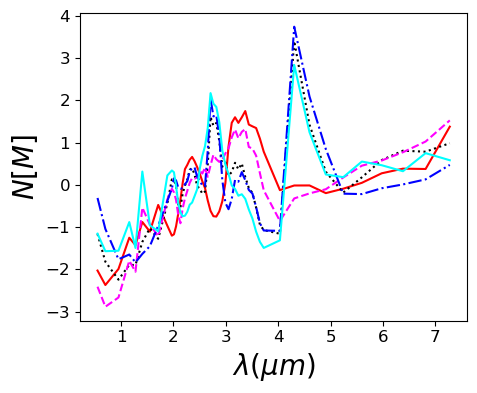}
\end{center}
\caption{Standardization and normalization effects on the transit spectra. Top left: five sample spectra from the database. Top right: spectra standardized according to eq.~(\ref{eq:xS_def}). Bottom left: subtracting the mean of spectrum has the effect of centering all of the original spectra around zero. Bottom right: spectra normalized according to eq.~(\ref{eq:xN_def}). }
\label{fig:preprocess_spectra}
\end{figure}

A visualization of the effect of standardization on the spectra is shown in the top right panel of Figure \ref{fig:preprocess_spectra}. The top left panel in the figure depicts five sample spectra, $M(\lambda)$, from the database. After applying the standardization (\ref{eq:xS_def}), the new standardized spectra, $S[M(\lambda)]$, are as shown in the top right panel.

\subsection{Normalization of the spectra}
\label{sec:normalization}

An alternative way to preprocess the data is to normalize each individual spectrum independently from the others. In what follows, we shall refer to this procedure as \textit{normalization} of the data. It has been shown to be beneficial in practice during the Ariel Machine Learning Data Challenge \citep{Ariel2022,unlu2023reproducing} and was theoretically motivated in \cite{2022ApJ...939...95M}.
The process of normalization centers the samples about their {\em spectral} means 
\begin{equation}
\bar{x}_i= \mathbb{E}_j [x_i^{(j)}] = \frac{1}{f}\sum_{j=1}^f x_i^{{(j)}}
\label{eq:meanx_norm}
\end{equation} 
and rescales them by the corresponding standard deviation of each spectrum
\begin{equation}
\sigma_{xi} = \sqrt{ \mathbb{E}_j \left[ \left(x_i^{(j)} - \bar{x}_i\right)^2 \right] } 
= \sqrt{\frac{1}{f}\sum_{j=1}^f  \left(x_i^{(j)} - \bar{x}_i\right)^2 }.
\label{eq:sigmax_norm}
\end{equation}
Here $\mathbb{E}_j[\cdot]$ represents the expectation value of input $\cdot$ over the distribution given by the index $j$. As a result, the variance of each sample is equal to one. The normalization procedure for the features, therefore, is defined by 
\begin{align}
x_{Ni}^{(j)} \equiv  N[x_i^{(j)}] \equiv 
\frac{x_i^{(j)}-\bar{x}_i}{\sigma_{xi}},
\label{eq:xN_def}
\end{align}
where $N[\cdot]$ denotes the normalization operator with respect to some input $\cdot$. Note that the averaging in equations (\ref{eq:meanx_norm}) and (\ref{eq:sigmax_norm}) is over the feature index $(j)$. The two steps involved in the normalization procedure are illustrated in the bottom two panels of Figure~\ref{fig:preprocess_spectra}. It should be noted that in this paper, the targets $y_i^{(k)}$ can only be standardized, whereas the features $x_i^{(j)}$ can either be standardized or normalized. In this analysis, normalization will be applied only to the spectral features, not to the auxiliary features.

The reason why normalization, as opposed to standardization, of the input features is useful in the case of exoplanet transit spectra is the following. Typically, in machine learning tasks, the features represent different types of quantities, with different physical units and, possibly, with very different numerical orders of magnitude. This variety hinders the training process, as the model focuses on the most numerically significant features and tends to ignore the others. This is when standardization is useful, as it equalizes the feature and forces the model to explore all features equally during training. However, in our case, all features represent the same type of physical quantity, and have therefore the same units and orders of magnitude. Therefore, we can normalize the inputs sample-wise rather than feature-wise while preserving the physical meaning. 

\begin{figure}[t]
    \begin{center}
        \includegraphics[width=0.49\columnwidth]{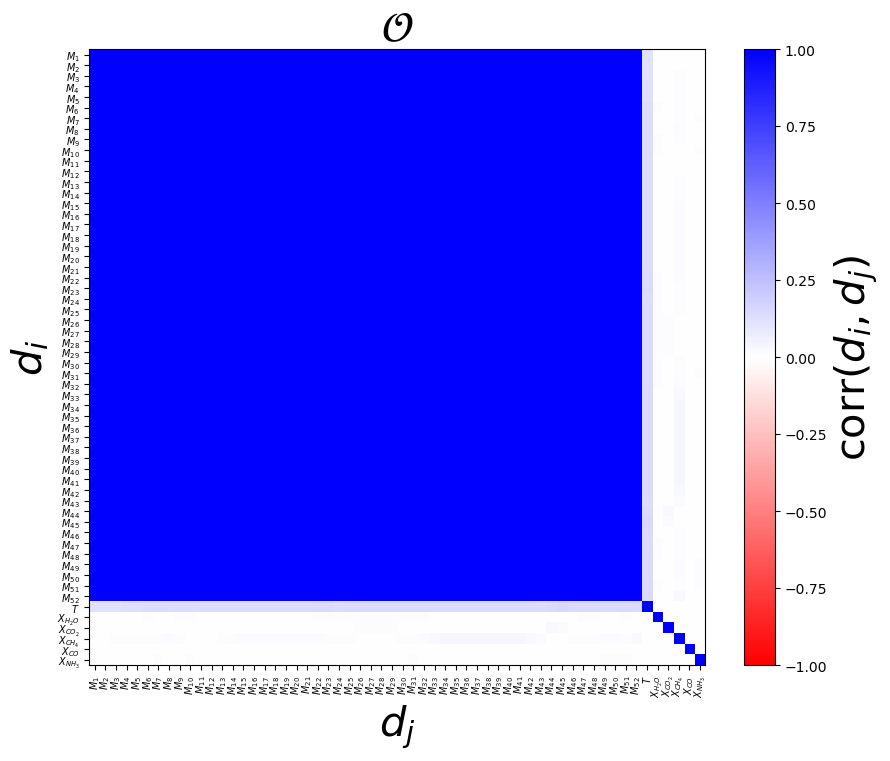} 
        \includegraphics[width=0.49\columnwidth]{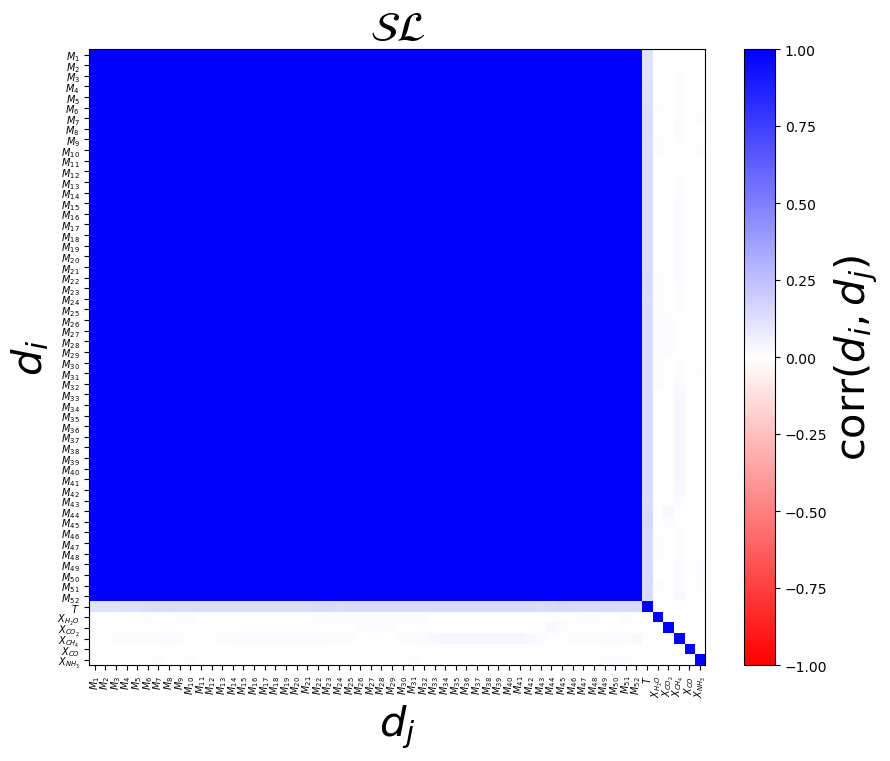}  \\
        \includegraphics[width=0.49\columnwidth]{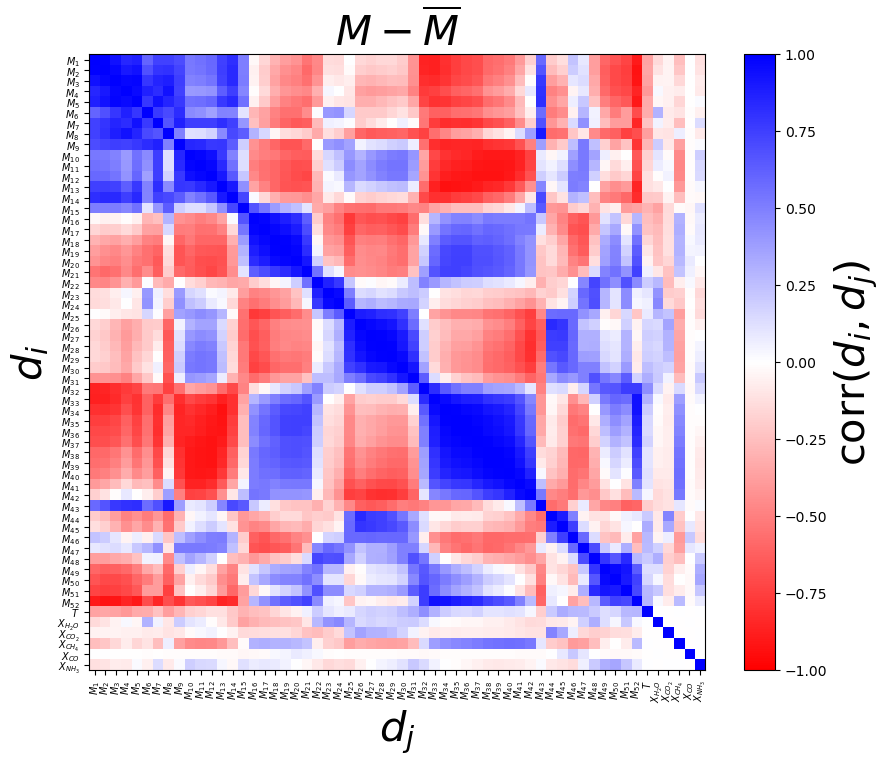} 
        \includegraphics[width=0.49\columnwidth]{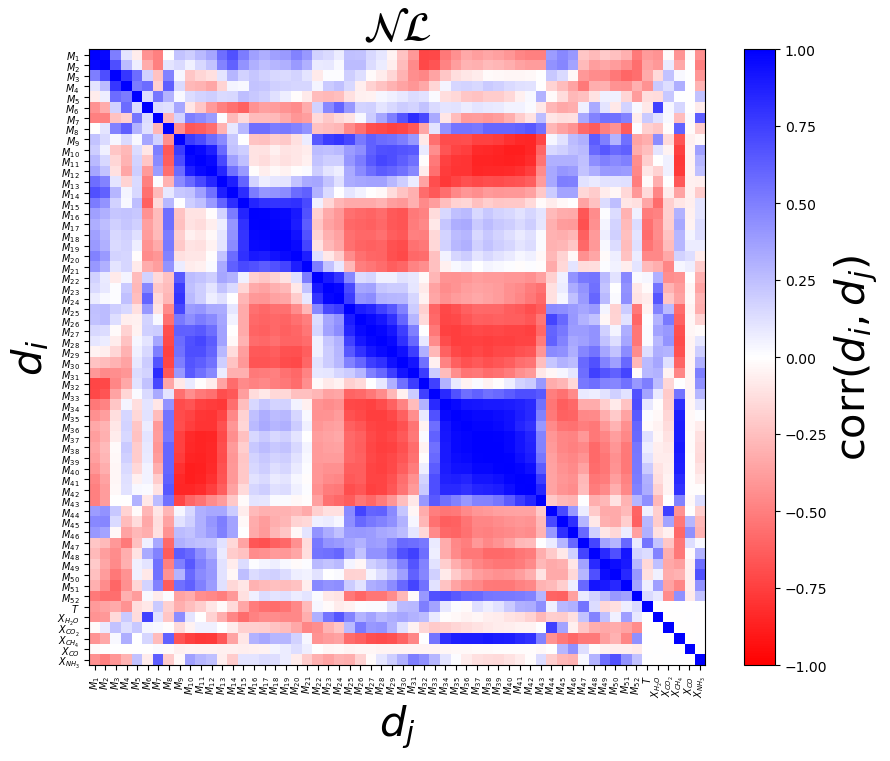}  
    \end{center}
    \caption{The correlation matrices among the original (top left), standardized (top right), centered by the sample means (bottom left), and normalized (bottom right) features. }
    \label{fig:cov_corr}
\end{figure}

In comparison to the original spectra, it is clear that standardization of the data leaves the relative distance between spectra the same, with all spectra positioned about a global mean of zero as shown in the upper right panel of Figure~\ref{fig:preprocess_spectra}. In contrast, normalization of the spectra centers each one around a local mean of zero, with both positive and negative entries for $x_{N}^{(j)}$, as shown in the lower right panel in Figure~\ref{fig:preprocess_spectra}. This effect can be seen in the covariance matrix relationship between two feature sets $v$ and $w$ defined by
\begin{align}
    \text{Cov}_{jk} (v,w) &= \mathbb{E}_{i}\left[ \left(v_i^{(j)} - \mathbb{E}_i [ v_i^{(j)} ]\right)\left( w_i^{(k)} - \mathbb{E}_i[ w_i^{(k)}] \right) \right].   
\end{align}
The covariance is a measure of how two variables, $v_i^{(j)}$ and $w_i^{(k)}$, indicated by the $j$-th and $k$-th indices, vary with respect to each other over a dataset. Here, we take the entire dataset, namely both the features $x$ and targets $y$, as a single dataset matrix, as depicted in Eq.~$\ref{eq:dataset}$, to compute the correlation between the variables
\begin{align}
    \text{Corr}_{jk} = \frac{\text{Cov}_{jk}}{\sqrt{\text{Cov}_{jj} \text{Cov}_{kk} }} = \frac{\text{Cov}_{jk}}{\sigma_j \sigma_k},
\label{eq:corr_def}
\end{align}
which normalizes the covariance matrix in the usual sense. The correlation matrix consists of ones along its diagonal showing maximum correlation between a variable and itself. Figure~\ref{fig:cov_corr} shows the correlation matrices for the case of the four different data representations shown in Figure~\ref{fig:preprocess_spectra}.
The original and standardized correlation matrices exhibit nearly identical behavior, showing the high correlation among the original spectral features, as noted in \cite{2022PSJ.....3..205M}, as well as among their standardized versions (top right panel). The correlation matrices for the centered features (lower left panel) and normalized features (lower right panel) contain much more expressivity in the variations among constituent variables. In the normalization case, the cross-correlations among the features and targets are clearly visible. This suggests that normalization may provide a more useful representation of the spectra for extracting information about the atmospheric composition and temperature.

\subsection{Training Data}
\label{sec:training_data}

For training the supervised machine learning methods, we used a random subset of $n = 60,000$ samples with a train/test split of $80/20$. In other words, this analysis used $48,000$ synthetic exoplanetary samples for training the models and $12,000$ samples for testing the performance of each model on unseen data. The six possible data configurations include 
\begin{enumerate}
    \item  $\mathcal{S} \equiv \{ \, S[M_i(\lambda_j)], \, S[y_i^{(k)}] \, \} = \{\,x_{Si}^{(j)}, y_{Si}^{(k)} \,\}$, where both the features and  targets have been standardized as in eqs.~(\ref{eq:xS_def}) and (\ref{eq:yS_def}).
    \item  $\mathcal{SL} \equiv \{ \, S[M_i(\lambda_j)], \, S[\{Y  _i^{(k=1)}, \log{Y  _i^{(k>1)}} \}] \, \}$, which is the same as $\mathcal{S}$, except the target variables for the chemical abundances are logarithmic, i.e., $y_i^{(k>1)} \to \log y_i^{(k>1)}$. 
    \item  $\mathcal{N} \equiv \{ \, N[M_i(\lambda_j)], \, S[y_i^{(k)}] \, \} = \{ \,x_{Ni}^{(j)}, y_{Si}^{(k)} \, \}$, where the features are normalized and the targets are standardized. 
    \item  $\mathcal{NL} \equiv \{ \, N[M_i(\lambda_j)], \, S[\{y_i^{(k=1)}, \log{y  _i^{(k>1)}} \}] \, \}$, which is the same as $\mathcal{N}$, except the target variables for the chemical abundances are logarithmic, i.e., $y_i^{(k>1)} \to \log y_i^{(k>1)}$. 
    \item  $\mathcal{NM} \equiv \{ \, \{N[M_i(\lambda_j)], \overline{M}_i, \sigma_i(M) \}, \, S[y _i^{(k)}] \, \}$,  which is the same as $\mathcal{N}$, but adding the spectral mean $\bar{x}_i=\bar{M}_i$ and the standard deviation $\sigma_{xi}=\sigma_i(M)$ as additional features. 
    \item  $\mathcal{NML} \equiv \{ \, \{N[M_i(\lambda_j)], \overline{M}_i, \sigma_i(M)  \}, \, S[\{y  _i^{(k=1)}, \ln{y  _i^{(k>1)}}  \}] \, \}$, which is the same as $\mathcal{NM}$, except the target variables for the chemical abundances are logarithmic, i.e., $y_i^{(k>1)} \to \log y_i^{(k>1)}$. 
\end{enumerate}
Here $\mathcal{S}$ denotes standardization of the inputs, $\mathcal{N}$ denotes normalization of the inputs, $\mathcal{L}$ denotes the use of the log concentrations in the targets, and $\mathcal{M}$ denotes the inclusion of the spectral means and standard deviations, which retains the full amount of information in the original data. For each configuration described here, the analysis to follow will only show the results for the performance of the already trained models on the testing sets.

\section{Methods}
\label{sec:methods}

Several well-known supervised regression algorithms will be explored in this analysis. These algorithms are \textit{supervised} because the data is labeled and the goal of the regression task is to predict the target variables, i.e. the labels. All algorithms, except XGB, will be deployed using the {\tt scikit-learn} machine learning library for the Python programming language \citep{scikit-learn}. The XGB method will be implemented using the {\tt xgboost} machine learning library \citep{Chen_2016}. The intention here is to test and exploit the capabilities of standard intuitive and explainable supervised machine learning methods without resorting to deep learning models often perceived as black-boxes. The supervised regression algorithms tested here include partial least squares (PLS), support vector machines (SVM), $k$ nearest neighbors (KNN), decision trees (DT), random forests (RF), and ensemble methods like voting (VOTE), stacking (STACK), and extreme gradient boosting (XGB).

\subsection{Partial Least Squares (PLS)}
\label{sec:pls}

PLS regression is a linear regression method in a reduced dimensionality latent space for both the features and the targets. In some sense, it is a generalization of the Principal Component Analysis to the case of labelled data --- the algorithm tries to maximize the covariance between the latent projections of the features and the targets. Further details about the PLS implementation can be found in \cite{Wegelin_2000} and \cite{scikit-learn}.

\subsection{Support Vector Machines (SVM)}
\label{sec:svm}

A SVM uses a sparse least-squares optimization technique by transforming the inputs and optimizing the least squares loss function to learn relevant support vectors for points which lie outside a small margin of the regression's output. Unlike the other methods presented here, SVMs only predict one-dimensional outputs, therefore, we will only consider a single fixed $k$, or column, of the targets $y_{i}^{(k)}$. More details about the method can be found in \cite{Bishop_ML, Hastie2001statisticallearning, Fan2005, Chen2006}.

\subsection{k-Nearest Neighbors (KNN)}
\label{sec:knn}

The KNN algorithm infers the targets $y$ of each of the inputs $x$ by finding the $k$ nearest neighbors of each $x$ in the input space and averaging the neighboring target values. This algorithm is simple yet requires a high density of points in all regions of the training input space in order to make accurate predictions.

\subsection{Decision Tree (DT)}
\label{sec:dt}

The DT algorithm is a least-squares minimization algorithm which makes distinct cuts in the input feature space and averages the target variables within each partition to make predictions. The cuts determine a tree structure which is learnable and easily interpretable. The method uses a greedy approach to choose the best cuts iteratively. More detail on the DT algorithm can be found in \cite{Bishop_ML, Breiman_1984, Hastie_2009}. 

\subsection{Random Forest (RF)}
\label{sec:rf}

The RF algorithm is an esemble method which utilizes several DTs each of which optimizes its splits by selecting a random subset of input features to choose from. The final prediction is averaged over the predictions of the individual trees \citep{Dietterich_1998}. The number of trees is a hyperparameter, and each tree is constructed using the procedure described in Section \ref{sec:dt}, except at each node a random subset of of the features is chosen to optimize the cut.

\subsection{Voting (VOTE) and Stacking (STACK)}
\label{sec:vote_and_stack}

The voting algorithm is an ensemble method which utilizes several weak learners, or regressors, $f^{(\alpha)}$ and averages their predictions $\mathbf{y}^{(\alpha)} = f^{(\alpha)}(\mathbf{x})$ over all regressors for each input $\mathbf{x}$ to make a final prediction. Each regressor $f^{(\alpha)}$ for $\alpha \in \{0,1, \dots,  m\}$ is optimized independently over the training data. The final prediction takes the form
\begin{align}
    \mathbf{y} = \frac{1}{m} \sum_{\alpha =1}^m f^{(\alpha)} (\mathbf{x}) = \frac{1}{m} \sum_{\alpha=1}^m \mathbf{y}^{(\alpha)}.
\end{align}
To improve the accuracy of the final predictions, one can also perform a weighted voting scheme which assigns a weight to each regressor's output.

The stacking algorithm is another ensemble method which also utilizes several weak learners $f^{(\alpha)}$ to make individual predictions. However, rather than simply averaging the predictions $\mathbf{y}^{(\alpha)}$ from each regressor $f^{(\alpha)}$, a final regressor $f^{(m+1)}$ is trained on the previous regressors to make a final prediction $\mathbf{y}^{(m+1)}$ as 
\begin{align}
    \mathbf{y}_{i}^{(m+1)} = f^{(m+1)} ( \mathbf{y}_i^{(\alpha)} ).
\end{align}

\subsection{Extreme Gradient Boosting (XGB)}
\label{sec:xgb}

Boosting in machine learning is a method for creating an ensemble consisting of weak learners $f^{(\alpha)}$ which, unlike the ensemble methods discussed in Sections \ref{sec:rf} and \ref{sec:vote_and_stack}, trains these regressors in sequence, where each regressor $f^{(\alpha)}$ is trained using weights associated with each data point that depend on the performance of the previous weak learners. To train the next regressor in the sequence, the algorithm increases the weights given to points which have inaccurate predictions associated with them based on the previous regressors' outputs. After each regressor has been optimized, their predictions $Y^{(\alpha)} = f^{(\alpha)}(X)$ are aggregated through a weighted majority voting procedure. Gradient boosting uses a perturbative approach to include the first and second order gradient statistics within the loss function to better optimize the weights corresponding to each point \citep{Friedman_2000, Chen_2016}. In particular, XGB iteratively trains an ensemble of random trees to optimize its predictions. 

\subsection{Hyperparameters}
\label{sec:hyperparameters}

Hyperparameters are model parameters which are defined by the user before the training of the algorithm. The choice of hyperparameters is left to the individual user. For all models, we optimized the respective hyperparameters to achieve best performance.

For the PLS regressor, the full length of features was used as the number of desired latent components with an early stopping tolerance of $10^{-5}$ for the covariance between the latent features. The feature dimension for the datasets 1-4 defined in Section $\ref{sec:training_data}$ was $52$, corresponding to all spectral wavelength bins. For datasets 5-6 the feature dimension was $54$, corresponding to all spectral wavelength bins plus the mean and standard deviation of each sample. 

For the SVM regressor, the regularization parameter was chosen as $C=500$, $\epsilon = 0.1$, and a radial basis function kernel was used with kernel coefficient $\gamma = 0.028$. 

For the KNN regressor, the number of neighbors was chosen as $k=6$. A ball tree algorithm with a leaf size of $30$ was used to rapidly find the nearest neighbors of each input data point.

For the DT regressor, a least-squares error objective was minimized with a random splitting criterion out of a random set of a maximum of $29$ features. The minimum samples required to split a node was $48$, and the minimum samples required at each leaf was $13$.

For the RF regressor, $115$ decision trees were constructed to optimize a least-squares error objective function. The same tree hyperparameters were used as those in the DT, except the random splitting was required to be out of a random set of a maximum of $27$ features, rather than $29$. 

For the voting regressor, the regressors used in the ensemble were SVM, KNN, DT, and RF regressors. Stacking used as base learners the SVM, KNN, and DT regressors and the RF regressor as the final estimator. 

For the XGB regressor, a maximum of $500$ trees were used with an early stopping criterion of $15$ rounds, if the loss did not decrease during these rounds. A learning rate of $0.1$ was used to optimize the model, and a random state of $42$ was used to initialize the model. To reduce each tree's complexity, a maximum tree depth, or longest path from the root node to a leaf node, of size 6 was used.

\section{Numerical Results}
\label{sec:analysis}

\subsection{Setup}

We perform 48 different numerical experiments, where each of the eight machine learning regressors from Section~\ref{sec:methods} is trained on each of the six types of training data described in Section~\ref{sec:training_data}. For easy reference, the results from the experiments are collected in an Appendix, Figures~\ref{fig:n_scatter}-\ref{fig:nml_pred}. Each figure caption has an identifier that corresponds to the type of preprocessing of the training data: 
$(\mathcal{S})$, $(\mathcal{SL})$, $(\mathcal{N})$, $(\mathcal{NL})$, $(\mathcal{NM})$, or $(\mathcal{NML})$. Note that in all cases the original spectral data is the same. In each figure, the eight rows show the results from the eight different trained regressors, as indicated on the leftmost panels.  It should be noted that, when training the three configurations $\mathcal{S},,\mathcal{N},\mathcal{NM}$ on the concentrations, negative concentration predictions resulted which were reset to the minimum value of $10^{-9}$ used to generate the dataset as can be seen in Figures \ref{fig:s_scatter}, \ref{fig:n_scatter}, and \ref{fig:nm_scatter} among the PLS, SVM, VOTE, and XGB methods.

An important advantage of ML-based exoplanet parameter retrievals is that the computational cost is amortized, i.e., the computational time for training the ML model is spent upfront and only once and is being leveraged each time the trained model is subsequently used. The predictions themselves from the trained regressor are obtained essentially instantaneously. The computational cost in seconds for training each of the eight ML methods with each of the six types of datasets is shown in Table \ref{table:comp_cost}. The size of the training sample was $48,000$ in each case. Table \ref{table:comp_cost} demonstrates that some methods are faster than others, and typical training times take from a second to a few hours. This is much faster than a typical exoplanetary atmospheric retrieval which, depending on its complexity, could takes several hours to several days for a single planet. 

\begin{table}[t]
  \caption{Computational cost (in seconds) of each regression algorithm training on $48000$ samples.}
  \label{table:comp_cost}
  \centering
        \begin{tabular}{ c  r r r r r r r  } 
        \hline
        \textbf{Model} & $\mathcal{S}$ & $\mathcal{SL}$ & $\mathcal{N}$ & $\mathcal{NL}$ & $\mathcal{NM}$  & $\mathcal{NML}$   \\
        \hline
        \textbf{PLS} & 3.3 & 5.8 & 3.3 &  5.2 &  7.3 & 4.6  \\         
        \textbf{SVM} & 1981 & 2774 & 1192 & 1375 & 1700  & 1702  \\ 
        \textbf{KNN} & 9.0 & 9.8 & 39.7  & 42.5 & 74.1 & 65.8  \\         
        \textbf{DTR} & 0.5 & 0.4 & 0.3 & 0.3 & 0.7  & 0.4   \\        
        \textbf{RFR} & 256.9  & 330.9 & 175.0 & 143.1 & 351.8  & 165.4   \\
        \textbf{VOTE} & 4253 & 6007 & 2256 & 2475  & 4506  & 3240   \\
        \textbf{STACK} & 6522 & 11265 & 4083 & 4514 & 4736  & 6030   \\
        \textbf{XGB} & 158.2 & 177.5 & 174.4 & 211.4 & 206.8  & 368.3   \\
        \hline
        \end{tabular}
\end{table}

\subsection{Predicted versus actual graphs and error estimates}

For each numerical experiment, three different types of plots are presented. First, Figures~\ref{fig:s_scatter}-\ref{fig:nml_scatter} 
show scatter plots of the model predictions over the test dataset for the 6 target variables from eq.~(\ref{eq:target_variables}) (on the $y$-axis) versus the true values $y_t$ of the target variables ($x$-axis). The colorbars indicate the absolute deviation of the prediction from the true value. Scatter plots which align along the diagonal $45^\circ$ line indicate accurate model predictions, while points away from that diagonal correspond to larger errors and poorly performing models. 

Second, the accuracy and the precision of the ML predictions are illustrated in Figures \ref{fig:s_true}-\ref{fig:nml_true}, 
which are constructed as follows. Each scatter plot from the first set of figures (Figures \ref{fig:s_scatter}-\ref{fig:nml_scatter}) 
is binned into quantile (equally-populated) bins along the $x$-axis, with about $520$ planets per bin. The average true value $\bar{y}_t$ of the target variable in each bin is plotted on the $x$-axis. The mean absolute deviation, $\overline{|y_t-y_p|}$, of the target prediction $y_p$ from the true target value $y_t$ for each bin, is plotted on the $y$-axis. The corresponding standard deviation of the quantity $|y_t-y_p|$ is shown as the error bar. Accurate models are characterized by small values of the bias $\overline{|y_t-y_p|}$, while models with high precision exhibit small error bars, which are measures of the model variance \citep{Yip_2021}.

The third and final set of plots, shown in Figures \ref{fig:s_pred}-\ref{fig:nml_pred}, 
contain an alternative representation of the quality of the model's performance in terms of accuracy and precision. This time, the quantile bins are formed using the {\em predicted} values shown along the $y$-axes of the panels in the first set of figures (Figures \ref{fig:s_scatter}-\ref{fig:nml_scatter}). 
Note that the predicted values depend on the ML method and on the preprocessing scheme, thus the test planets are distributed differently across individual figures and rows. Then, in analogy to Figures \ref{fig:s_true}-\ref{fig:nml_true}, 
the average values $\bar{y}_p$ of the predictions within each quantile bin are plotted on the $y$-axis in Figures \ref{fig:s_pred}-\ref{fig:nml_pred}, 
while the $x$-axis shows the corresponding mean (depicted with the circle symbol) and standard deviation (indicated with the error bar) of the absolute deviation $|y_t-y_p|$ of the target prediction $y_p$ from the true values $y_t$.

The two set of plots, those in Figures \ref{fig:s_true}-\ref{fig:nml_true},
and those in Figures \ref{fig:s_pred}-\ref{fig:nml_pred}, answer complementary, and somewhat orthogonal questions about the accuracy of the models. First, given a true value for the target parameter, Figures \ref{fig:s_true}-\ref{fig:nml_true} illustrate the range of predictions which could be obtained by running different ML models or preprocessing schemes. In contrast, Figures \ref{fig:s_pred}-\ref{fig:nml_pred} answer a different question: given that we obtained a certain value of $y_p$ from the ML analysis, what is the range of plausible true values $y_t$?

\subsection{Discussion}
\label{sec:discusstion}

Evaluating the performance of a ML regression model for a complex multi-dimensional task could be rather subtle. There are a number of global evaluation metrics which attempt to quantify performance in terms of a single parameter, e.g., mean absolute error (MAE), root mean squared error (RMSE), mean absolute percentage error (MAPE), coefficient of determination (R-squared) and others. However, the performance may vary significantly across the parameter space, for example, due to uneven representation in the training data, or different behavior of the target function throughout the parameter space. A predicted versus actual scatter plot like those in Figures~\ref{fig:s_scatter}-\ref{fig:nml_scatter} is a common tool to visually assess the performance of the regression model globally throughout the parameter space.

The results shown in Figures~\ref{fig:s_scatter}-\ref{fig:nml_scatter} can be used to answer several questions. First by comparing the same row (ML method) across different figures, we can see the effect of using different preprocessing schemes. For example, we note that models tend to perform better when they are trained on data normalized as in Section~\ref{sec:normalization} as opposed to data standardized as in Section~\ref{sec:standardization}. This observation goes against the standard ML common lore, but has been confirmed in the ARIEL machine learning data challenges \citep{Ariel2022,2023arXiv230909337A,unlu2023reproducing} and in related ML studies of detection of anomalous exoplanet chemistry \citep{Forestano_anomaly_2023}. 

Comparing Figure~\ref{fig:n_scatter} to Figure~\ref{fig:nl_scatter}
(or Figure~\ref{fig:nm_scatter} to Figure~\ref{fig:nml_scatter}), we notice that using logarithmic values for the chemical abundances in the training data helps the models at both low and high values of the corresponding abundance. At the same time, the comparison of Figure~\ref{fig:nm_scatter} to Figure~\ref{fig:n_scatter} reveals that using the mean and the standard deviation of the spectrum as additional features during the training does not improve the predictions. This can be understood in terms of the transverse decomposition analysis in \cite{2022ApJ...939...95M}, which analyzed the information content in the transmission spectra.

By comparing the different rows in each individual figure among Figures~\ref{fig:s_scatter}-\ref{fig:nml_scatter}, we can deduce the relative performance of the eight ML methods. In the following we focus on Figure~\ref{fig:nl_scatter}, which corresponds to the best preprocessing choice, as already discussed above.

We observe that at large values of the chemical abundances most models perform well, yet the most accurate predictions are made by SVM and the ensemble methods (XGB, VOTE, STACK). At low abundances, the quality of the predictions in general deteriorates, which is not surprising, since the spectral signatures are very weak. In particular, at low abundances there is a noticeable deviation from the diagonal trend --- a ``knee'' develops around $X\sim 10^{-6}-10^{-7}$, depending on the particular chemical. It is desirable to have this knee as low as possible --- this would imply better sensitivity of the model in terms of its ability to detect very low abundances of trace chemicals.

Overall, the best performing models across all datasets were XGB and SVM, followed closely by STACK and VOTE. Note that the method of random forest (RF) has been often used in the literature for exoplanet parameter retrievals, but in this analysis is outperformed by most of the other methods considered here. As expected, all models had a difficult time predicting $X_{CO}$, although SVM and XGB still showed a decent performance.

The conclusions from the predicted-versus-actual graphs in Figures~\ref{fig:s_scatter}-\ref{fig:nml_scatter} are supported by the results shown in the subsequent Figures~\ref{fig:s_true}-\ref{fig:nml_pred}. In particular, Figure~\ref{fig:nl_true} demonstrates that both the bias and the variance of the predictions are very low for XGB and SVM. Note that the popular RF method has difficulties at both low and high values of the chemical abundances. We also note that for all methods, the temperature is well captured at low values, and the errors increase with the temperature. This can be attributed to the fact that there were very few planets with high tempteratures in the training data --- see the left panel in Figure~\ref{fig:FM}. This motivates the development of a more comprehensive database in which all parameter ranges are properly sampled.

Figures~\ref{fig:s_true}-\ref{fig:nml_true} (see also Figures 3, 10 and 11 in \cite{Yip_2021}) present the bias and variance of the model as a function of the true input parameter. This representation is useful for explainability, i.e., understanding what the model is doing ``under the hood", but is of limited value in practice since in the case of a real planet, the true values of the parameters are not known a priori. This is where the alternative representation shown in Figures~\ref{fig:s_pred}-\ref{fig:nml_pred} has a greater practical value, since they inform the experimenter of the range of possible true values $y_t$, given the predicted value $y_p$. As already discussed, the predicted values are more accurate at high abundances and at low temperatures. The best results are offered by XGB and SVM, when trained with the $\mathcal{NL}$ dataset.

\begin{figure}
    \begin{center}
        \includegraphics[width=0.24\columnwidth]{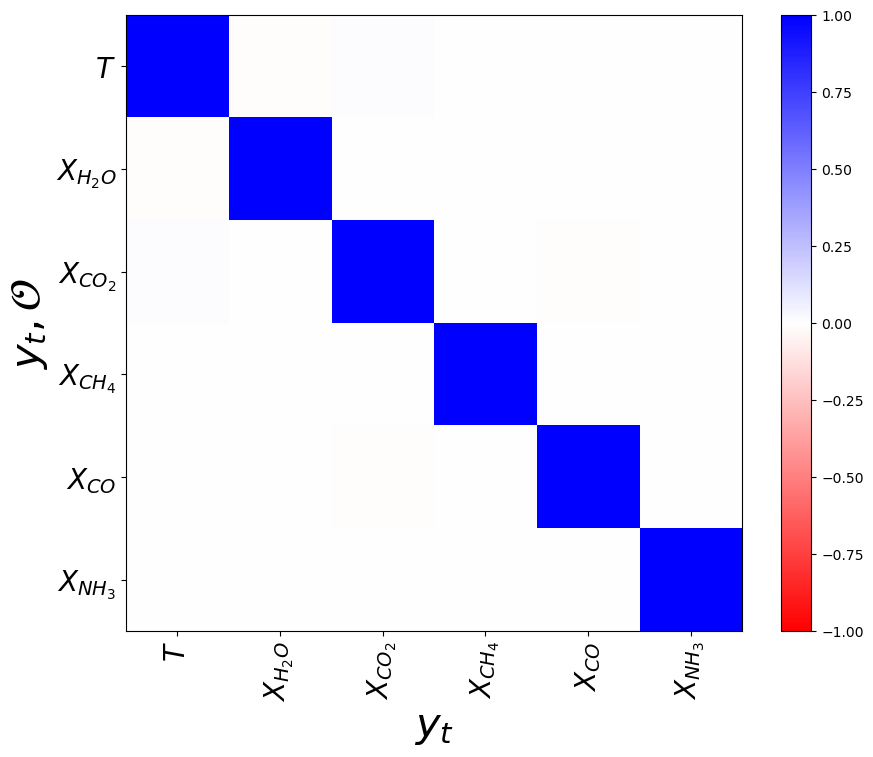}
        \includegraphics[width=0.24\columnwidth]{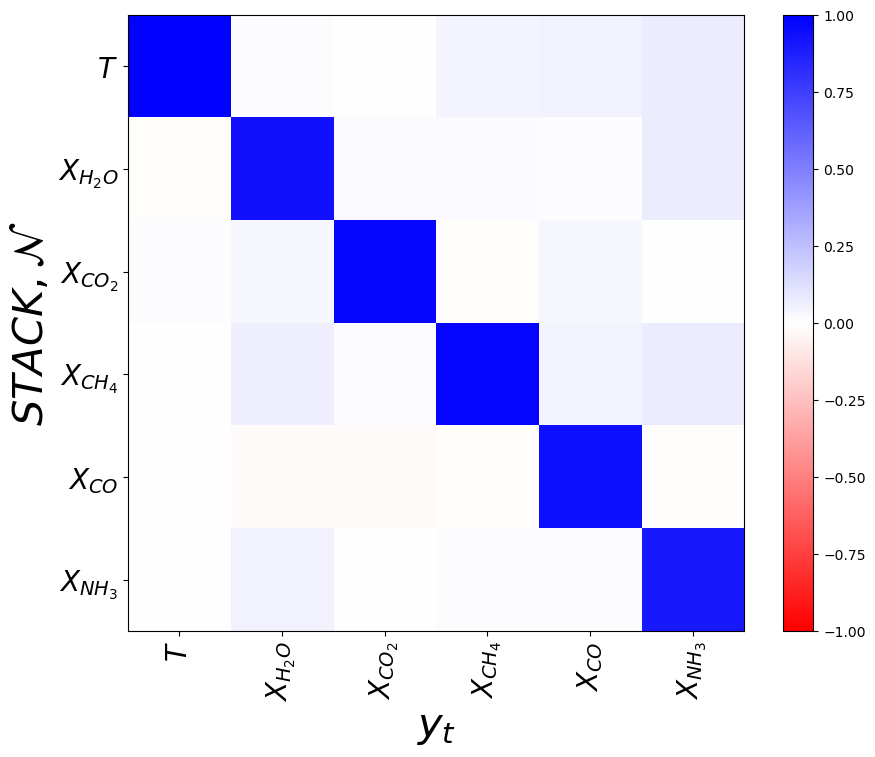}
        \includegraphics[width=0.24\columnwidth]{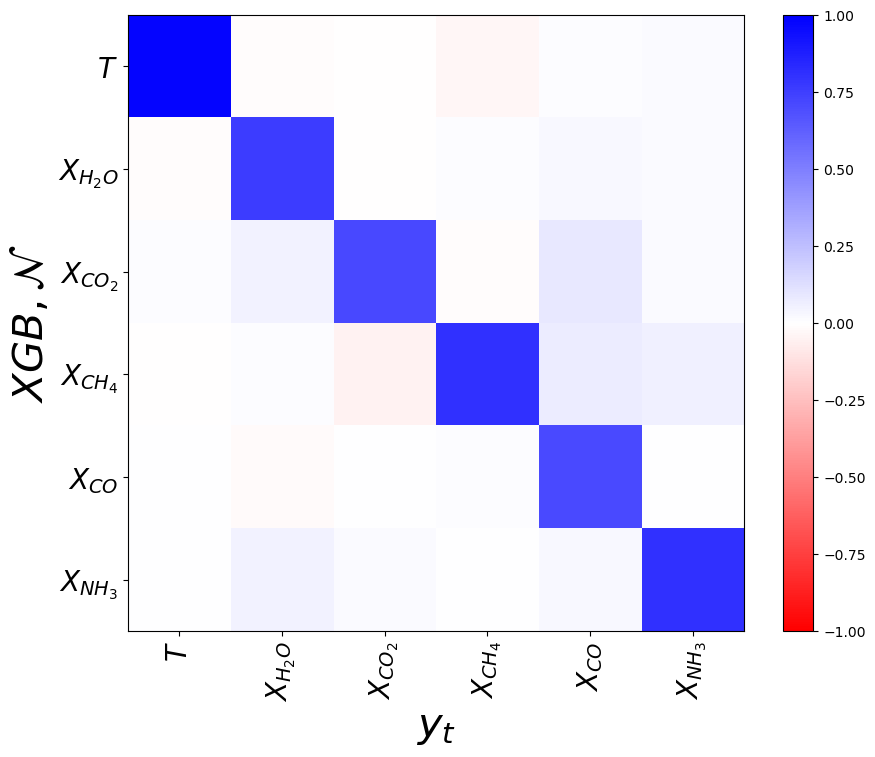}
        \includegraphics[width=0.24\columnwidth]{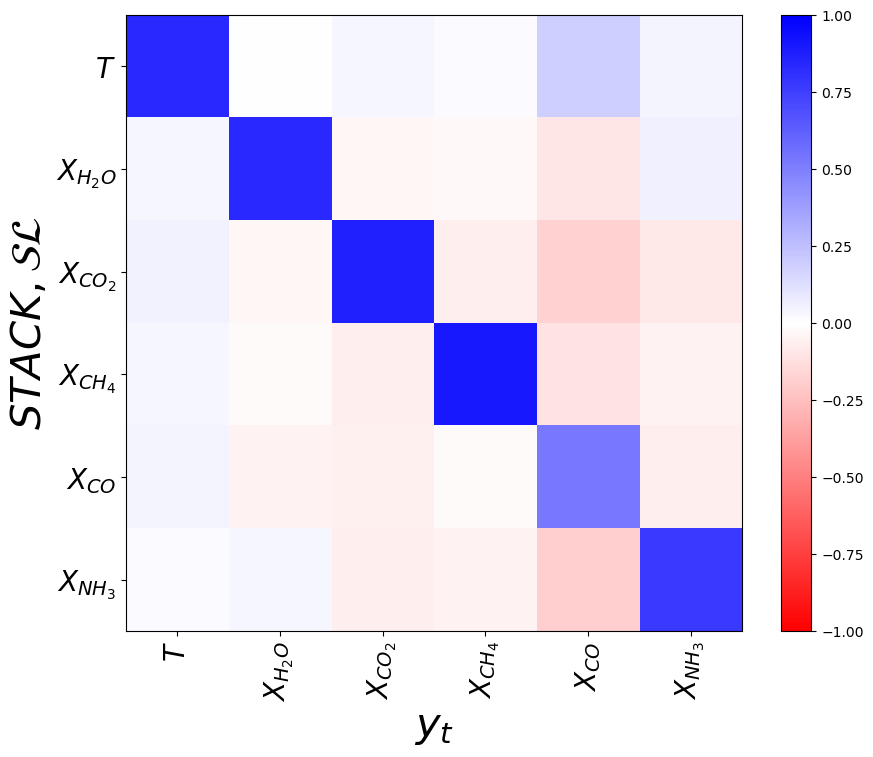} \\
        \includegraphics[width=0.24\columnwidth]{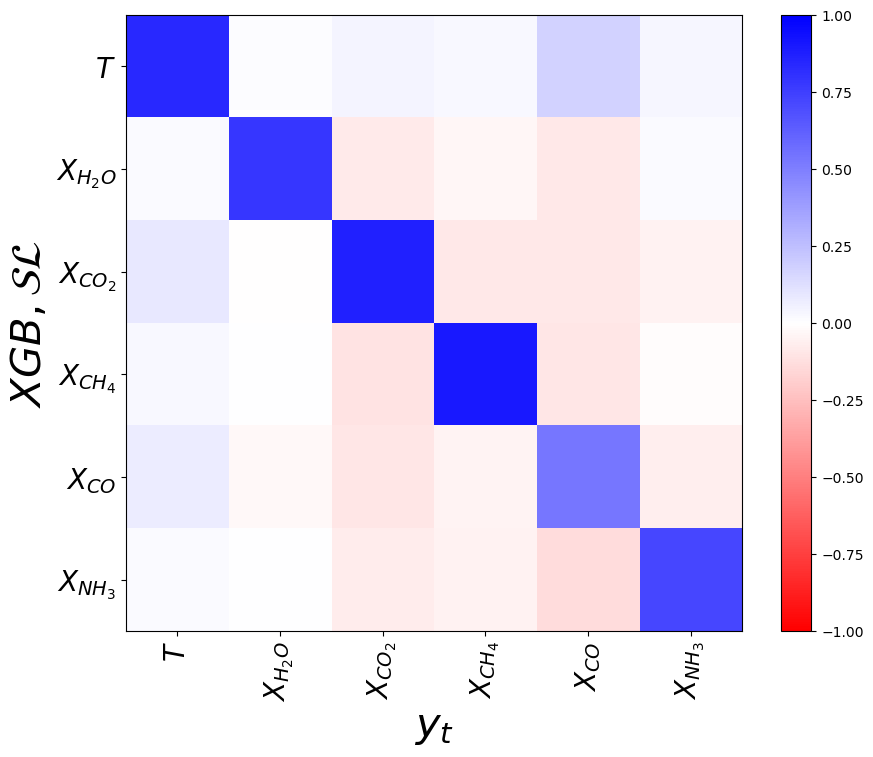}
        \includegraphics[width=0.24\columnwidth]{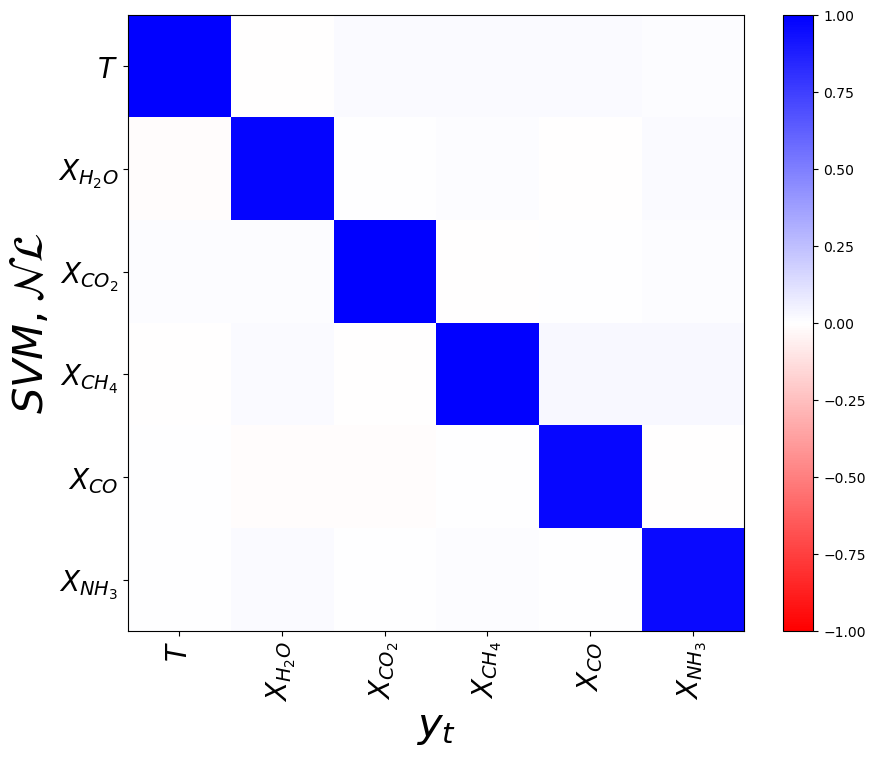}
        \includegraphics[width=0.24\columnwidth]{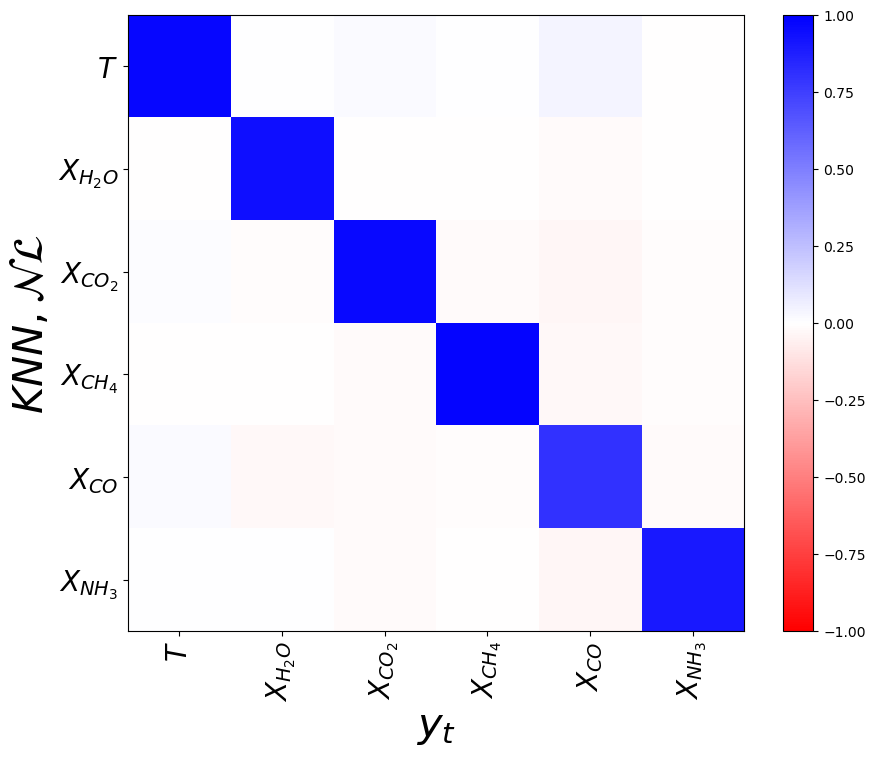}
        \includegraphics[width=0.24\columnwidth]{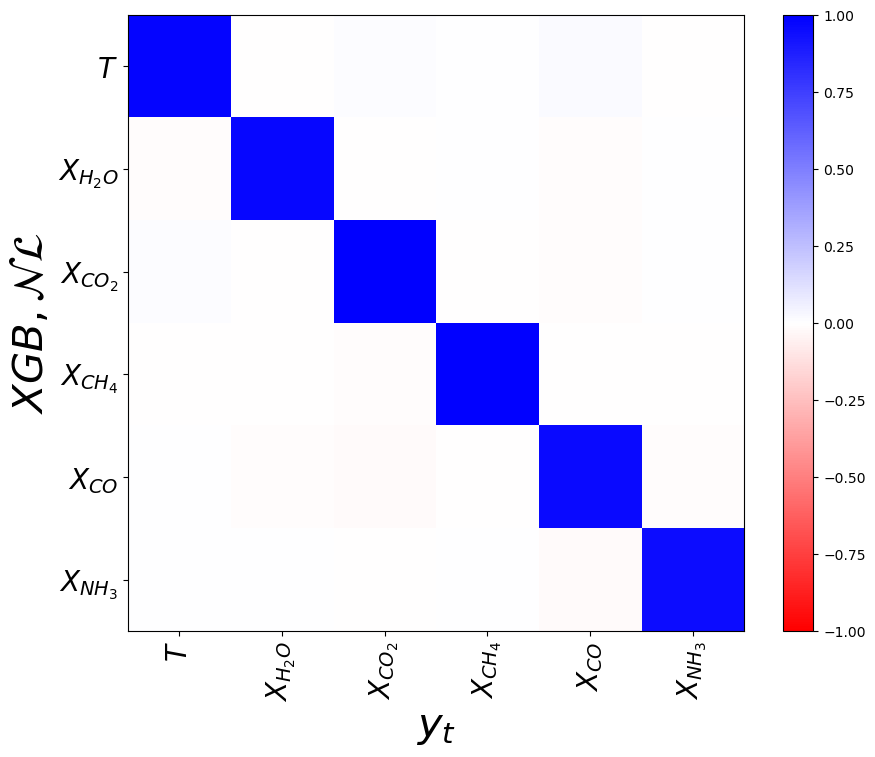}
    \end{center}
    \label{fig:correlation_outputs}
    \caption{Correlation matrices of the predicted and true target values for a few selected pairings of regressors and preprocessing techniques. For reference, the baseline calculation of the true target correlations is provided in the top left plot.}
\end{figure}

Figure~\ref{fig:correlation_outputs} provides yet another way of judging the preformance of the ML methods and pre-processing techniques. It relies on the observation that the target parameters were independently sampled when generating the database. Therefore, any correlations among the predicted values are spurious and the predictions from a well-trained model should be uncorrelated as well. Figure~\ref{fig:correlation_outputs} shows the correlation matrices of the predicted and true target values for a few selected pairings of regressors and preprocessing techniques. The baseline calculation for the true target parameters among themselves is provided in the top left plot. Thus, an identity matrix is desired as the features only exhibit strong correlations amongst themselves. Once again, the SVM and XGB models, trained on the {\cal NL} dataset, show best performance.

\section{Case study: WASP-39b}
\label{sec:WASP39b}

As a demonstration of the model performance on real data, we evaluate the predictions from the top performing model and preprocessed dataset configuration, i.e. the XGB regressor which uses the $\mathcal{NL}$ dataset of normalized spectra and log chemical abundances, on WASP-39b spectral data from the James Webb Space Telescope (JWST) \citep{2023Rustamkulov,2024Powell}. The two original spectra were combined to produce a single spectrum in the range from 0.51 $\mu$m to 11.4 $\mu$m to cover the entire spectral range of the training dataset. Using the $\texttt{TauRex FluxBinner}$ we rebin this spectrum down to the same exact 52 spectral wavelength values used in our synthetic dataset. 
\begin{figure}
    \begin{center}
        \includegraphics[width=0.6\columnwidth]{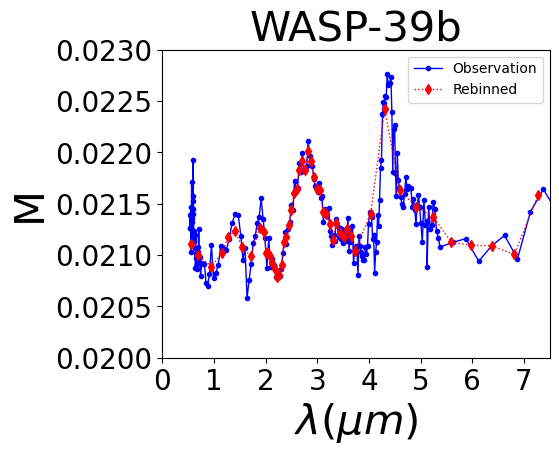}
    \end{center}
    \label{fig:wasp39b_spectra}
    \caption{The original observed JWST and rebinned WASP-39b spectra.}
\end{figure}
Figure~\ref{fig:wasp39b_spectra} depicts the original (blue dots) and the rebinned (red diamonds) spectra. To predict the atmospheric composition of the exoplanet WASP-39b, we train the XGB regressor on the entire $\mathcal{NL}$ dataset consisting of 105,887 planet instances and then apply the trained model to the similarly normalized rebinned WASP-39b spectrum. The results can be found in Table~\ref{table:conc_wasp39b} alongside those from traditional retrievals recorded in the literature: the results for $X_{CO_2}$ and $X_{CO}$ are based on the Tiberius and Eureka pipelines \citep{Constantinou_2023}, while the results for $X_{CH_4}$ \citep{2023Rustamkulov,Ahrer_2023} and $X_{NH_3}$ \citep{Alderson_2023} are based on the PICASO pipeline. We note that, while the presence of water in WASP-39b is generally accepted, the exact amount of the water abundance is an open subject, since the values reported by different studies are not consistent, even when the same observational dataset is used (see, e.g., Table 6 in \cite{2019AJ....158..144K}).
\begin{table}
  \caption{Predicted target values from XGB on a reduced resolution experimental spectra of WASP-39b in comparison to values extracted in the literature.}
  \label{table:conc_wasp39b}
  \centering
        \begin{tabular}{ c c c c } 
        \hline
        \textbf{$X_i$ \textbackslash Model} & $\mathcal{N}$ XGB & Retrieval  & Reference \\
        \hline
        \textbf{log $X_{H_2O}$} & $-6.68$ & $-5.94\pm 0.61$ & \cite{Tsiaras_2018}\\ 
        \textbf{log $X_{CO_2}$} & $-4.50$ & $-6.59$ to $-4.16$  & \cite{Constantinou_2023}\\
        \textbf{log $X_{CH_4}$} & $-7.98$ & $<-5.3$ & \cite{Ahrer_2023}\\        
        \textbf{log $X_{CO}$}   & $-2.80$  & $-4.25$ to $-2.58$ & \cite{Constantinou_2023}\\
        \textbf{log $X_{NH_3}$} & $-10.36$  &  $<-6$ & \cite{Alderson_2023}\\  
        \hline
        \end{tabular}
\end{table}
\begin{figure}[t]
    \begin{center}
        \includegraphics[width=1.0\columnwidth]{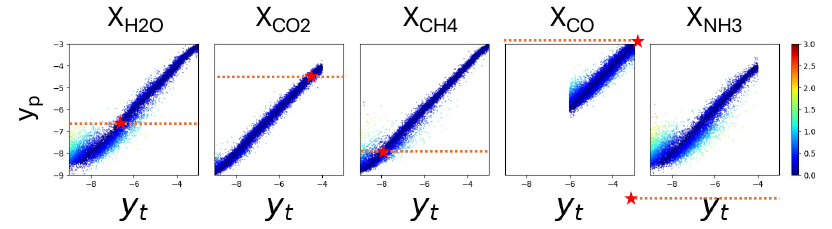}
    \end{center}
    \label{fig:wasp39b_XGB_predictions}
    \caption{XGB ($\mathcal{NL}$) model prediction for WASP 39b put in the context of the model performance on the test data set (last row in Figure~\ref{fig:nl_scatter}).}
\end{figure}
The predictions from the XGB regressor shown in Table~\ref{table:conc_wasp39b} are contained within the ranges reported in the literature. We can also correlate the results in Table~\ref{table:conc_wasp39b} to our previous discussion in Section~\ref{sec:discusstion}. In particular, let's focus on the last row in Figure~\ref{fig:nl_scatter}, which we reproduce in Figure~\ref{fig:wasp39b_XGB_predictions}, together with the predictions from Table~\ref{table:conc_wasp39b}, shown with the horizontal red dotted lines. The star marks the intersection with the diagonal $45^\circ$ line. The predicted value for $CO_2$ is well within the range for optimal model performance. The prediction for $H_2O$ is right around the ``knee'' and is expected to be relatively uncertain. The predictions for $CH_4$ and $NH_3$ are well below their respective ``knees'', which suggests that only upper limits can be placed on those abundances. Note that the predictions for the $CO$ and $NH_3$ abundances are just outside the sampled parameter range in the training data, which could happen as a result of the model extrapolation.

\section{Summary and outlook}
\label{sec:summary}

In machine learning (statistical inference) it is generally accepted that no method preforms better than {\em all} other methods under {\em all} possible circumstances (no free lunch theorem). This is why a large variety of ML methods have been developed in the ML community. When faced with a specific problem, one should perform an exhaustive and comprehensive comparison of all the different ML techniques.

In this paper, we benchmarked a number of popular regression techniques on a dataset which is taylored to the open questions in the exoplanet community. We studied both different ML models and different ways to preprocess the transit spectra data contributing to a total of 48 combinations. We showed that choosing the right preprocessing method is just as important as choosing the right model. In addition, we also discussed the uncertainties of the model predictions over the full parameter space. Our results demonstrated that ML regressors, and in particular, the XGB and SVM regressors, are capable of reliably reproducing the planetary parameters. In particular, we were able to reproduce the values for the chemical abundances which were found by previous analyses of the WASP-39b transit spectrum.

Our analysis involved standard regression methods and did not invoke any deep learning architectures, which are left for future work. The regression methods considered here have solid mathematical foundation and are easily interpretable. Future improvements could invoke physics-motivated preprocessing including the knowledge of symmetries \citep{Forestano_2023,Roman:2023ypv,Forestano:2023qcy, FORESTANO2023138266} and known degeneracies \citep{Griffith2014,Heng2017,2022ApJ...933...79W,Matchev2021analytical}. Models with a priori identified symmetries can use equivariant layers in their structure to construct more accurate and physically consistent models \citep{Gerken2023,Batzner2022,lim2022equivariant}.

\textit{\large Acknowledgements:}
The work of KTM is supported by the Shelby Endowment for Distinguished Faculty at the University of Alabama.


\software{
{\tt jupyter} \citep{Kluyver2016},
{\tt matplotlib} \citep{Hunter:2007ouj},
{\tt numpy} \citep{vanderWalt2011},
{\tt plotly} \citep{plotly},
{\tt scikit-learn} \citep{scikit-learn},
{\tt scipy} \citep{Scipy2020}.
}

\section*{Data Availability}

The data underlying this article are described in \citep{Changeat_2022,Yip_competition} and publicly available at \url{https://doi.org/10.5281/zenodo.6770103}.



\bibliography{references}{}

\begin{thebibliography}{}
\expandafter\ifx\csname natexlab\endcsname\relax\def\natexlab#1{#1}\fi
\providecommand{\url}[1]{\href{#1}{#1}}
\providecommand{\dodoi}[1]{doi:~\href{http://doi.org/#1}{\nolinkurl{#1}}}
\providecommand{\doeprint}[1]{\href{http://ascl.net/#1}{\nolinkurl{http://ascl.net/#1}}}
\providecommand{\doarXiv}[1]{\href{https://arxiv.org/abs/#1}{\nolinkurl{https://arxiv.org/abs/#1}}}

\bibitem[{Ahrer {et~al.}(2023)Ahrer, Stevenson, Mansfield, Moran, Brande, Morello, Murray, Nikolov, Petit dit de~la Roche, Schlawin, Wheatley, Zieba, Batalha, Damiano, Goyal, Lendl, Lothringer, Mukherjee, Ohno, Batalha, Battley, Bean, Beatty, Benneke, Berta-Thompson, Carter, Cubillos, Daylan, Espinoza, Gao, Gibson, Gill, Harrington, Hu, Kreidberg, Lewis, Line, L{\'o}pez-Morales, Parmentier, Powell, Sing, Tsai, Wakeford, Welbanks, Alam, Alderson, Allen, Anderson, Barstow, Bayliss, Bell, Blecic, Bryant, Burleigh, Carone, Casewell, Changeat, Chubb, Crossfield, Crouzet, Decin, D{\'e}sert, Feinstein, Flagg, Fortney, Gizis, Heng, Iro, Kempton, Kendrew, Kirk, Knutson, Komacek, Lagage, Leconte, Lustig-Yaeger, MacDonald, Mancini, May, Mayne, Miguel, Mikal-Evans, Molaverdikhani, Palle, Piaulet, Rackham, Redfield, Rogers, Roy, Rustamkulov, Shkolnik, Sotzen, Taylor, Tremblin, Tucker, Turner, de~Val-Borro, Venot, \& Zhang}]{Ahrer_2023}
Ahrer, E.-M., Stevenson, K.~B., Mansfield, M., {et~al.} 2023, Nature, 614, 653, \dodoi{10.1038/s41586-022-05590-4}

\bibitem[{{Al-Refaie} {et~al.}(2021){Al-Refaie}, {Changeat}, {Waldmann}, \& {Tinetti}}]{Taurex3}
{Al-Refaie}, A.~F., {Changeat}, Q., {Waldmann}, I.~P., \& {Tinetti}, G. 2021, \apj, 917, 37, \dodoi{10.3847/1538-4357/ac0252}

\bibitem[{Alderson {et~al.}(2023)Alderson, Wakeford, Alam, Batalha, Lothringer, Adams~Redai, Barat, Brande, Damiano, Daylan, Espinoza, Flagg, Goyal, Grant, Hu, Inglis, Lee, Mikal-Evans, Ramos-Rosado, Roy, Wallack, Batalha, Bean, Benneke, Berta-Thompson, Carter, Changeat, Col{\'o}n, Crossfield, D{\'e}sert, Foreman-Mackey, Gibson, Kreidberg, Line, L{\'o}pez-Morales, Molaverdikhani, Moran, Morello, Moses, Mukherjee, Schlawin, Sing, Stevenson, Taylor, Aggarwal, Ahrer, Allen, Barstow, Bell, Blecic, Casewell, Chubb, Crouzet, Cubillos, Decin, Feinstein, Fortney, Harrington, Heng, Iro, Kempton, Kirk, Knutson, Krick, Leconte, Lendl, MacDonald, Mancini, Mansfield, May, Mayne, Miguel, Nikolov, Ohno, Palle, Parmentier, Petit dit de~la Roche, Piaulet, Powell, Rackham, Redfield, Rogers, Rustamkulov, Tan, Tremblin, Tsai, Turner, de~Val-Borro, Venot, Welbanks, Wheatley, \& Zhang}]{Alderson_2023}
Alderson, L., Wakeford, H.~R., Alam, M.~K., {et~al.} 2023, Nature, 614, 664, \dodoi{10.1038/s41586-022-05591-3}

\bibitem[{{Ard{\'e}vol Mart{\'\i}nez} {et~al.}(2024){Ard{\'e}vol Mart{\'\i}nez}, {Min}, {Huppenkothen}, {Kamp}, \& {Palmer}}]{2024A&A...681L..14A}
{Ard{\'e}vol Mart{\'\i}nez}, F., {Min}, M., {Huppenkothen}, D., {Kamp}, I., \& {Palmer}, P.~I. 2024, \aap, 681, L14, \dodoi{10.1051/0004-6361/202348367}

\bibitem[{{Ardevol Martinez} {et~al.}(2022){Ardevol Martinez}, {Min}, {Kamp}, \& {Palmer}}]{Ardevol2022}
{Ardevol Martinez}, F., {Min}, M., {Kamp}, I., \& {Palmer}, P.~I. 2022, arXiv e-prints, arXiv:2203.01236.
\newblock \doarXiv{2203.01236}

\bibitem[{{Aubin} {et~al.}(2023){Aubin}, {Cuesta-Lazaro}, {Tregidga}, {Via{\~n}a}, {Garraffo}, {Gordon}, {L{\'o}pez-Morales}, {Hargreaves}, {Makhnev}, {Drake}, {Finkbeiner}, \& {Cargile}}]{2023arXiv230909337A}
{Aubin}, M., {Cuesta-Lazaro}, C., {Tregidga}, E., {et~al.} 2023, arXiv e-prints, arXiv:2309.09337, \dodoi{10.48550/arXiv.2309.09337}

\bibitem[{Batzner {et~al.}(2022)Batzner, Musaelian, Sun, Geiger, Mailoa, Kornbluth, Molinari, Smidt, \& Kozinsky}]{Batzner2022}
Batzner, S., Musaelian, A., Sun, L., {et~al.} 2022, Nature Communications, 13, 2453, \dodoi{10.1038/s41467-022-29939-5}

\bibitem[{Benneke \& Seager(2012)}]{Benneke_2012}
Benneke, B., \& Seager, S. 2012, The Astrophysical Journal, 753, 100, \dodoi{10.1088/0004-637X/753/2/100}

\bibitem[{Bishop(2006)}]{Bishop_ML}
Bishop, C.~M. 2006, Pattern Recognition and Machine Learning (Information Science and Statistics) (Berlin, Heidelberg: Springer-Verlag)

\bibitem[{{Blecic} {et~al.}(2021){Blecic}, {Harrington}, {Cubillos}, {Bowman}, {Rojo}, {Stemm}, {Challener}, {Himes}, {Foster}, {Dobbs-Dixon}, {Foster}, {Lust}, {Blumenthal}, {Bruce}, \& {Loredo}}]{Blecic2021}
{Blecic}, J., {Harrington}, J., {Cubillos}, P.~E., {et~al.} 2021, arXiv e-prints, arXiv:2104.12525.
\newblock \doarXiv{2104.12525}

\bibitem[{Breiman {et~al.}(1984)Breiman, Friedman, Olshen, \& J.}]{Breiman_1984}
Breiman, L., Friedman, J., Olshen, R.~A., \& J., S.~C. 1984, Classification and Regression Trees (1st ed.). (Chapman and Hall/CRC.), \dodoi{https://doi.org/10.1201/9781315139470}

\bibitem[{{Changeat} {et~al.}(2020){Changeat}, {Al-Refaie}, {Mugnai}, {Edwards}, {Waldmann}, {Pascale}, \& {Tinetti}}]{2020AJ....160...80C}
{Changeat}, Q., {Al-Refaie}, A., {Mugnai}, L.~V., {et~al.} 2020, \aj, 160, 80, \dodoi{10.3847/1538-3881/ab9a53}

\bibitem[{{Changeat} \& {Yip}(2023)}]{Changeat_2022}
{Changeat}, Q., \& {Yip}, K.~H. 2023, RAS Techniques and Instruments, 2, 45, \dodoi{10.1093/rasti/rzad001}

\bibitem[{{Charbonneau} {et~al.}(2000){Charbonneau}, {Brown}, {Latham}, \& {Mayor}}]{Charbonneau2000}
{Charbonneau}, D., {Brown}, T.~M., {Latham}, D.~W., \& {Mayor}, M. 2000, \apjl, 529, L45, \dodoi{10.1086/312457}

\bibitem[{Chen {et~al.}(2006)Chen, Fan, \& Lin}]{Chen2006}
Chen, P.-H., Fan, R.-E., \& Lin, C.-J. 2006, IEEE Transactions on Neural Networks, 17, 893, \dodoi{10.1109/TNN.2006.875973}

\bibitem[{Chen \& Guestrin(2016)}]{Chen_2016}
Chen, T., \& Guestrin, C. 2016, in Proceedings of the 22nd ACM SIGKDD International Conference on Knowledge Discovery and Data Mining, KDD '16 (New York, NY, USA: Association for Computing Machinery), 785–794, \dodoi{10.1145/2939672.2939785}

\bibitem[{{Cobb} {et~al.}(2019){Cobb}, {Himes}, {Soboczenski}, {Zorzan}, {O'Beirne}, {G{\"u}ne{\c{s}} Baydin}, {Gal}, {Domagal-Goldman}, {Arney}, {Angerhausen}, \& {2018 NASA FDL Astrobiology Team}}]{Cobb2019}
{Cobb}, A.~D., {Himes}, M.~D., {Soboczenski}, F., {et~al.} 2019, \aj, 158, 33, \dodoi{10.3847/1538-3881/ab2390}

\bibitem[{Constantinou {et~al.}(2023)Constantinou, Madhusudhan, \& Gandhi}]{Constantinou_2023}
Constantinou, S., Madhusudhan, N., \& Gandhi, S. 2023, The Astrophysical Journal Letters, 943, L10, \dodoi{10.3847/2041-8213/acaead}

\bibitem[{Cubillos {et~al.}(2013)Cubillos, Harrington, Madhusudhan, Stevenson, Hardy, Blecic, Anderson, Hardin, \& Campo}]{Cubillos_2013}
Cubillos, P., Harrington, J., Madhusudhan, N., {et~al.} 2013, The Astrophysical Journal, 768, 42, \dodoi{10.1088/0004-637X/768/1/42}

\bibitem[{{Cubillos} {et~al.}(2021){Cubillos}, {Harrington}, {Blecic}, {Himes}, {Rojo}, {Loredo}, {Lust}, {Challener}, {Foster}, {Stemm}, {Foster}, \& {Blumenthal}}]{Cubillos2021}
{Cubillos}, P.~E., {Harrington}, J., {Blecic}, J., {et~al.} 2021, arXiv e-prints, arXiv:2104.12524.
\newblock \doarXiv{2104.12524}

\bibitem[{Dietterich(1998)}]{Dietterich_1998}
Dietterich, T.~G. 1998, Neural Computation, 10, 1895, \dodoi{10.1162/089976698300017197}

\bibitem[{{Edwards} {et~al.}(2019{\natexlab{a}}){Edwards}, {Mugnai}, {Tinetti}, {Pascale}, \& {Sarkar}}]{2019AJ....157..242E}
{Edwards}, B., {Mugnai}, L., {Tinetti}, G., {Pascale}, E., \& {Sarkar}, S. 2019{\natexlab{a}}, \aj, 157, 242, \dodoi{10.3847/1538-3881/ab1cb9}

\bibitem[{{Edwards} \& {Tinetti}(2022)}]{2022AJ....164...15E}
{Edwards}, B., \& {Tinetti}, G. 2022, \aj, 164, 15, \dodoi{10.3847/1538-3881/ac6bf9}

\bibitem[{{Edwards} {et~al.}(2019{\natexlab{b}}){Edwards}, {Rice}, {Zingales}, {Tessenyi}, {Waldmann}, {Tinetti}, {Pascale}, {Savini}, \& {Sarkar}}]{2019ExA....47...29E}
{Edwards}, B., {Rice}, M., {Zingales}, T., {et~al.} 2019{\natexlab{b}}, Experimental Astronomy, 47, 29, \dodoi{10.1007/s10686-018-9611-4}

\bibitem[{{Falco} {et~al.}(2022){Falco}, {Zingales}, {Pluriel}, \& {Leconte}}]{2022A&A...658A..41F}
{Falco}, A., {Zingales}, T., {Pluriel}, W., \& {Leconte}, J. 2022, \aap, 658, A41, \dodoi{10.1051/0004-6361/202141940}

\bibitem[{Fan {et~al.}(2005)Fan, Chen, \& Lin}]{Fan2005}
Fan, R.-E., Chen, P.-H., \& Lin, C.-J. 2005, J. Mach. Learn. Res., 6, 1889–1918

\bibitem[{{Fisher} {et~al.}(2020){Fisher}, {Hoeijmakers}, {Kitzmann}, {M{\'a}rquez-Neila}, {Grimm}, {Sznitman}, \& {Heng}}]{Fisher2020}
{Fisher}, C., {Hoeijmakers}, H.~J., {Kitzmann}, D., {et~al.} 2020, \aj, 159, 192, \dodoi{10.3847/1538-3881/ab7a92}

\bibitem[{Forestano {et~al.}(2023{\natexlab{a}})Forestano, Matchev, Matcheva, Roman, Unlu, \& Verner}]{Forestano_2023}
Forestano, R.~T., Matchev, K.~T., Matcheva, K., {et~al.} 2023{\natexlab{a}}, Machine Learning: Science and Technology, 4, 025027, \dodoi{10.1088/2632-2153/acd989}

\bibitem[{Forestano {et~al.}(2023{\natexlab{b}})Forestano, Matchev, Matcheva, Roman, Unlu, \& Verner}]{Forestano:2023qcy}
---. 2023{\natexlab{b}}, Physics Letters B, 844, \dodoi{10.1016/j.physletb.2023.138086}

\bibitem[{Forestano {et~al.}(2023{\natexlab{c}})Forestano, Matchev, Matcheva, Roman, Unlu, \& Verner}]{FORESTANO2023138266}
---. 2023{\natexlab{c}}, Physics Letters B, 847, 138266, \dodoi{https://doi.org/10.1016/j.physletb.2023.138266}

\bibitem[{Forestano {et~al.}(2023{\natexlab{d}})Forestano, Matchev, Matcheva, \& Unlu}]{Forestano_anomaly_2023}
Forestano, R.~T., Matchev, K.~T., Matcheva, K., \& Unlu, E.~B. 2023{\natexlab{d}}, The Astrophysical Journal, 958, 106, \dodoi{10.3847/1538-4357/ad0047}

\bibitem[{Friedman {et~al.}(2000)Friedman, Hastie, \& Tibshirani}]{Friedman_2000}
Friedman, J., Hastie, T., \& Tibshirani, R. 2000, The Annals of Statistics, 28, 337 , \dodoi{10.1214/aos/1016218223}

\bibitem[{Gandhi \& Madhusudhan(2018)}]{Gandhi_2018}
Gandhi, S., \& Madhusudhan, N. 2018, Monthly Notices of the Royal Astronomical Society, 474, 271–, \dodoi{https://doi.org/10.1093/mnras/stx2748}

\bibitem[{Gerken {et~al.}(2023)Gerken, Aronsson, Carlsson, Linander, Ohlsson, Petersson, \& Persson}]{Gerken2023}
Gerken, J.~E., Aronsson, J., Carlsson, O., {et~al.} 2023, Artificial Intelligence Review, 56, 14605, \dodoi{10.1007/s10462-023-10502-7}

\bibitem[{{Greene} {et~al.}(2016){Greene}, {Line}, {Montero}, {Fortney}, {Lustig-Yaeger}, \& {Luther}}]{2016ApJ...817...17G}
{Greene}, T.~P., {Line}, M.~R., {Montero}, C., {et~al.} 2016, \apj, 817, 17, \dodoi{10.3847/0004-637X/817/1/17}

\bibitem[{{Griffith}(2014)}]{Griffith2014}
{Griffith}, C.~A. 2014, Philosophical Transactions of the Royal Society of London Series A, 372, 20130086, \dodoi{10.1098/rsta.2013.0086}

\bibitem[{{Guzm{\'a}n-Mesa} {et~al.}(2020){Guzm{\'a}n-Mesa}, {Kitzmann}, {Fisher}, {Burgasser}, {Hoeijmakers}, {M{\'a}rquez-Neila}, {Grimm}, {Mandell}, {Sznitman}, \& {Heng}}]{Guzman2020}
{Guzm{\'a}n-Mesa}, A., {Kitzmann}, D., {Fisher}, C., {et~al.} 2020, \aj, 160, 15, \dodoi{10.3847/1538-3881/ab9176}

\bibitem[{{Haldemann} {et~al.}(2023){Haldemann}, {Ksoll}, {Walter}, {Alibert}, {Klessen}, {Benz}, {Koethe}, {Ardizzone}, \& {Rother}}]{Haldemann2022}
{Haldemann}, J., {Ksoll}, V., {Walter}, D., {et~al.} 2023, \aap, 672, A180, \dodoi{10.1051/0004-6361/202243230}

\bibitem[{{Harrington} {et~al.}(2021){Harrington}, {Himes}, {Cubillos}, {Blecic}, {Rojo}, {Challener}, {Lust}, {Bowman}, {Blumenthal}, {Dobbs-Dixon}, {Foster}, {Foster}, {Green}, {Loredo}, {McIntyre}, {Stemm}, \& {Wright}}]{Harrington2021}
{Harrington}, J., {Himes}, M.~D., {Cubillos}, P.~E., {et~al.} 2021, arXiv e-prints, arXiv:2104.12522.
\newblock \doarXiv{2104.12522}

\bibitem[{Hastie {et~al.}(2001)Hastie, Tibshirani, \& Friedman}]{Hastie2001statisticallearning}
Hastie, T., Tibshirani, R., \& Friedman, J. 2001, The Elements of Statistical Learning, Springer Series in Statistics (New York, NY, USA: Springer New York Inc.)

\bibitem[{{Hayes} {et~al.}(2020){Hayes}, {Kerins}, {Awiphan}, {McDonald}, {Morgan}, {Chuanraksasat}, {Komonjinda}, {Sanguansak}, {Kittara}, \& {SPEARNet Collaboration}}]{Hayes2020}
{Hayes}, J.~J.~C., {Kerins}, E., {Awiphan}, S., {et~al.} 2020, \mnras, 494, 4492, \dodoi{10.1093/mnras/staa978}

\bibitem[{{Heng} \& {Kitzmann}(2017)}]{Heng2017}
{Heng}, K., \& {Kitzmann}, D. 2017, \mnras, 470, 2972, \dodoi{10.1093/mnras/stx1453}

\bibitem[{Hunter(2007)}]{Hunter:2007ouj}
Hunter, J.~D. 2007, Comput. Sci. Eng., 9, 90, \dodoi{10.1109/MCSE.2007.55}

\bibitem[{Inc.(2015)}]{plotly}
Inc., P.~T. 2015, Collaborative data science,  Montreal, QC: Plotly Technologies Inc.
\newblock \url{https://plot.ly}

\bibitem[{{Kirk} {et~al.}(2019){Kirk}, {L{\'o}pez-Morales}, {Wheatley}, {Weaver}, {Skillen}, {Louden}, {McCormac}, \& {Espinoza}}]{2019AJ....158..144K}
{Kirk}, J., {L{\'o}pez-Morales}, M., {Wheatley}, P.~J., {et~al.} 2019, \aj, 158, 144, \dodoi{10.3847/1538-3881/ab397d}

\bibitem[{{Kitzmann} {et~al.}(2020){Kitzmann}, {Heng}, {Oreshenko}, {Grimm}, {Apai}, {Bowler}, {Burgasser}, \& {Marley}}]{Kitzmann2020}
{Kitzmann}, D., {Heng}, K., {Oreshenko}, M., {et~al.} 2020, \apj, 890, 174, \dodoi{10.3847/1538-4357/ab6d71}

\bibitem[{Kluyver {et~al.}(2016)Kluyver, Ragan-Kelley, P{\'e}rez, Granger, Bussonnier, Frederic, Kelley, Hamrick, Grout, Corlay, Ivanov, Avila, Abdalla, Willing, \& development team}]{Kluyver2016}
Kluyver, T., Ragan-Kelley, B., P{\'e}rez, F., {et~al.} 2016, in Positioning and Power in Academic Publishing: Players, Agents and Agendas, ed. F.~Loizides \& B.~Scmidt (IOS Press), 87--90.
\newblock \url{https://eprints.soton.ac.uk/403913/}

\bibitem[{Lim \& Nelson(2022)}]{lim2022equivariant}
Lim, L.-H., \& Nelson, B.~J. 2022, What is an equivariant neural network?
\newblock \doarXiv{2205.07362}

\bibitem[{Line {et~al.}(2013)Line, Wolf, Zhang, Knutson, Kammer, Ellison, Deroo, Crisp, \& Yung}]{Line_2013}
Line, M.~R., Wolf, A.~S., Zhang, X., {et~al.} 2013, The Astrophysical Journal, 775, 137, \dodoi{10.1088/0004-637X/775/2/137}

\bibitem[{{Lueber} {et~al.}(2025){Lueber}, {Karchev}, {Fisher}, {Heim}, {Trotta}, \& {Heng}}]{2025ApJ...984L..32L}
{Lueber}, A., {Karchev}, K., {Fisher}, C., {et~al.} 2025, \apjl, 984, L32, \dodoi{10.3847/2041-8213/adc7aa}

\bibitem[{{MacDonald} \& {Lewis}(2022)}]{2022ApJ...929...20M}
{MacDonald}, R.~J., \& {Lewis}, N.~K. 2022, \apj, 929, 20, \dodoi{10.3847/1538-4357/ac47fe}

\bibitem[{Madhusudhan(2018)}]{Madhusudhan2018}
Madhusudhan, N. 2018, Atmospheric Retrieval of Exoplanets, ed. H.~J. Deeg \& J.~A. Belmonte (Cham: Springer International Publishing), 1--30, \dodoi{10.1007/978-3-319-30648-3_104-1}

\bibitem[{Madhusudhan \& Seager(2009)}]{Madhusudhan_2009}
Madhusudhan, N., \& Seager, S. 2009, The Astrophysical Journal, 707, 24, \dodoi{10.1088/0004-637X/707/1/24}

\bibitem[{{M{\'a}rquez-Neila} {et~al.}(2018){M{\'a}rquez-Neila}, {Fisher}, {Sznitman}, \& {Heng}}]{Marquez2018}
{M{\'a}rquez-Neila}, P., {Fisher}, C., {Sznitman}, R., \& {Heng}, K. 2018, Nature Astronomy, 2, 719, \dodoi{10.1038/s41550-018-0504-2}

\bibitem[{{Matchev} {et~al.}(2022{\natexlab{a}}){Matchev}, {Matcheva}, \& {Roman}}]{2022PSJ.....3..205M}
{Matchev}, K.~T., {Matcheva}, K., \& {Roman}, A. 2022{\natexlab{a}}, The Planetary Sciences Journal, 3, 205, \dodoi{10.3847/PSJ/ac880b}

\bibitem[{{Matchev} {et~al.}(2022{\natexlab{b}}){Matchev}, {Matcheva}, \& {Roman}}]{2022ApJ...939...95M}
---. 2022{\natexlab{b}}, \apj, 939, 95, \dodoi{10.3847/1538-4357/ac82f3}

\bibitem[{{Matchev} {et~al.}(2022{\natexlab{c}}){Matchev}, {Matcheva}, \& {Roman}}]{Matchev2021analytical}
---. 2022{\natexlab{c}}, \apj, 930, 33, \dodoi{10.3847/1538-4357/ac610c}

\bibitem[{Mugnai {et~al.}(2020)Mugnai, Pascale, Edwards, Papageorgiou, \& Sarkar}]{Mugnai_2020}
Mugnai, L.~V., Pascale, E., Edwards, B., Papageorgiou, A., \& Sarkar, S. 2020, Experimental Astronomy, 50, 303–328, \dodoi{10.1007/s10686-020-09676-7}

\bibitem[{{Nixon} \& {Madhusudhan}(2020)}]{Nixon2020}
{Nixon}, M.~C., \& {Madhusudhan}, N. 2020, \mnras, 496, 269, \dodoi{10.1093/mnras/staa1150}

\bibitem[{Oreshenko {et~al.}(2017)Oreshenko, Lavie, Grimm, Tsai, Malik, Demory, Mordasini, Alibert, Benz, Quanz, Trotta, \& Heng}]{Oreshenko_2017}
Oreshenko, M., Lavie, B., Grimm, S.~L., {et~al.} 2017, The Astrophysical Journal Letters, 847, L3, \dodoi{10.3847/2041-8213/aa8acf}

\bibitem[{{Oreshenko} {et~al.}(2020){Oreshenko}, {Kitzmann}, {M{\'a}rquez-Neila}, {Malik}, {Bowler}, {Burgasser}, {Sznitman}, {Fisher}, \& {Heng}}]{Oreshenko2020}
{Oreshenko}, M., {Kitzmann}, D., {M{\'a}rquez-Neila}, P., {et~al.} 2020, \aj, 159, 6, \dodoi{10.3847/1538-3881/ab5955}

\bibitem[{Pedregosa {et~al.}(2011)Pedregosa, Varoquaux, Gramfort, Michel, Thirion, Grisel, Blondel, Prettenhofer, Weiss, Dubourg, Vanderplas, Passos, Cournapeau, Brucher, Perrot, \& Duchesnay}]{scikit-learn}
Pedregosa, F., Varoquaux, G., Gramfort, A., {et~al.} 2011, Journal of Machine Learning Research, 12, 2825

\bibitem[{{Pluriel}(2023)}]{2023RemS...15..635P}
{Pluriel}, W. 2023, Remote Sensing, 15, 635, \dodoi{10.3390/rs15030635}

\bibitem[{{Pluriel} {et~al.}(2022){Pluriel}, {Leconte}, {Parmentier}, {Zingales}, {Falco}, {Selsis}, \& {Bord{\'e}}}]{2022A&A...658A..42P}
{Pluriel}, W., {Leconte}, J., {Parmentier}, V., {et~al.} 2022, \aap, 658, A42, \dodoi{10.1051/0004-6361/202141943}

\bibitem[{Powell {et~al.}(2024)Powell, Feinstein, Lee, Zhang, Tsai, Taylor, Kirk, Bell, Barstow, Gao, Bean, Blecic, Chubb, Crossfield, Jordan, Kitzmann, Moran, Morello, Moses, Welbanks, Yang, Zhang, Ahrer, Bello-Arufe, Brande, Casewell, Crouzet, Cubillos, Demory, Dyrek, Flagg, Hu, Inglis, Jones, Kreidberg, L{\'o}pez-Morales, Lagage, Meier~Vald{\'e}s, Miguel, Parmentier, Piette, Rackham, Radica, Redfield, Stevenson, Wakeford, Aggarwal, Alam, Batalha, Batalha, Benneke, Berta-Thompson, Brady, Caceres, Carter, D{\'e}sert, Harrington, Iro, Line, Lothringer, MacDonald, Mancini, Molaverdikhani, Mukherjee, Nixon, Oza, Palle, Rustamkulov, Sing, Steinrueck, Venot, Wheatley, \& Yurchenko}]{2024Powell}
Powell, D., Feinstein, A.~D., Lee, E. K.~H., {et~al.} 2024, Nature, 626, 979, \dodoi{10.1038/s41586-024-07040-9}

\bibitem[{Roman {et~al.}(2023)Roman, Forestano, Matchev, Matcheva, \& Unlu}]{Roman:2023ypv}
Roman, A., Forestano, R.~T., Matchev, K.~T., Matcheva, K., \& Unlu, E.~B. 2023, Symmetry, 15, \dodoi{10.3390/sym15071352}

\bibitem[{Rustamkulov {et~al.}(2023)Rustamkulov, Sing, Mukherjee, May, Kirk, Schlawin, Line, Piaulet, Carter, Batalha, Goyal, L{\'o}pez-Morales, Lothringer, MacDonald, Moran, Stevenson, Wakeford, Espinoza, Bean, Batalha, Benneke, Berta-Thompson, Crossfield, Gao, Kreidberg, Powell, Cubillos, Gibson, Leconte, Molaverdikhani, Nikolov, Parmentier, Roy, Taylor, Turner, Wheatley, Aggarwal, Ahrer, Alam, Alderson, Allen, Banerjee, Barat, Barrado, Barstow, Bell, Blecic, Brande, Casewell, Changeat, Chubb, Crouzet, Daylan, Decin, D{\'e}sert, Mikal-Evans, Feinstein, Flagg, Fortney, Harrington, Heng, Hong, Hu, Iro, Kataria, Kempton, Krick, Lendl, Lillo-Box, Louca, Lustig-Yaeger, Mancini, Mansfield, Mayne, Miguel, Morello, Ohno, Palle, Petit dit de~la Roche, Rackham, Radica, Ramos-Rosado, Redfield, Rogers, Shkolnik, Southworth, Teske, Tremblin, Tucker, Venot, Waalkes, Welbanks, Zhang, \& Zieba}]{2023Rustamkulov}
Rustamkulov, Z., Sing, D.~K., Mukherjee, S., {et~al.} 2023, Nature, 614, 659, \dodoi{10.1038/s41586-022-05677-y}

\bibitem[{{Schneider}(1994)}]{Schneider1994}
{Schneider}, J. 1994, \apss, 212, 321, \dodoi{10.1007/BF00984535}

\bibitem[{{Soboczenski} {et~al.}(2018){Soboczenski}, {Himes}, {O'Beirne}, {Zorzan}, {Gunes Baydin}, {Cobb}, {Gal}, {Angerhausen}, {Mascaro}, {Arney}, \& {Domagal-Goldman}}]{Soboczenski2018}
{Soboczenski}, F., {Himes}, M.~D., {O'Beirne}, M.~D., {et~al.} 2018, arXiv e-prints, arXiv:1811.03390.
\newblock \doarXiv{1811.03390}

\bibitem[{{Tinetti} {et~al.}(2021){Tinetti}, {Eccleston}, {Haswell}, {Lagage}, {Leconte}, {L{\"u}ftinger}, {Micela}, {Min}, {Pilbratt}, {Puig}, {Swain}, {Testi}, {Turrini}, {Vandenbussche}, {Rosa Zapatero Osorio}, {Aret}, {Beaulieu}, {Buchhave}, {Ferus}, {Griffin}, {Guedel}, {Hartogh}, {Machado}, {Malaguti}, {Pall{\'e}}, {Rataj}, {Ray}, {Ribas}, {Szab{\'o}}, {Tan}, {Werner}, {Ratti}, {Scharmberg}, {Salvignol}, {Boudin}, {Halain}, {Haag}, {Crouzet}, {Kohley}, {Symonds}, {Renk}, {Caldwell}, {Abreu}, {Alonso}, {Amiaux}, {Berth{\'e}}, {Bishop}, {Bowles}, {Carmona}, {Coffey}, {Colom{\'e}}, {Crook}, {D{\'e}sjonqueres}, {D{\'\i}az}, {Drummond}, {Focardi}, {G{\'o}mez}, {Holmes}, {Krijger}, {Kovacs}, {Hunt}, {Machado}, {Morgante}, {Ollivier}, {Ottensamer}, {Pace}, {Pagano}, {Pascale}, {Pearson}, {M{\o}ller Pedersen}, {Pniel}, {Roose}, {Savini}, {Stamper}, {Szirovicza}, {Szoke}, {Tosh}, {Vilardell}, {Barstow}, {Borsato}, {Casewell}, {Changeat}, {Charnay}, {Civi{\v{s}}}, {Coud{\'e} du Foresto}, {Coustenis}, {Cowan},
  {Danielski}, {Demangeon}, {Drossart}, {Edwards}, {Gilli}, {Encrenaz}, {Kiss}, {Kokori}, {Ikoma}, {Morales}, {Mendon{\c{c}}a}, {Moneti}, {Mugnai}, {Garc{\'\i}a Mu{\~n}oz}, {Helled}, {Kama}, {Miguel}, {Nikolaou}, {Pagano}, {Panic}, {Rengel}, {Rickman}, {Rocchetto}, {Sarkar}, {Selsis}, {Tennyson}, {Tsiaras}, {Venot}, {Vida}, {Waldmann}, {Yurchenko}, {Szab{\'o}}, {Zellem}, {Al-Refaie}, {Perez Alvarez}, {Anisman}, {Arhancet}, {Ateca}, {Baeyens}, {Barnes}, {Bell}, {Benatti}, {Biazzo}, {B{\l}{\k{e}}cka}, {Bonomo}, {Bosch}, {Bossini}, {Bourgalais}, {Brienza}, {Brucalassi}, {Bruno}, {Caines}, {Calcutt}, {Campante}, {Canestrari}, {Cann}, {Casali}, {Casas}, {Cassone}, {Cara}, {Carmona}, {Carone}, {Carrasco}, {Changeat}, {Chioetto}, {Cortecchia}, {Czupalla}, {Chubb}, {Ciaravella}, {Claret}, {Claudi}, {Codella}, {Garcia Comas}, {Cracchiolo}, {Cubillos}, {Da Peppo}, {Decin}, {Dejabrun}, {Delgado-Mena}, {Di Giorgio}, {Diolaiti}, {Dorn}, {Doublier}, {Doumayrou}, {Dransfield}, {Dumaye}, {Dunford}, {Jimenez Escobar}, {Van
  Eylen}, {Farina}, {Fedele}, {Fern{\'a}ndez}, {Fleury}, {Fonte}, {Fontignie}, {Fossati}, {Funke}, {Galy}, {Garai}, {Garc{\'\i}a}, {Garc{\'\i}a-Rigo}, {Garufi}, {Germano Sacco}, {Giacobbe}, {G{\'o}mez}, {Gonzalez}, {Gonzalez-Galindo}, {Grassi}, {Griffith}, {Guarcello}, {Goujon}, {Gressier}, {Grzegorczyk}, {Guillot}, {Guilluy}, {Hargrave}, {Hellin}, {Herrero}, {Hills}, {Horeau}, {Ito}, {Jessen}, {Kabath}, {K{\'a}lm{\'a}n}, {Kawashima}, {Kimura}, {Kn{\'\i}{\v{z}}ek}, {Kreidberg}, {Kruid}, {Kruijssen}, {Kubel{\'\i}k}, {Lara}, {Lebonnois}, {Lee}, {Lefevre}, {Lichtenberg}, {Locci}, {Lombini}, {Sanchez Lopez}, {Lorenzani}, {MacDonald}, {Magrini}, {Maldonado}, {Marcq}, {Migliorini}, {Modirrousta-Galian}, {Molaverdikhani}, {Molinari}, {Molli{\`e}re}, {Moreau}, {Morello}, {Morinaud}, {Morvan}, {Moses}, {Mouzali}, {Nakhjiri}, {Naponiello}, {Narita}, {Nascimbeni}, {Nikolaou}, {Noce}, {Oliva}, {Palladino}, {Papageorgiou}, {Parmentier}, {Peres}, {P{\'e}rez}, {Perez-Hoyos}, {Perger}, {Cecchi Pestellini}, {Petralia},
  {Philippon}, {Piccialli}, {Pignatari}, {Piotto}, {Podio}, {Polenta}, {Preti}, {Pribulla}, {Lopez Puertas}, {Rainer}, {Reess}, {Rimmer}, {Robert}, {Rosich}, {Rossi}, {Rust}, {Saleh}, {Sanna}, {Schisano}, {Schreiber}, {Schwartz}, {Scippa}, {Seli}, {Shibata}, {Simpson}, {Shorttle}, {Skaf}, {Skup}, {Sobiecki}, {Sousa}, {Sozzetti}, {{\v{S}}poner}, {Steiger}, {Tanga}, {Tackley}, {Taylor}, {Tecza}, {Terenzi}, {Tremblin}, {Tozzi}, {Triaud}, {Trompet}, {Tsai}, {Tsantaki}, {Valencia}, {Carine Vandaele}, {Van der Swaelmen}, {Adibekyan}, {Vasisht}, {Vazan}, {Del Vecchio}, {Waltham}, {Wawer}, {Widemann}, {Wolkenberg}, {Hou Yip}, {Yung}, {Zilinskas}, {Zingales}, \& {Zuppella}}]{2021arXiv210404824T}
{Tinetti}, G., {Eccleston}, P., {Haswell}, C., {et~al.} 2021, arXiv e-prints, arXiv:2104.04824, \dodoi{10.48550/arXiv.2104.04824}

\bibitem[{Trevor~Hastie(2009)}]{Hastie_2009}
Trevor~Hastie, Robert~Tibshirani, J.~F. 2009, The Elements of Statistical Learning (New York, New York: Springer)

\bibitem[{{Tsiaras} {et~al.}(2018){Tsiaras}, {Waldmann}, {Zingales}, {Rocchetto}, {Morello}, {Damiano}, {Karpouzas}, {Tinetti}, {McKemmish}, {Tennyson}, \& {Yurchenko}}]{Tsiaras_2018}
{Tsiaras}, A., {Waldmann}, I.~P., {Zingales}, T., {et~al.} 2018, The Astronomical Journal, 155, 156, \dodoi{10.3847/1538-3881/aaaf75}

\bibitem[{Unlu {et~al.}(2023)Unlu, Forestano, Matchev, \& Matcheva}]{unlu2023reproducing}
Unlu, E.~B., Forestano, R.~T., Matchev, K.~T., \& Matcheva, K. 2023, Reproducing Bayesian Posterior Distributions for Exoplanet Atmospheric Parameter Retrievals with a Machine Learning Surrogate Model.
\newblock \doarXiv{2310.10521}

\bibitem[{van~der Walt {et~al.}(2011)van~der Walt, Colbert, \& Varoquaux}]{vanderWalt2011}
van~der Walt, S., Colbert, S.~C., \& Varoquaux, G. 2011, Computing in Science Engineering, 13, 22, \dodoi{10.1109/MCSE.2011.37}

\bibitem[{Virtanen {et~al.}(2020)Virtanen, Gommers, Oliphant, Haberland, Reddy, Cournapeau, Burovski, Peterson, Weckesser, Bright, {van der Walt}, Brett, Wilson, Millman, Mayorov, Nelson, Jones, Kern, Larson, Carey, Polat, Feng, Moore, {VanderPlas}, Laxalde, Perktold, Cimrman, Henriksen, Quintero, Harris, Archibald, Ribeiro, Pedregosa, {van Mulbregt}, \& {SciPy 1.0 Contributors}}]{Scipy2020}
Virtanen, P., Gommers, R., Oliphant, T.~E., {et~al.} 2020, Nature Methods, 17, 261, \dodoi{10.1038/s41592-019-0686-2}

\bibitem[{{Waldmann}(2016)}]{Waldmann2016ApJ}
{Waldmann}, I.~P. 2016, \apj, 820, 107, \dodoi{10.3847/0004-637X/820/2/107}

\bibitem[{Waldmann {et~al.}(2015)Waldmann, Tinetti, Rocchetto, Barton, Yurchenko, \& Tennyson}]{Waldmann_2015}
Waldmann, I.~P., Tinetti, G., Rocchetto, M., {et~al.} 2015, The Astrophysical Journal, 802, 107, \dodoi{10.1088/0004-637X/802/2/107}

\bibitem[{Wegelin(2000)}]{Wegelin_2000}
Wegelin, J. 2000, Technical report

\bibitem[{{Welbanks} \& {Madhusudhan}(2021)}]{Welbanks2021}
{Welbanks}, L., \& {Madhusudhan}, N. 2021, \apj, 913, 114, \dodoi{10.3847/1538-4357/abee94}

\bibitem[{{Welbanks} \& {Madhusudhan}(2022)}]{2022ApJ...933...79W}
---. 2022, \apj, 933, 79, \dodoi{10.3847/1538-4357/ac6df1}

\bibitem[{{Yip} {et~al.}(2024){Yip}, {Changeat}, {Al-Refaie}, \& {Waldmann}}]{2024ApJ...961...30Y}
{Yip}, K.~H., {Changeat}, Q., {Al-Refaie}, A., \& {Waldmann}, I.~P. 2024, \apj, 961, 30, \dodoi{10.3847/1538-4357/ad063f}

\bibitem[{Yip {et~al.}(2021)Yip, Changeat, Nikolaou, Morvan, Edwards, Waldmann, \& Tinetti}]{Yip_2021}
Yip, K.~H., Changeat, Q., Nikolaou, N., {et~al.} 2021, The Astronomical Journal, 162, 195, \dodoi{10.3847/1538-3881/ac1744}

\bibitem[{{Yip} {et~al.}(2022){Yip}, {Waldmann}, {Changeat}, {Morvan}, {Al-Refaie}, {Edwards}, {Nikolaou}, {Tsiaras}, {Alves de Oliveira}, {Lagage}, {Jenner}, {Y-K. Cho}, {Thiyagalingam}, \& {Tinetti}}]{Yip_competition}
{Yip}, K.~H., {Waldmann}, I.~P., {Changeat}, Q., {et~al.} 2022, arXiv e-prints, arXiv:2206.14642.
\newblock \doarXiv{2206.14642}

\bibitem[{Yip {et~al.}(2023)Yip, Changeat, Waldmann, Unlu, Forestano, Roman, Matcheva, Matchev, Stefanov, Podsztavek, Morvan, Nikolaou, Al-Refaie, Jenner, Johnson, Tsiaras, Edwards, Catarina, Thiyagalingam, Lagage, Cho, \& Tinetti}]{Ariel2022}
Yip, K.~H., Changeat, Q., Waldmann, I., {et~al.} 2023, Proceedings of Machine Learning Research, 220, 1.
\newblock \url{https://proceedings.mlr.press/v220/yip23a.html}

\bibitem[{{Zingales} \& {Waldmann}(2018)}]{Zingales2018}
{Zingales}, T., \& {Waldmann}, I.~P. 2018, \aj, 156, 268, \dodoi{10.3847/1538-3881/aae77c}

\end{thebibliography}
\bibliographystyle{aasjournal}

\appendix
In this appendix, we collect the 18 figures introduced and discussed in Section~\ref{sec:analysis}.


\begin{figure}[t]
\begin{center}
\includegraphics[width=0.8\columnwidth]{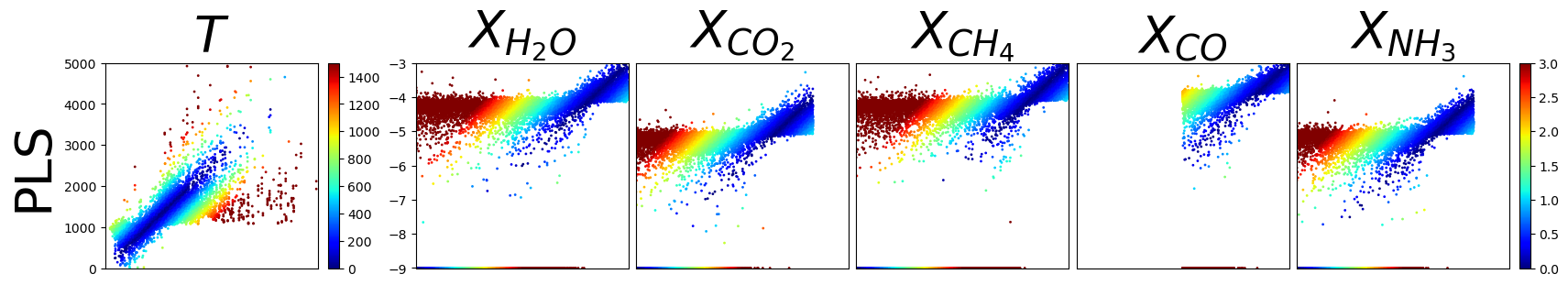}
\includegraphics[width=0.8\columnwidth]{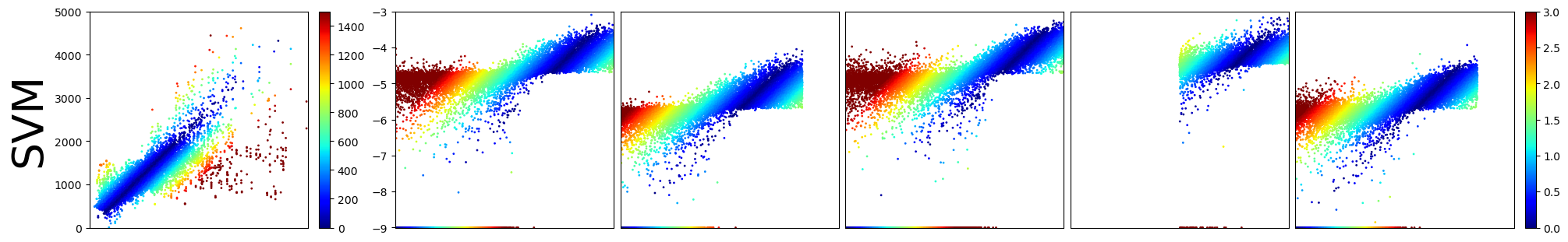}
\includegraphics[width=0.8\columnwidth]{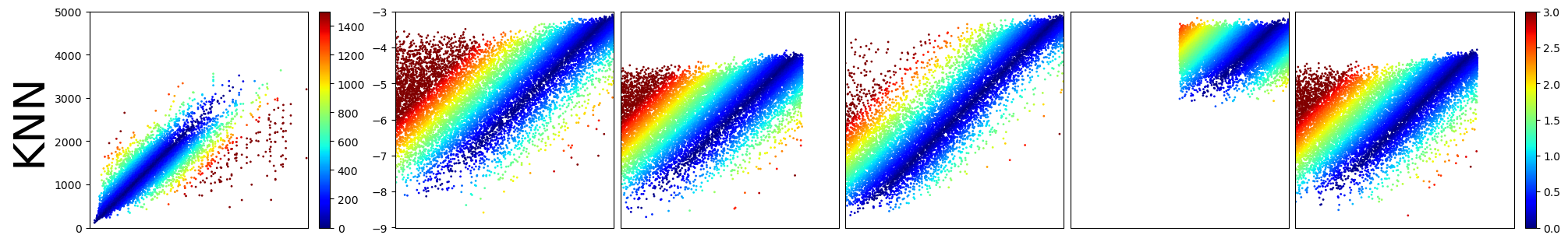}
\includegraphics[width=0.8\columnwidth]{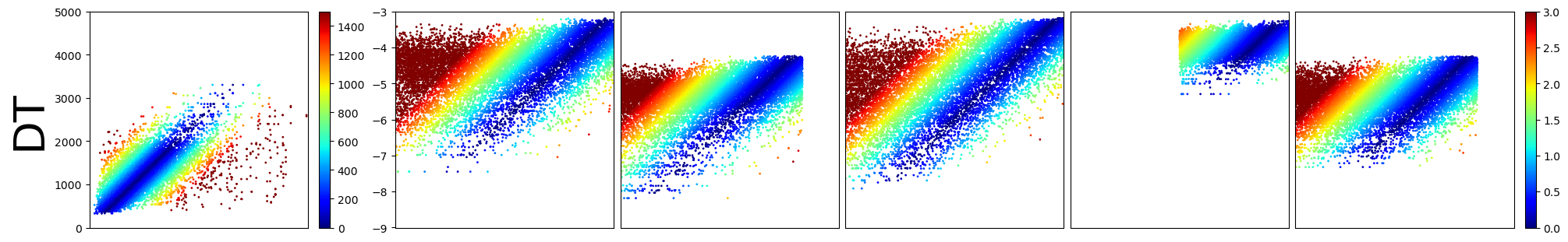}
\includegraphics[width=0.8\columnwidth]{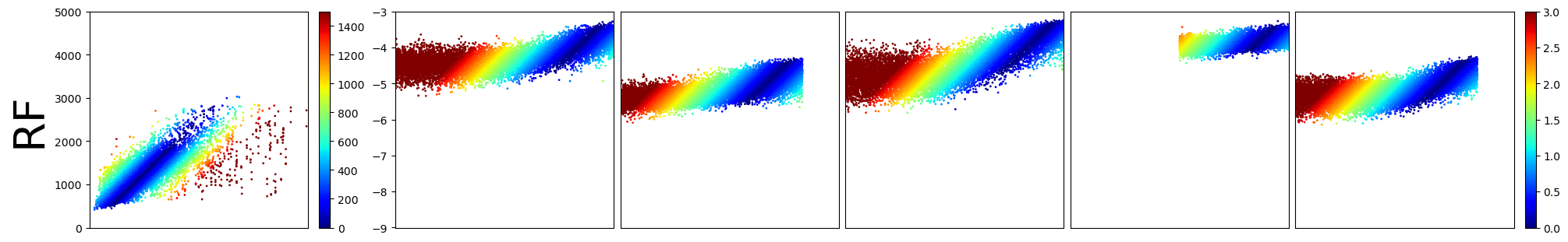}
\includegraphics[width=0.8\columnwidth]{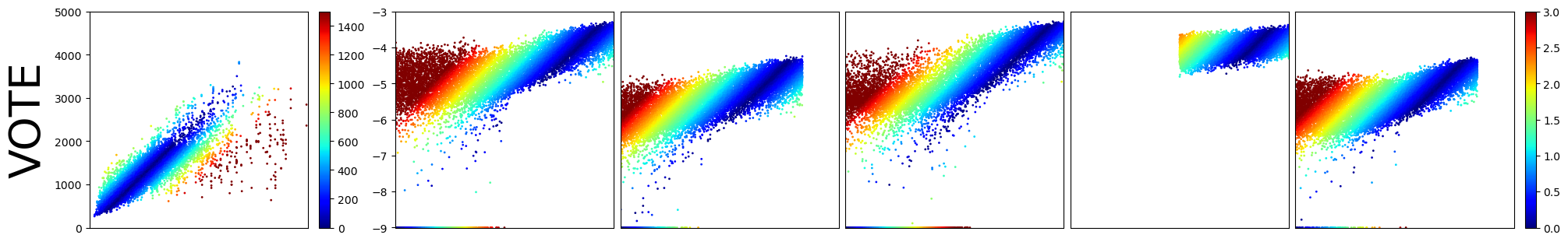}
\includegraphics[width=0.8\columnwidth]{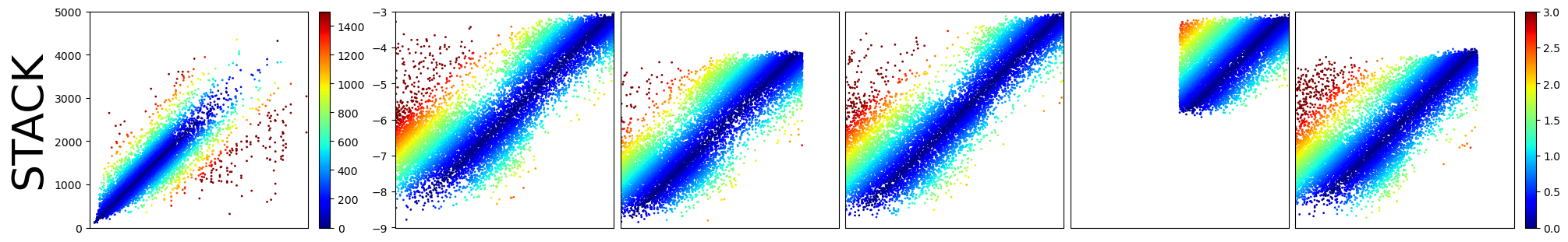}
\includegraphics[width=0.8\columnwidth]{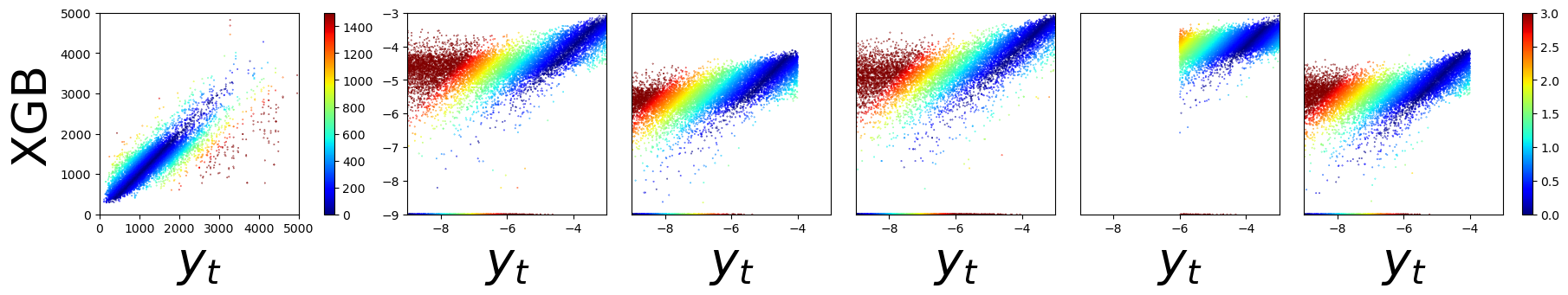}
\end{center}
    \caption{($\mathcal{S}$) 
    Scatter plots of the model predictions for the 6 target variables from eq.~(\ref{eq:target_variables}) ($y$-axis) versus the true values $y_t$ of the target variables ($x$-axis). The model is trained and tested on the standardized spectral data $\{S[M],S[y]\}$ (see Section~\ref{sec:training_data}). The colorbars indicate the absolute deviation of the prediction from the true value.
    }
    \label{fig:s_scatter}
\end{figure}

\begin{figure}[t]
\begin{center}
\includegraphics[width=0.8\columnwidth]{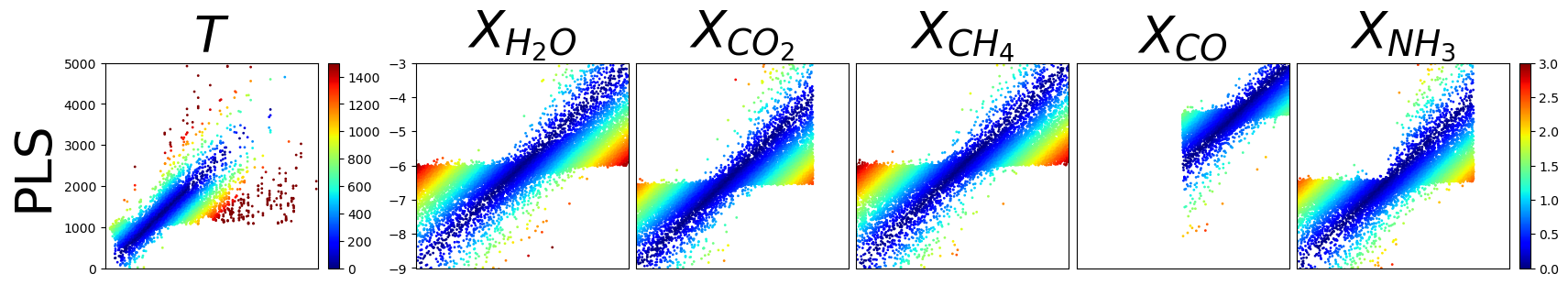}
\includegraphics[width=0.8\columnwidth]{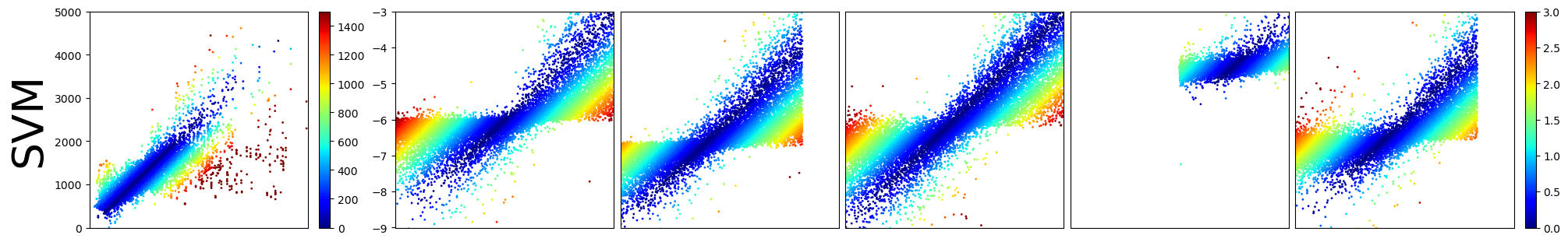}
\includegraphics[width=0.8\columnwidth]{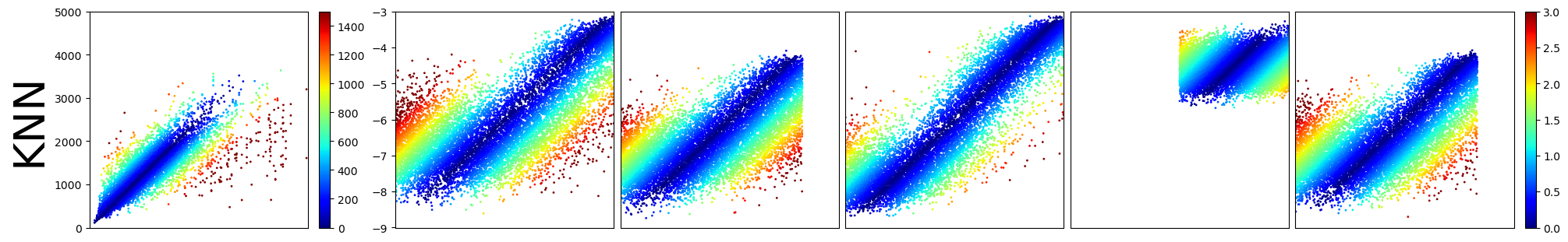}
\includegraphics[width=0.8\columnwidth]{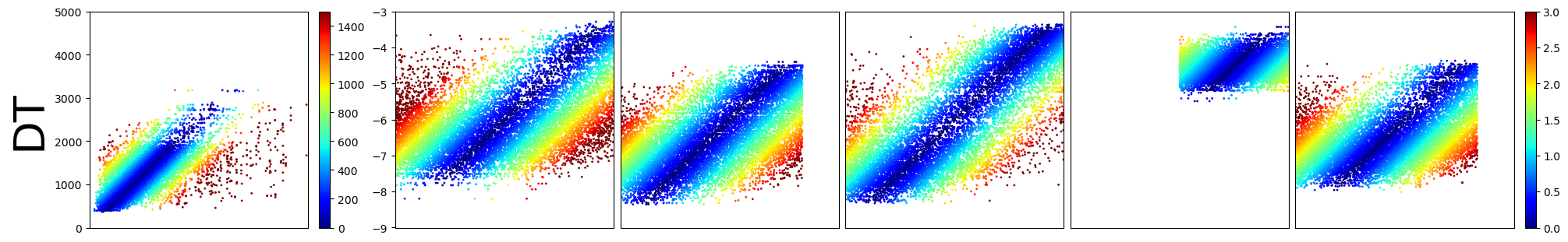}
\includegraphics[width=0.8\columnwidth]{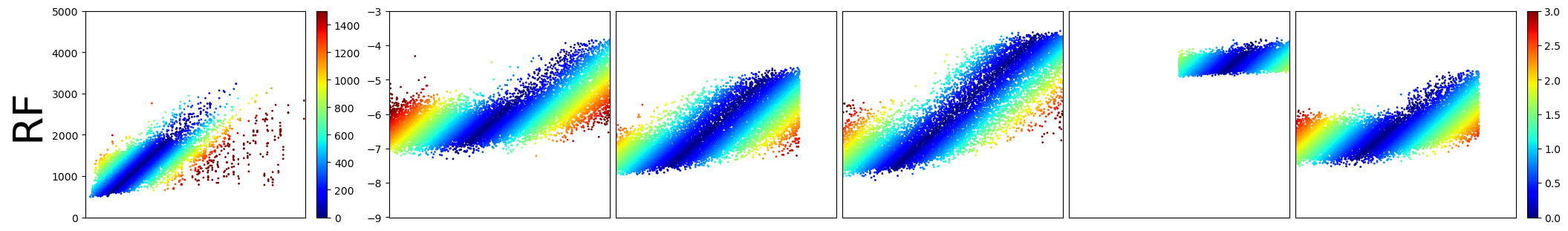}
\includegraphics[width=0.8\columnwidth]{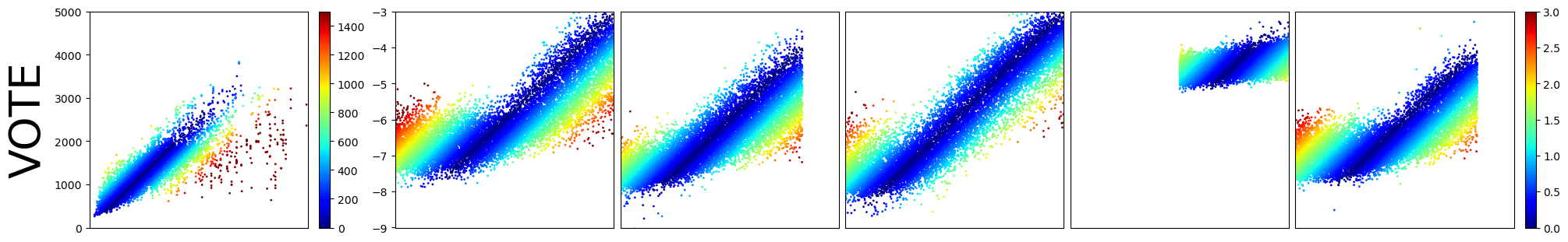}
\includegraphics[width=0.8\columnwidth]{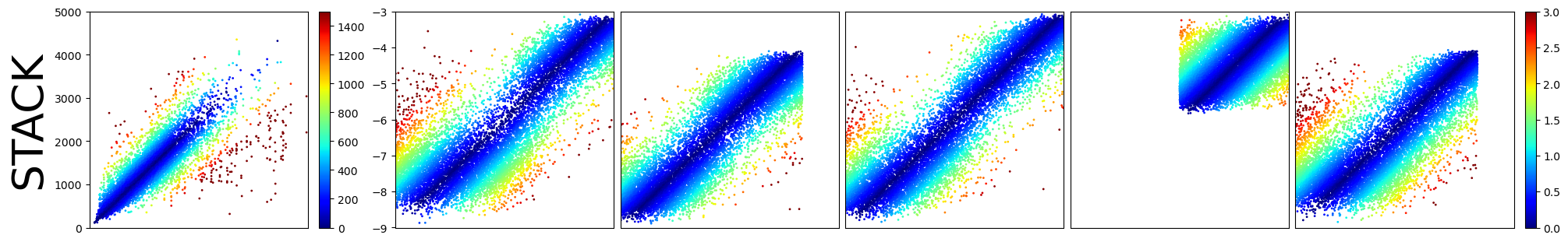}
\includegraphics[width=0.8\columnwidth]{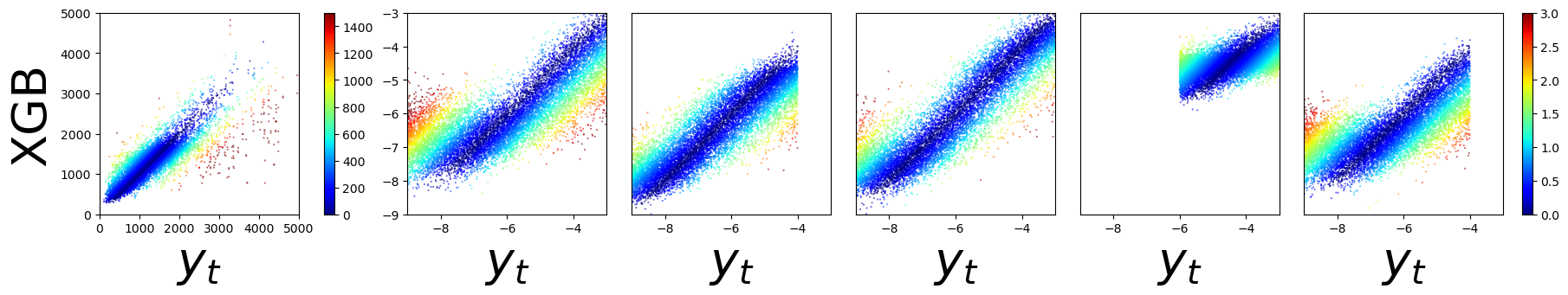}
\end{center}
    \caption{($\mathcal{SL}$) 
    The same as Figure~\ref{fig:s_scatter}, but using log concentrations, $\text{log}(X)$, as target variables during training and testing.
    }
    \label{fig:sl_scatter}
\end{figure}

\begin{figure}[t]
\begin{center}
\includegraphics[width=0.8\columnwidth]{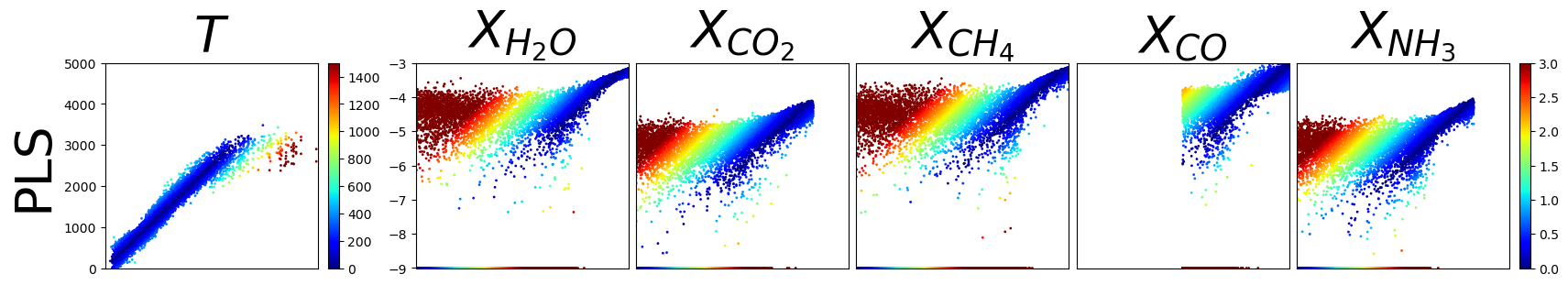}
\includegraphics[width=0.8\columnwidth]{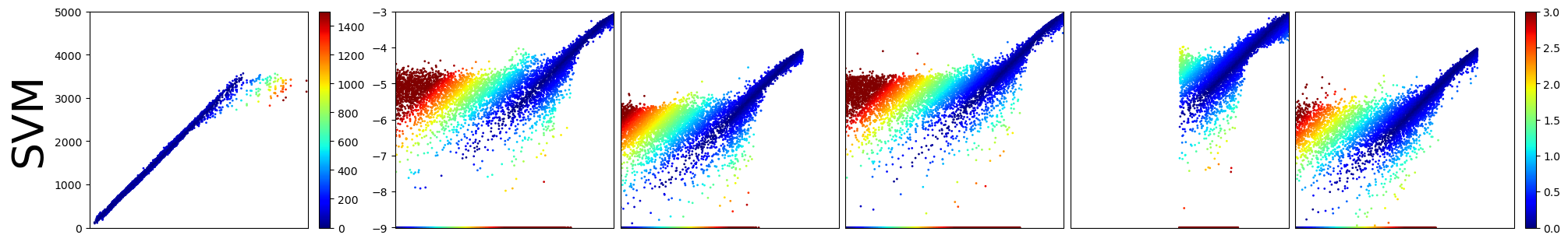}
\includegraphics[width=0.8\columnwidth]{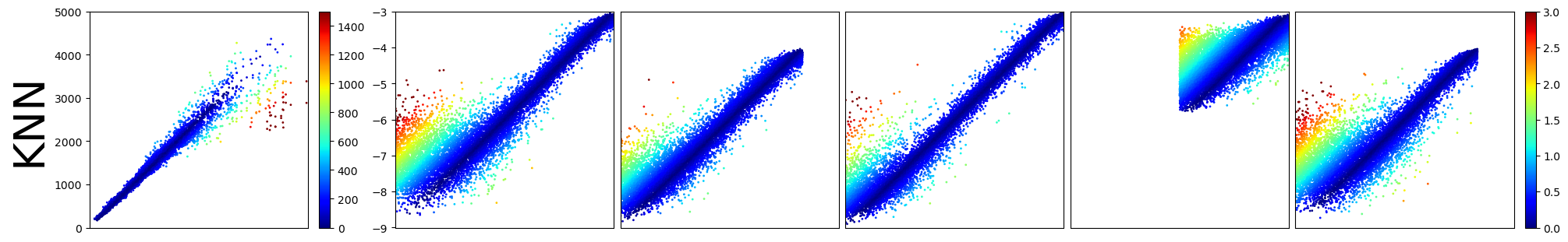}
\includegraphics[width=0.8\columnwidth]{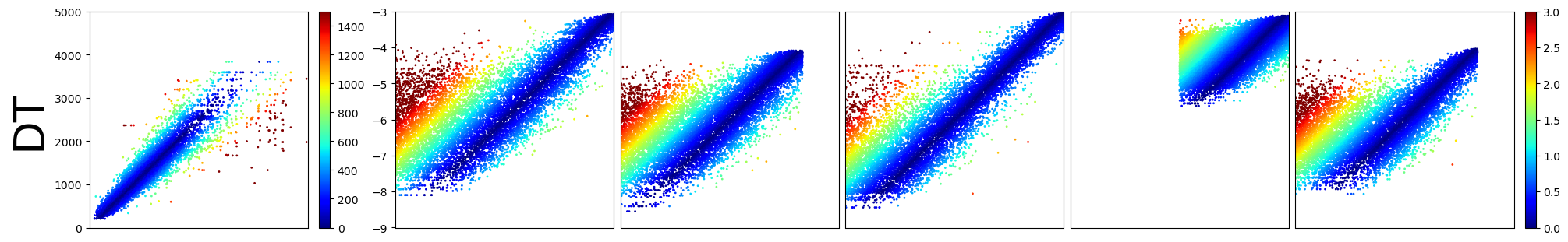}
\includegraphics[width=0.8\columnwidth]{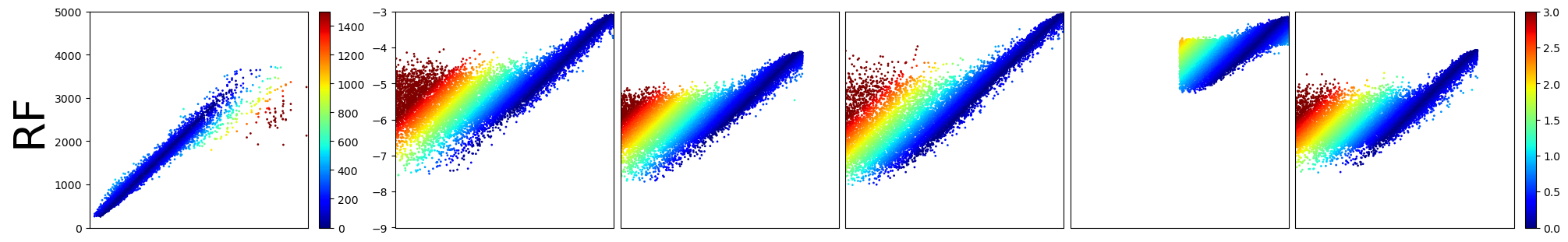}
\includegraphics[width=0.8\columnwidth]{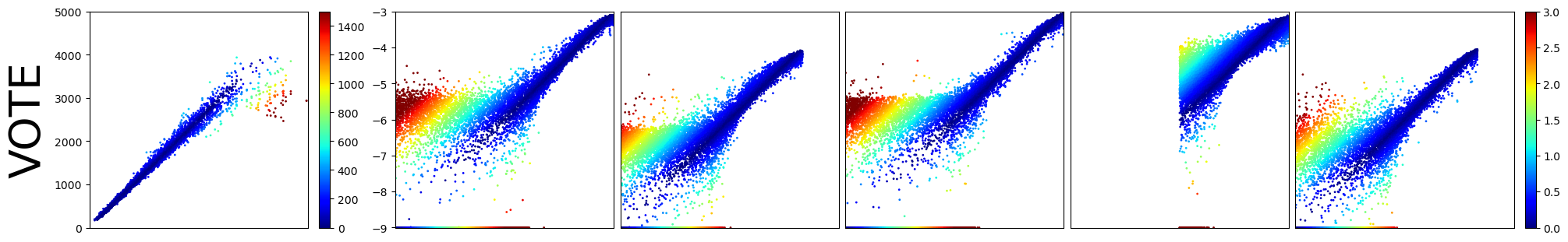}
\includegraphics[width=0.8\columnwidth]{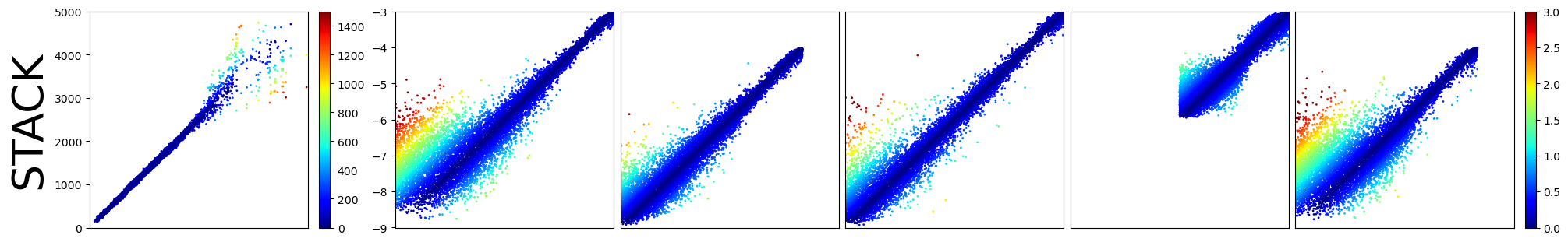}
\includegraphics[width=0.8\columnwidth]{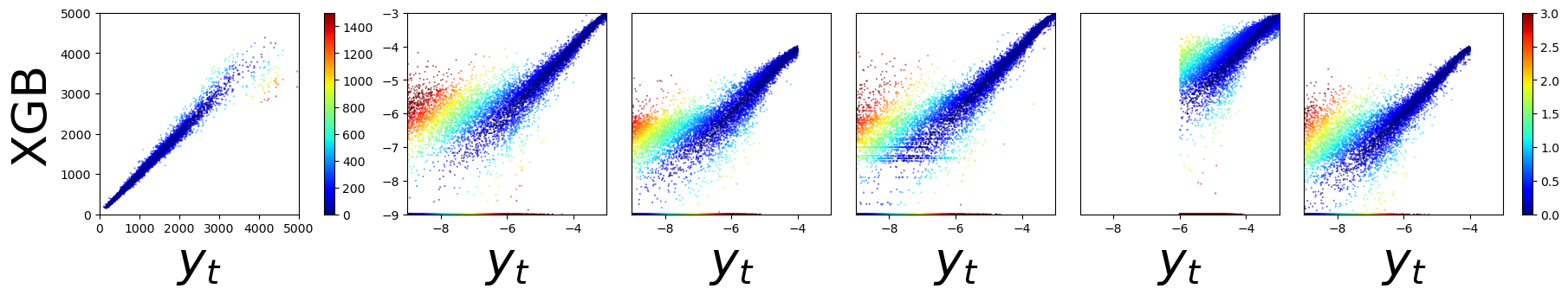}
\end{center}
    \caption{($\mathcal{N}$) The same as Figure~\ref{fig:s_scatter}, but using the normalized spectral data $N[M]$ for training and testing.
    }
    \label{fig:n_scatter}
\end{figure}

\begin{figure}[t]
\begin{center}
\includegraphics[width=0.8\columnwidth]{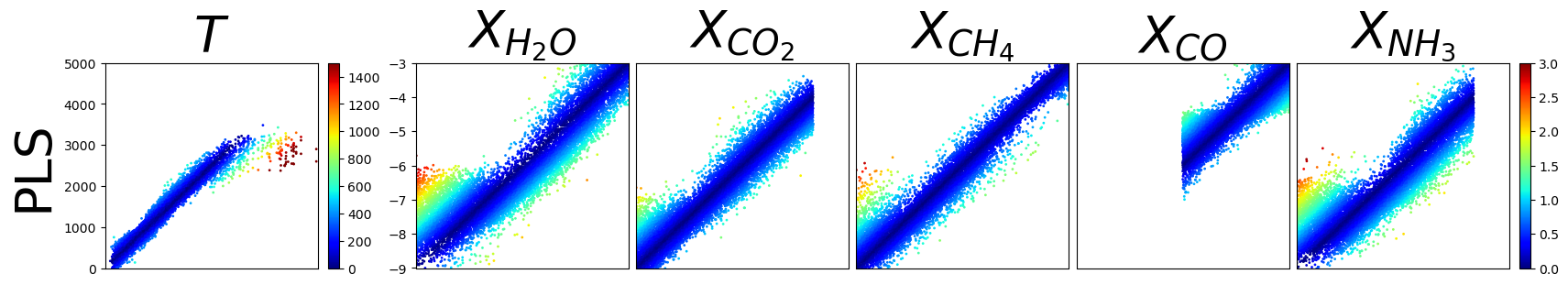}
\includegraphics[width=0.8\columnwidth]{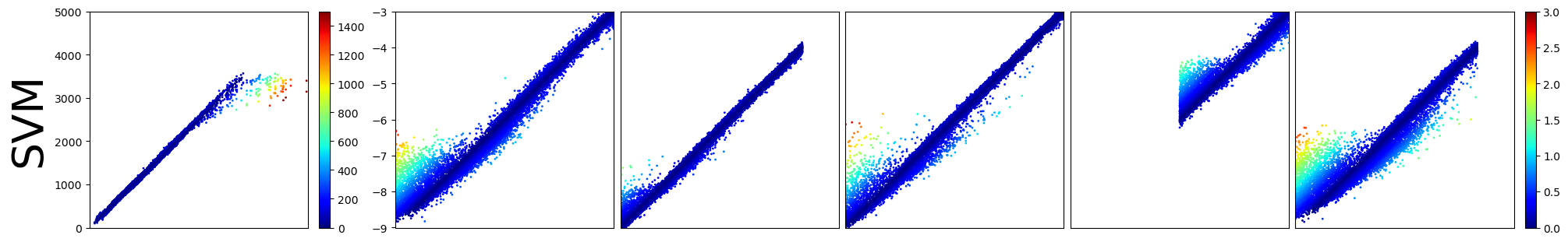}
\includegraphics[width=0.8\columnwidth]{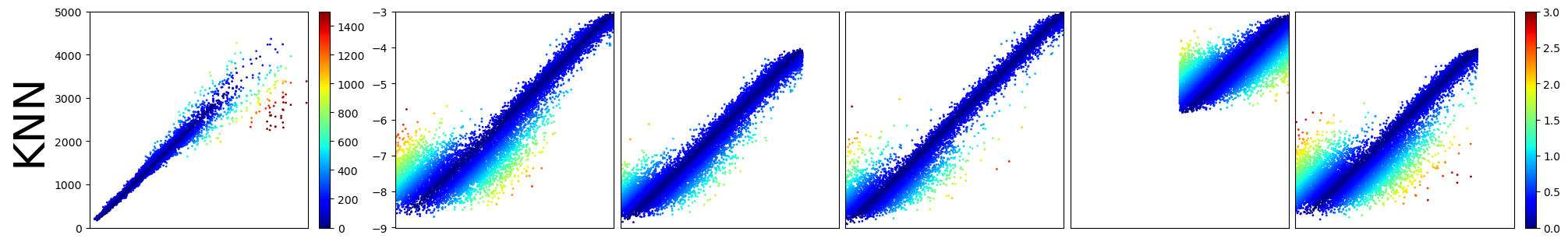}
\includegraphics[width=0.8\columnwidth]{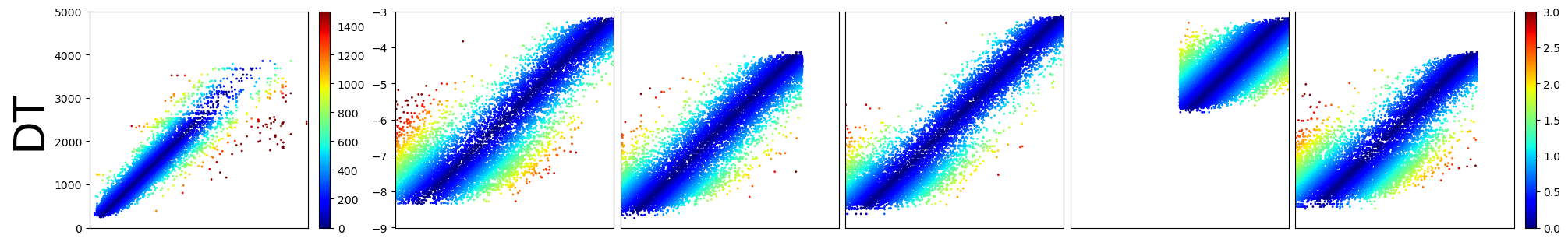}
\includegraphics[width=0.8\columnwidth]{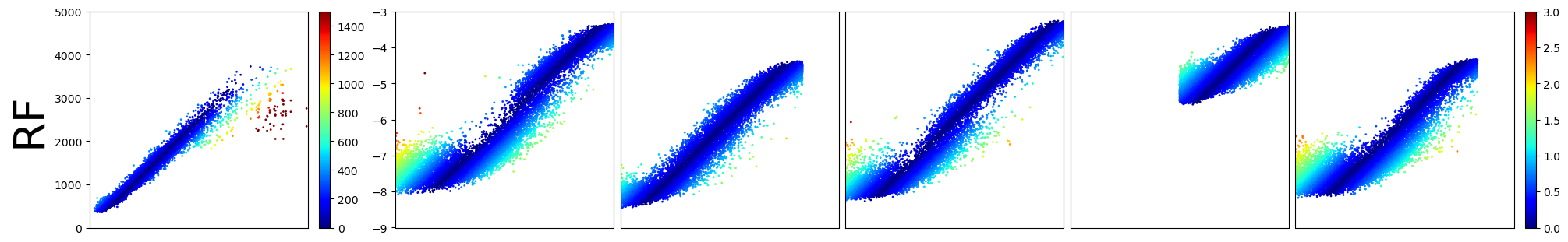}
\includegraphics[width=0.8\columnwidth]{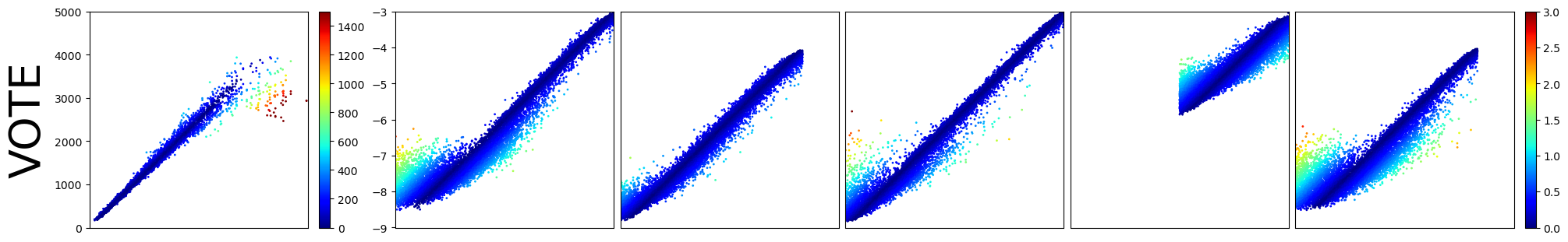}
\includegraphics[width=0.8\columnwidth]{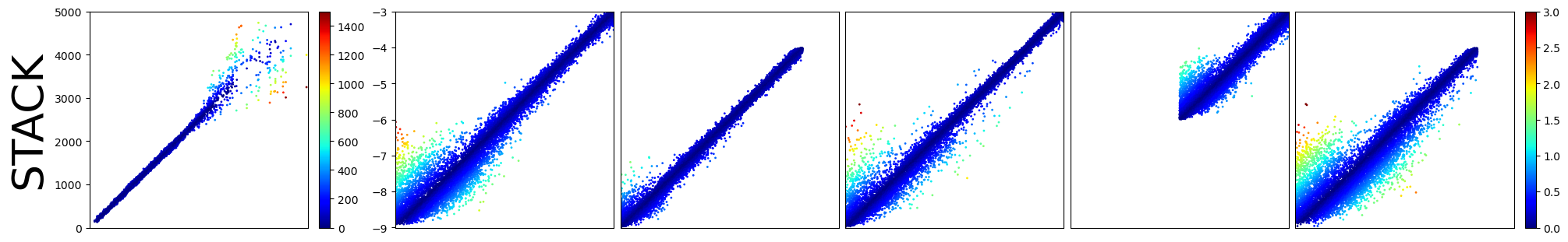}
\includegraphics[width=0.8\columnwidth]{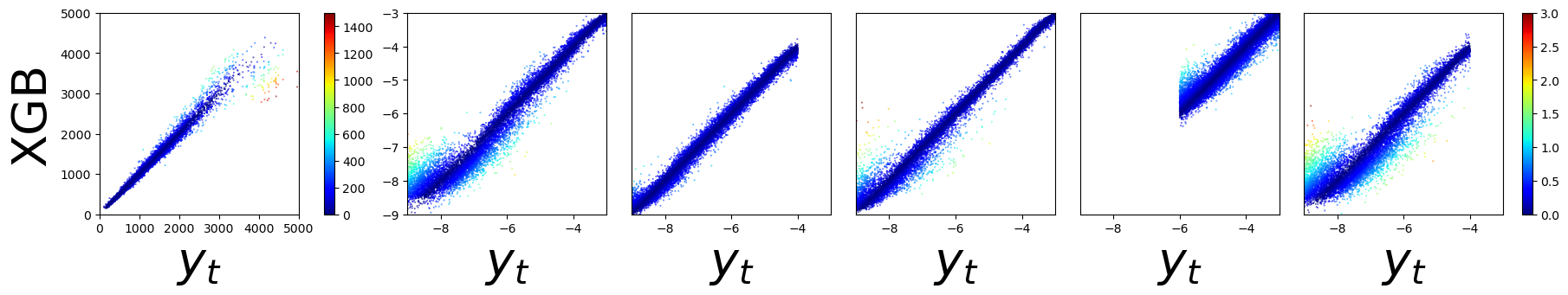}
\end{center}
    \caption{($\mathcal{NL}$) The same as Figure~\ref{fig:n_scatter}, but using log concentrations, $\text{log}(X)$, as target variables during training and testing.
    }
    \label{fig:nl_scatter}
\end{figure}

\begin{figure}[t]
\begin{center}
\includegraphics[width=0.8\columnwidth]{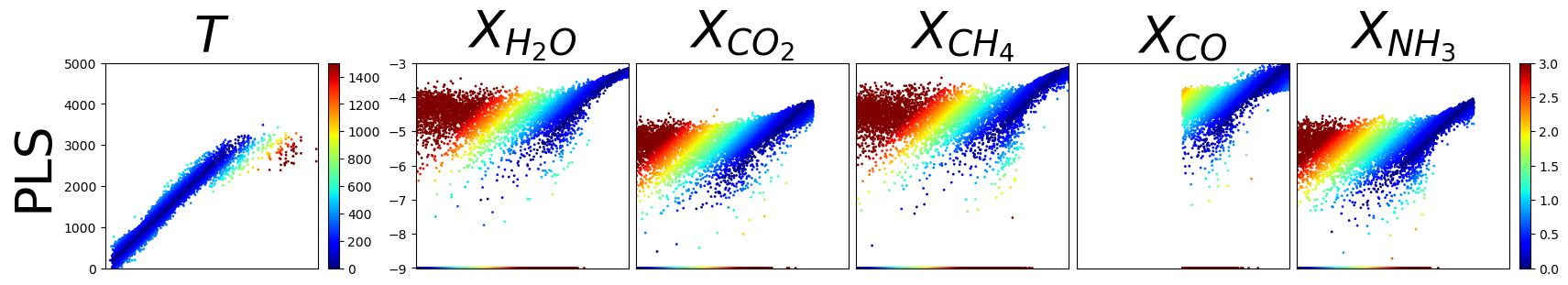}
\includegraphics[width=0.8\columnwidth]{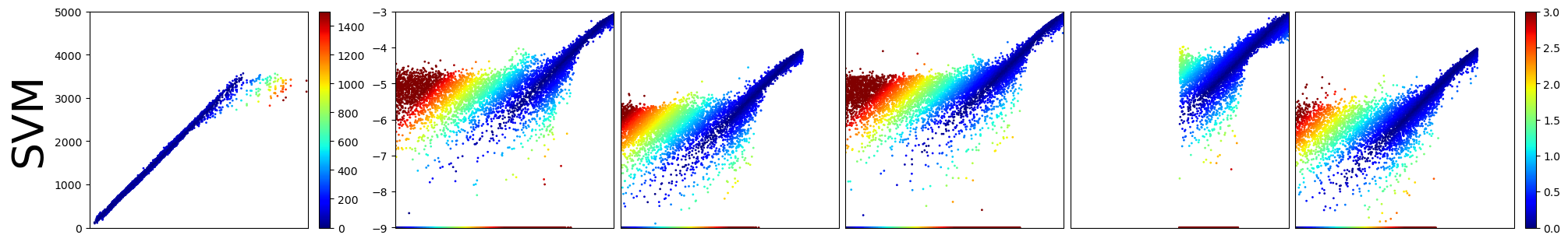}
\includegraphics[width=0.8\columnwidth]{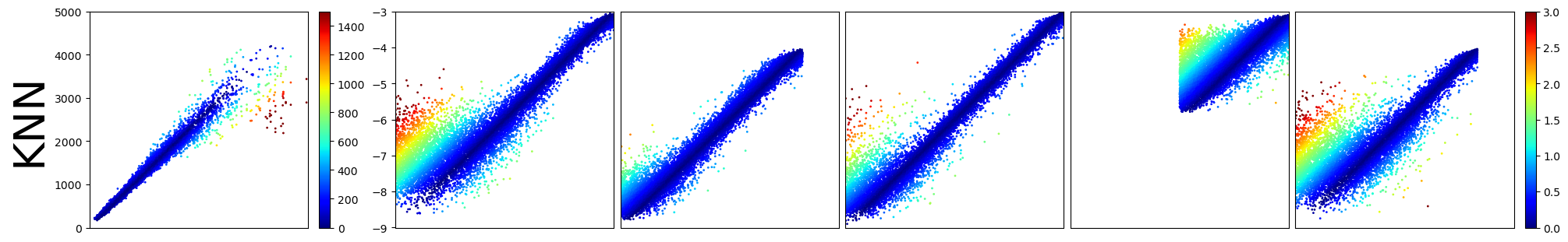}
\includegraphics[width=0.8\columnwidth]{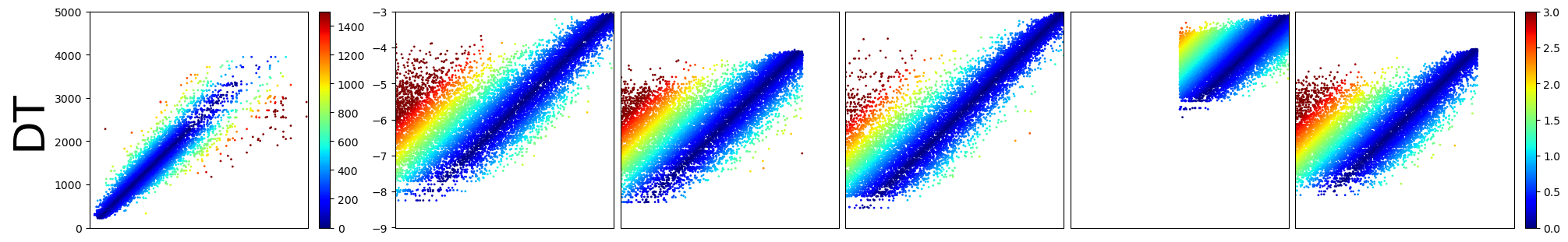}
\includegraphics[width=0.8\columnwidth]{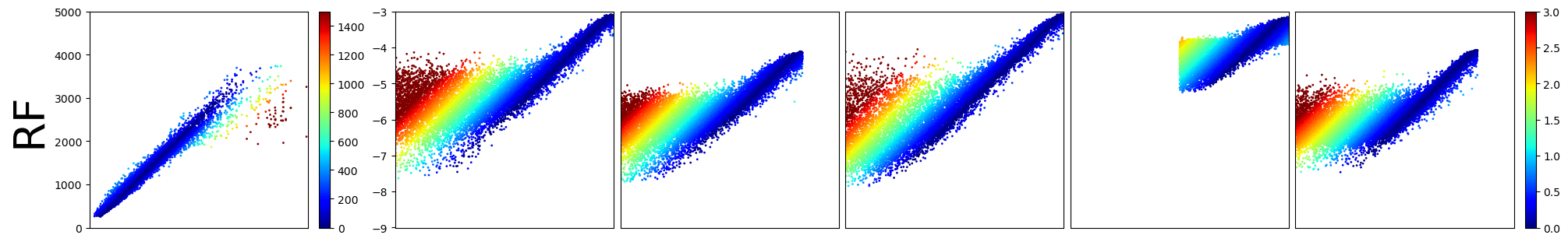}
\includegraphics[width=0.8\columnwidth]{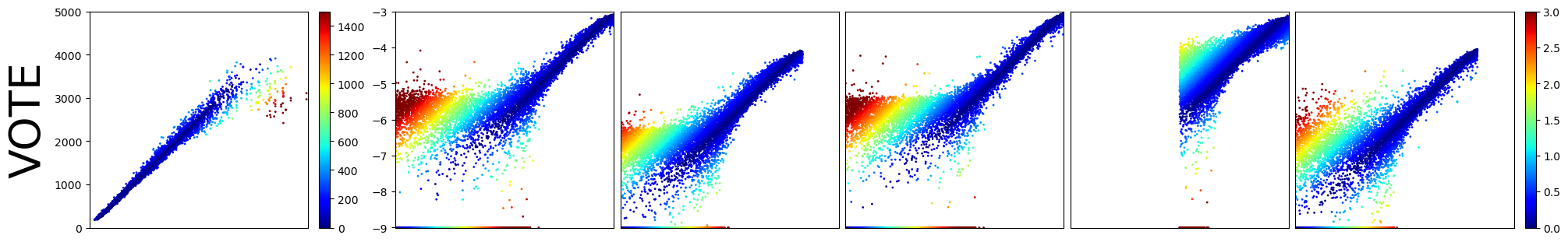}
\includegraphics[width=0.8\columnwidth]{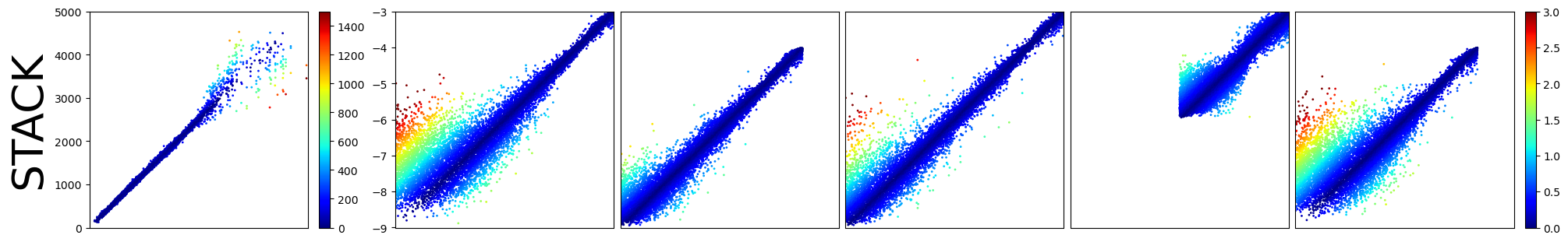}
\includegraphics[width=0.8\columnwidth]{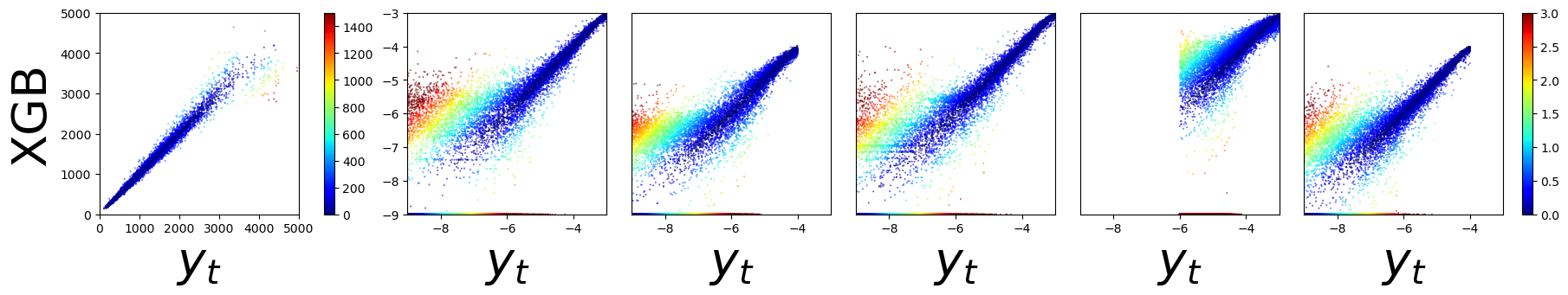}
\end{center}
    \caption{($\mathcal{NM}$) The same as Figure~\ref{fig:n_scatter}, but using in addition the spectral mean and standard deviation as feature variables.
    }
    \label{fig:nm_scatter}
\end{figure}

\begin{figure}[t]
\begin{center}
\includegraphics[width=0.8\columnwidth]{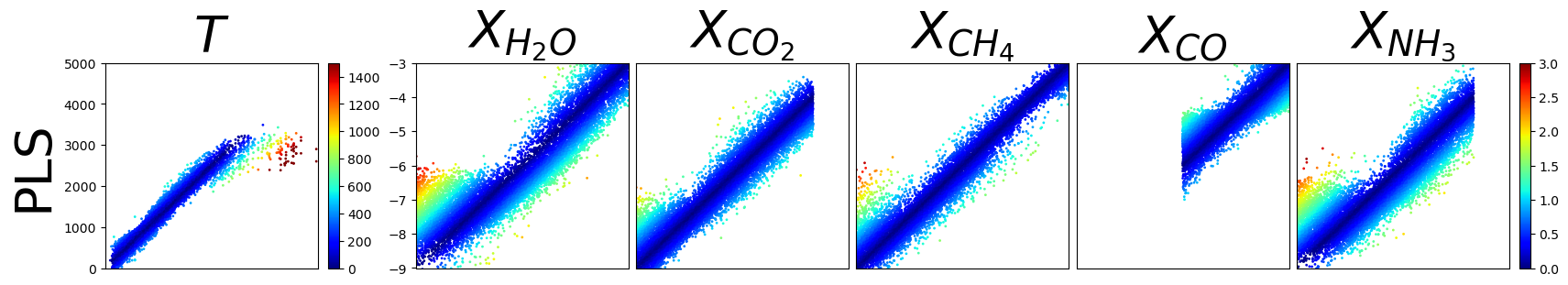}
\includegraphics[width=0.8\columnwidth]{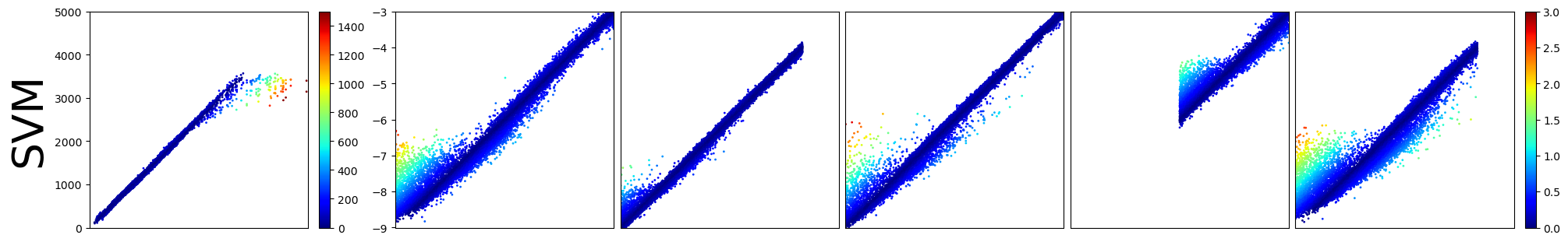}
\includegraphics[width=0.8\columnwidth]{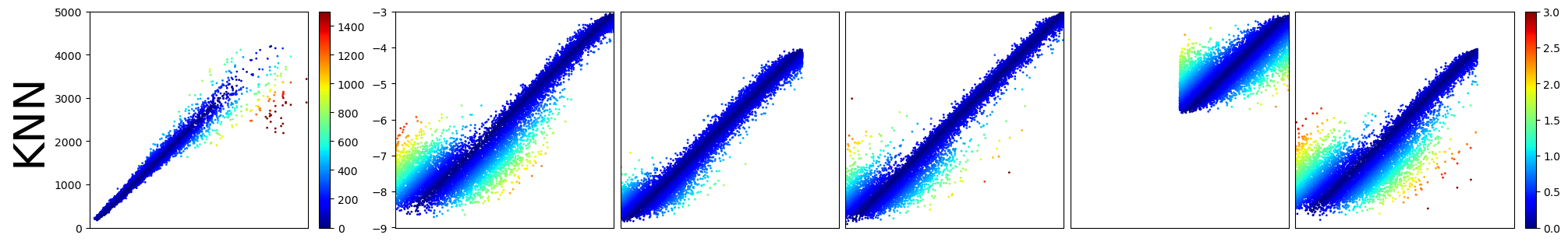}
\includegraphics[width=0.8\columnwidth]{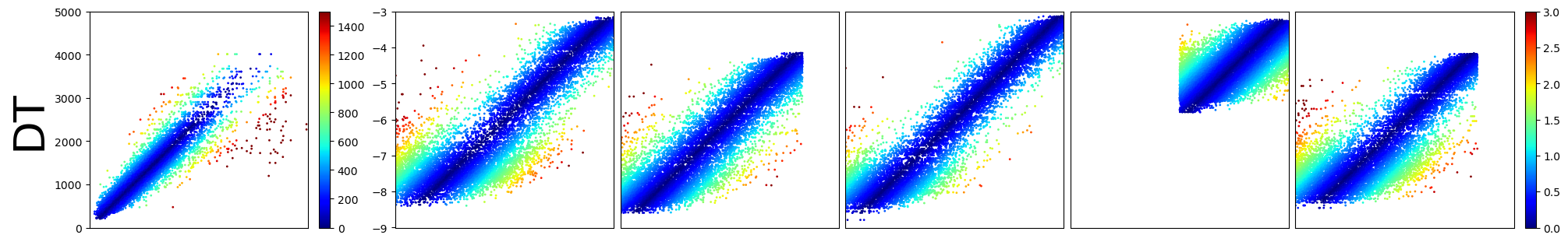}
\includegraphics[width=0.8\columnwidth]{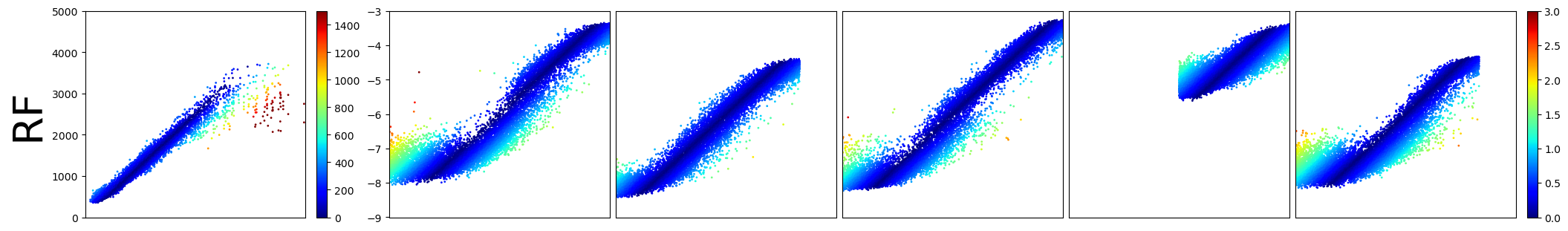}
\includegraphics[width=0.8\columnwidth]{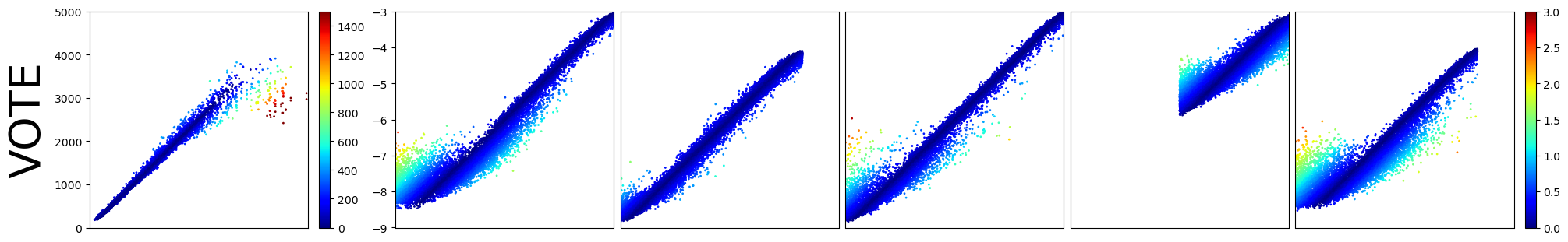}
\includegraphics[width=0.8\columnwidth]{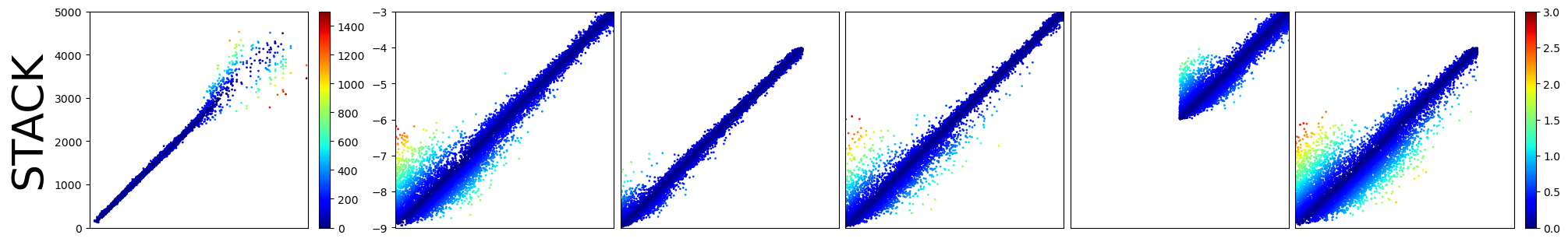}
\includegraphics[width=0.8\columnwidth]{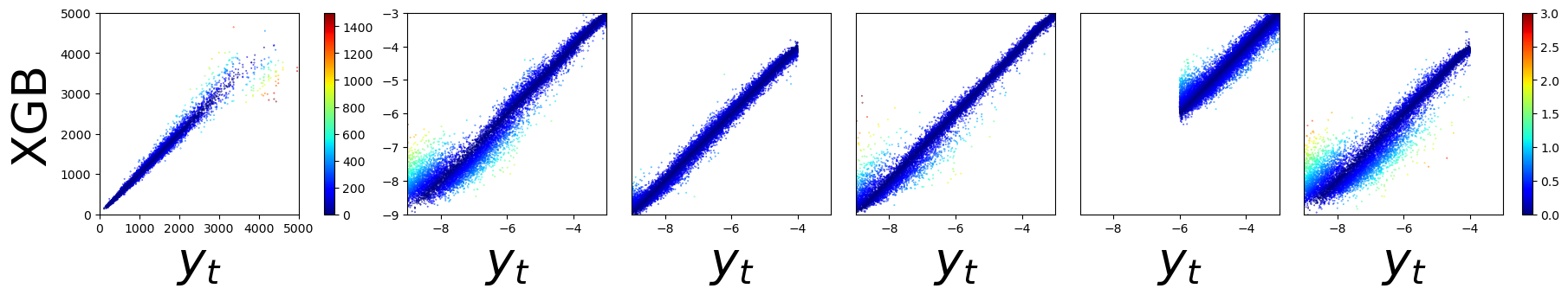}
\end{center}
    \caption{($\mathcal{NML}$) The same as Figure~\ref{fig:nm_scatter}, but using log concentrations, $\text{log}(X)$, as target variables during training and testing.
    }
    \label{fig:nml_scatter}
\end{figure}


\begin{figure}[t]
\begin{center}
\includegraphics[width=0.85\columnwidth]{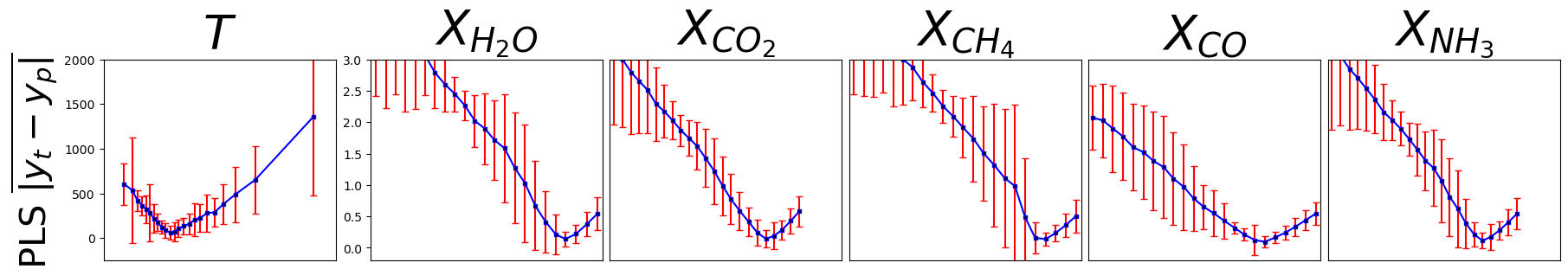}
\includegraphics[width=0.85\columnwidth]{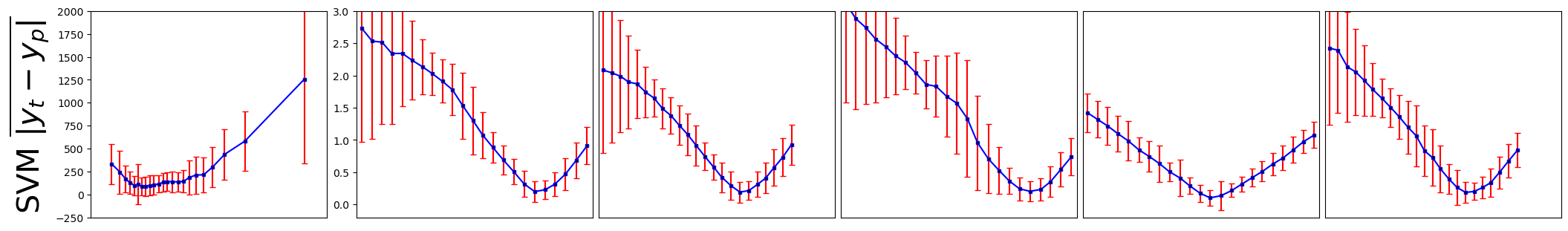}
\includegraphics[width=0.85\columnwidth]{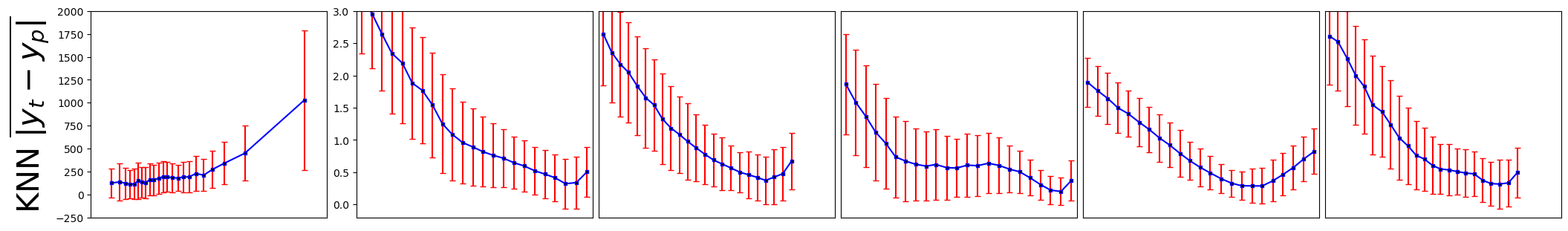}
\includegraphics[width=0.85\columnwidth]{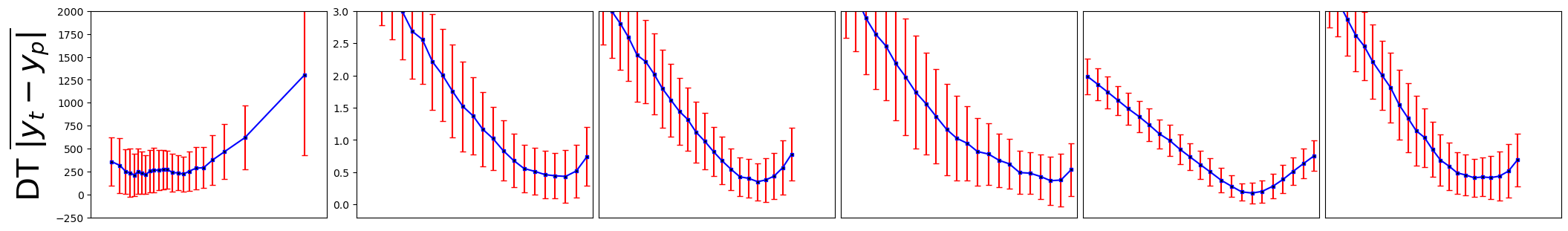}
\includegraphics[width=0.85\columnwidth]{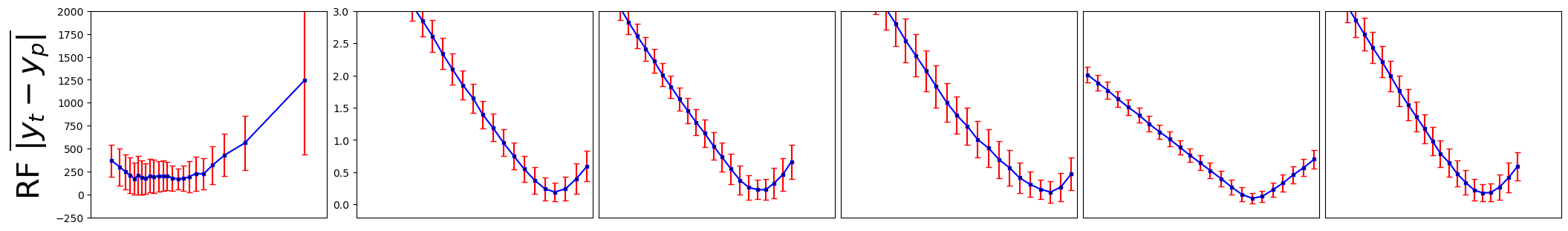}
\includegraphics[width=0.85\columnwidth]{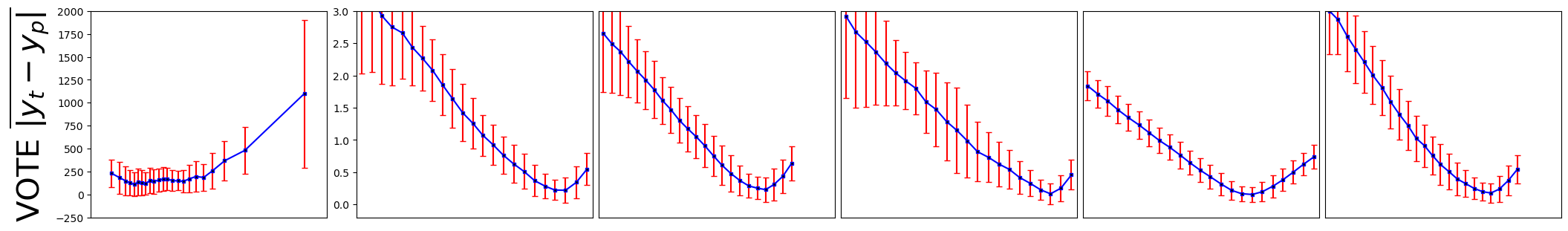}
\includegraphics[width=0.85\columnwidth]{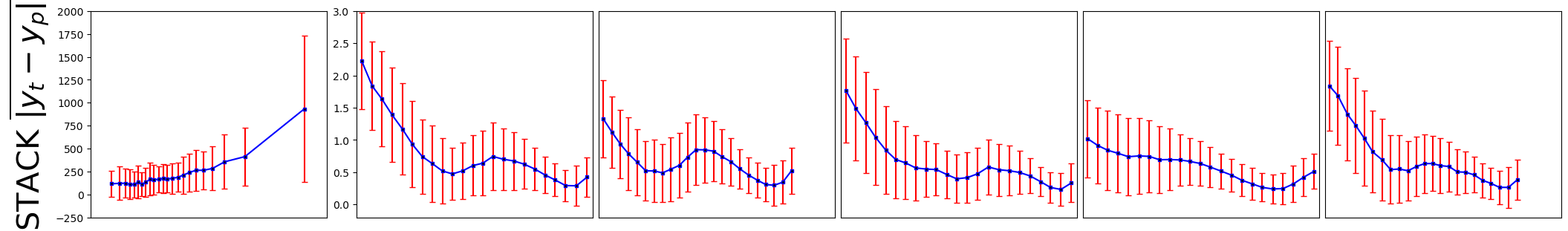}
\includegraphics[width=0.85\columnwidth]{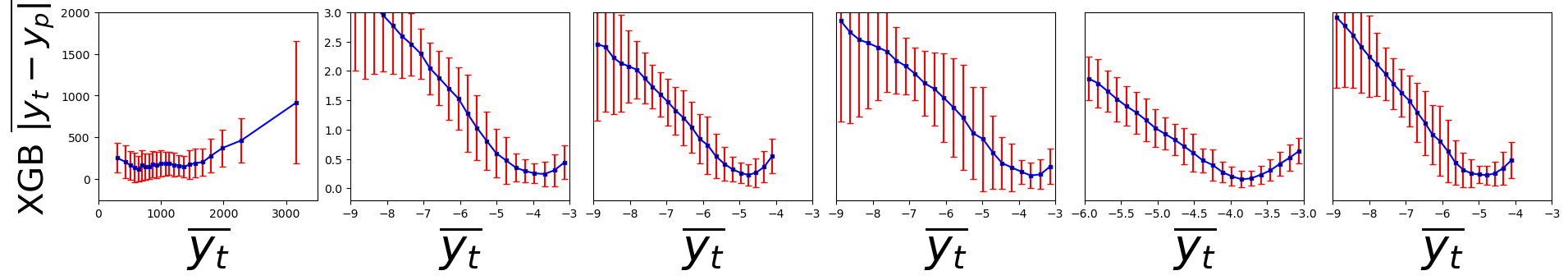}
\end{center}
    \caption{($\mathcal{S}$) Plots of the average absolute error $\overline{|y_t-y_p|}$, versus the average true value $\bar{y}_t$ per $y_t$ quantile bin. The model is trained and tested on the standardized spectral data $\{S[M],S[y]\}$ (see Section~\ref{sec:training_data}). The error bars indicate the corresponding standard deviations of the quantity $|y_t-y_p|$ within each bin. The quantile bins were formed based on the sorted true target values $y_t$.
    }
    \label{fig:s_true}
\end{figure}

\begin{figure}[t]
\begin{center}
\includegraphics[width=0.85\columnwidth]{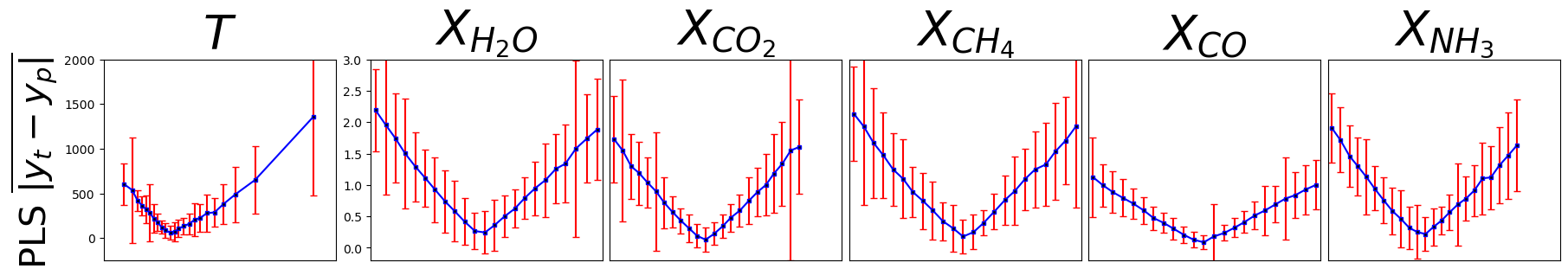}
\includegraphics[width=0.85\columnwidth]{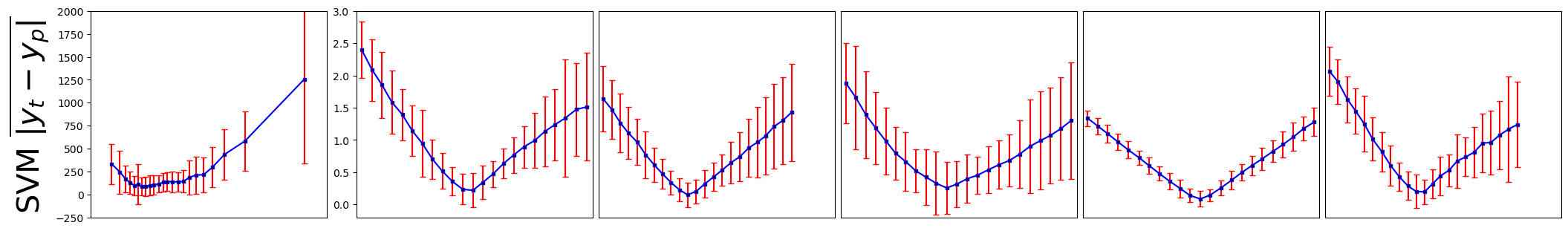}
\includegraphics[width=0.85\columnwidth]{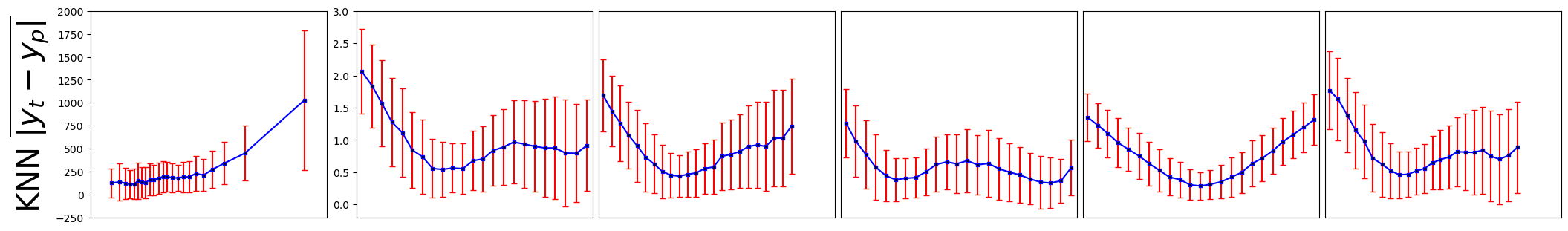}
\includegraphics[width=0.85\columnwidth]{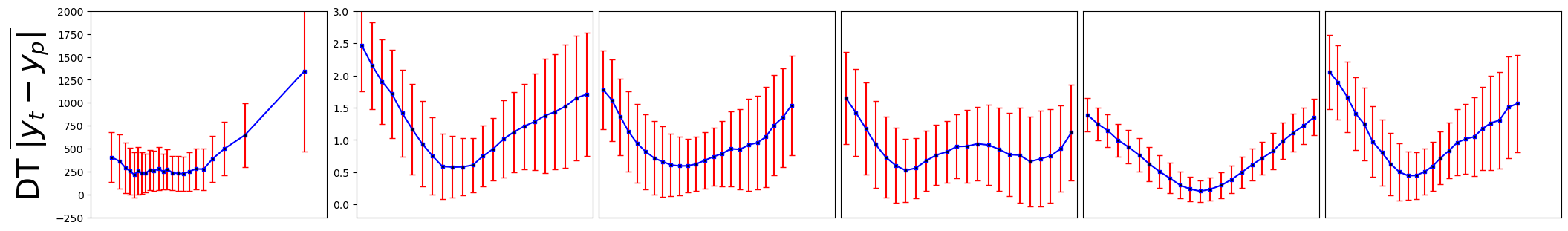}
\includegraphics[width=0.85\columnwidth]{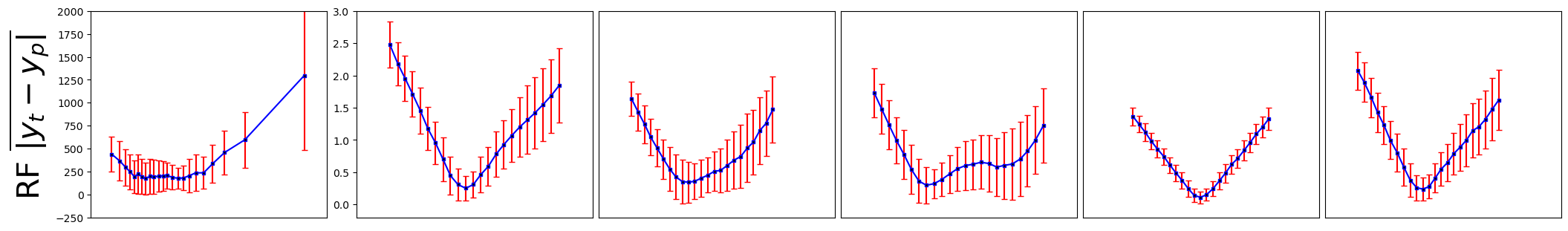}
\includegraphics[width=0.85\columnwidth]{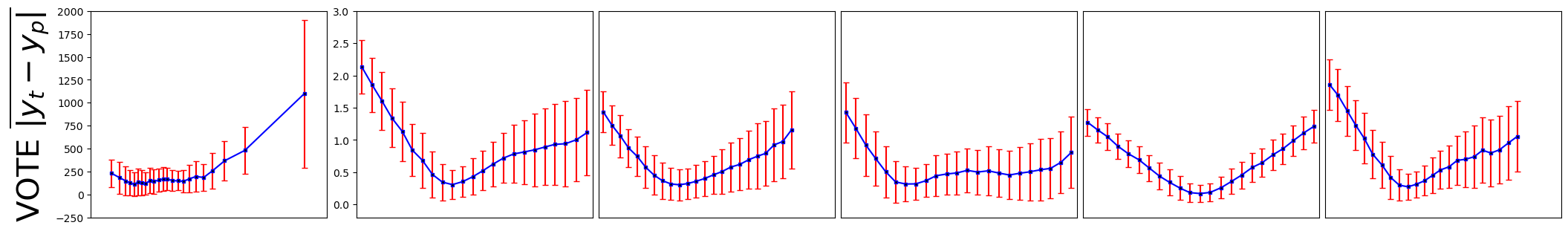}
\includegraphics[width=0.85\columnwidth]{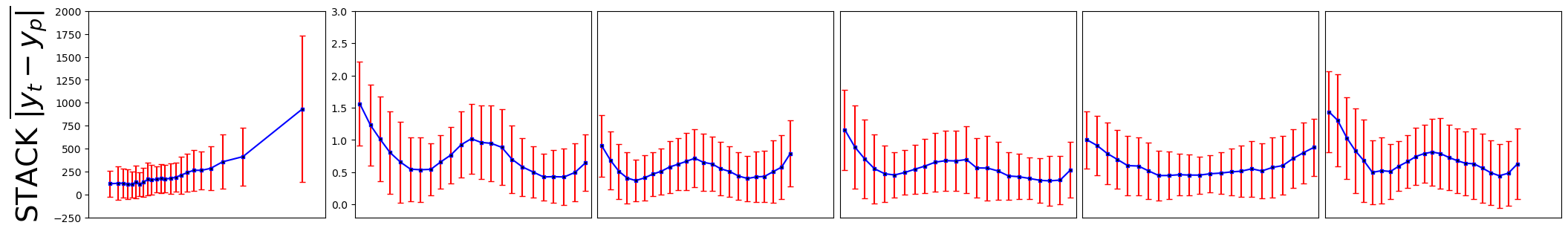}
\includegraphics[width=0.85\columnwidth]{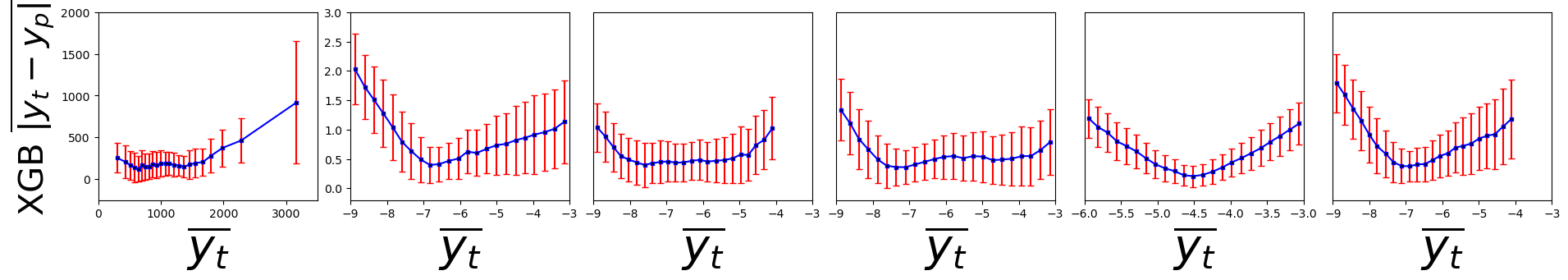}
\end{center}
    \caption{($\mathcal{SL}$) 
    The same as Figure~\ref{fig:s_true}, but using log concentrations, $\text{log}(X)$, as target variables during training and testing.
    }
    \label{fig:sl_true}
\end{figure}

\begin{figure}[t]
\begin{center}
\includegraphics[width=0.85\columnwidth]{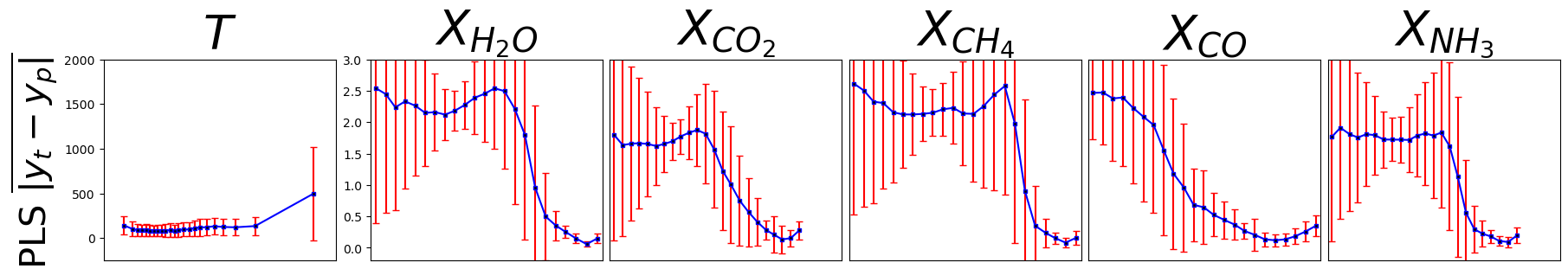}
\includegraphics[width=0.85\columnwidth]{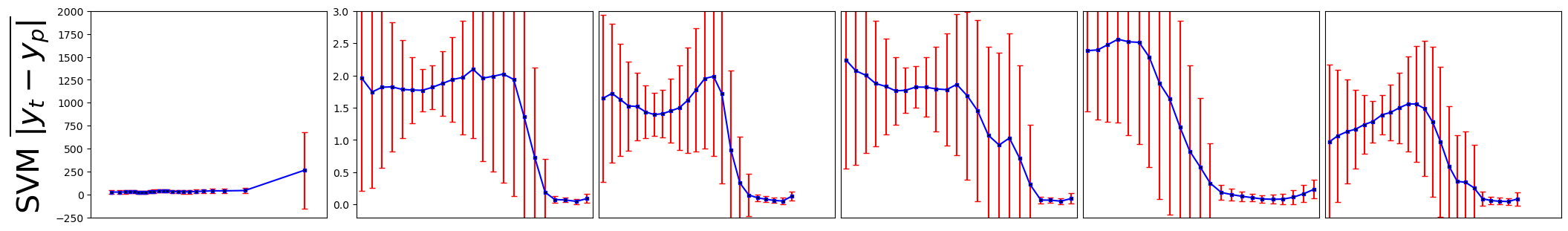}
\includegraphics[width=0.85\columnwidth]{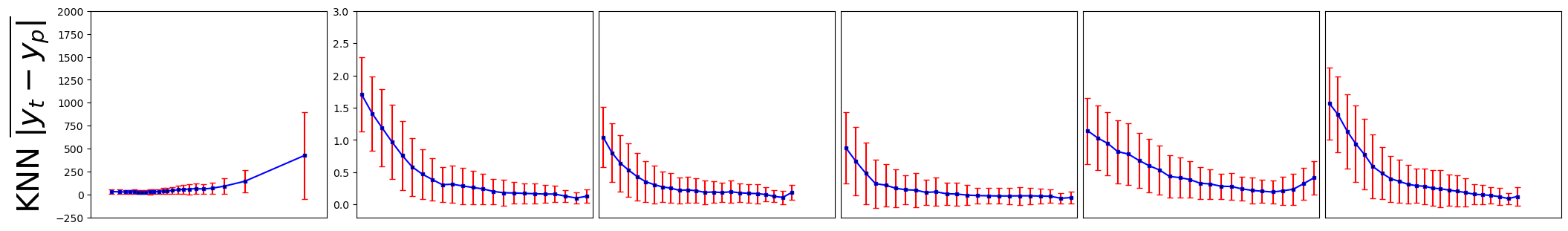}
\includegraphics[width=0.85\columnwidth]{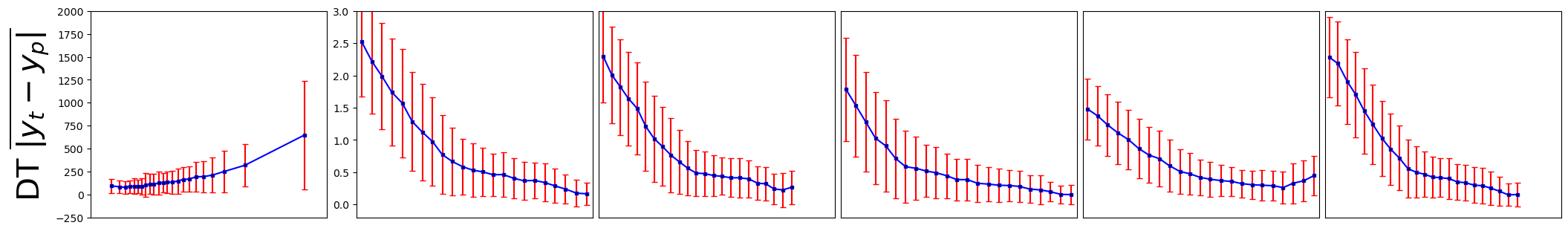}
\includegraphics[width=0.85\columnwidth]{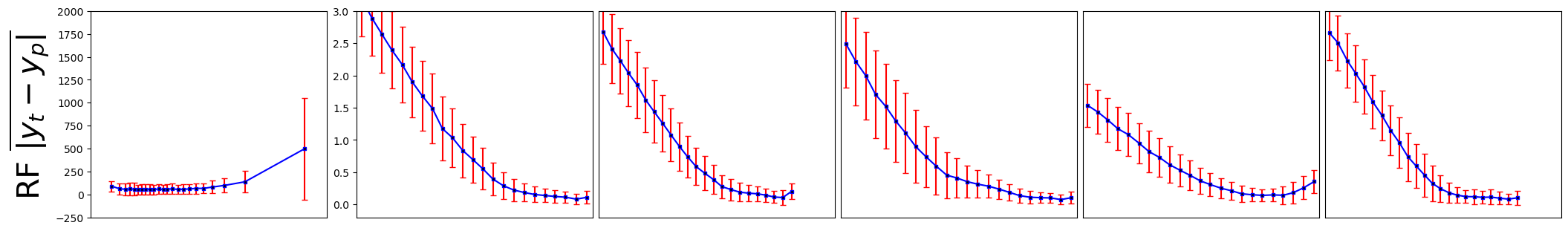}
\includegraphics[width=0.85\columnwidth]{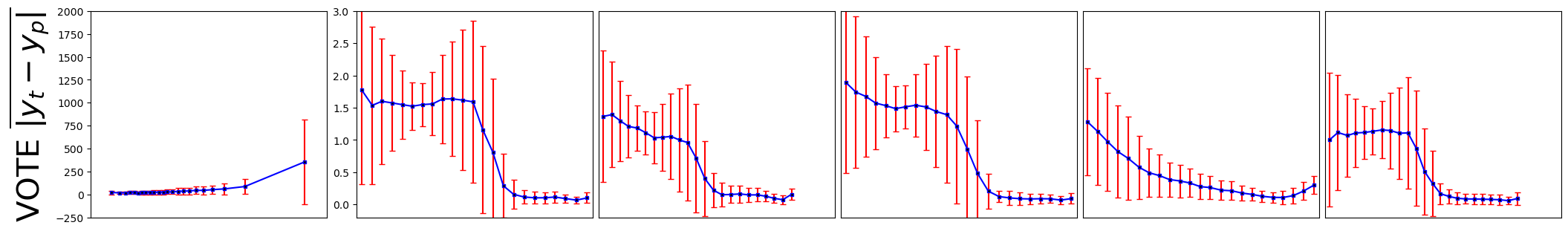}
\includegraphics[width=0.85\columnwidth]{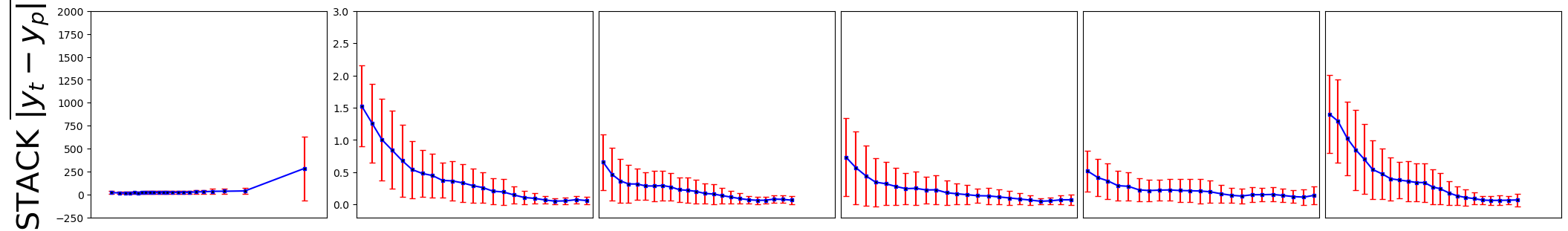}
\includegraphics[width=0.85\columnwidth]{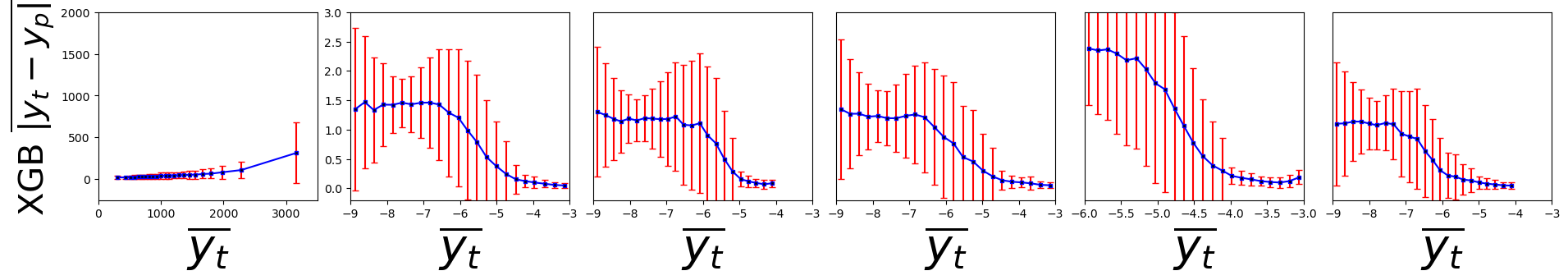}
\end{center}
    \caption{($\mathcal{N}$) The same as Figure~\ref{fig:s_true}, but using the normalized spectral data $N[M]$ for training and testing.
    }
    \label{fig:n_true}
\end{figure}

\begin{figure}[t]
\begin{center}
\includegraphics[width=0.85\columnwidth]{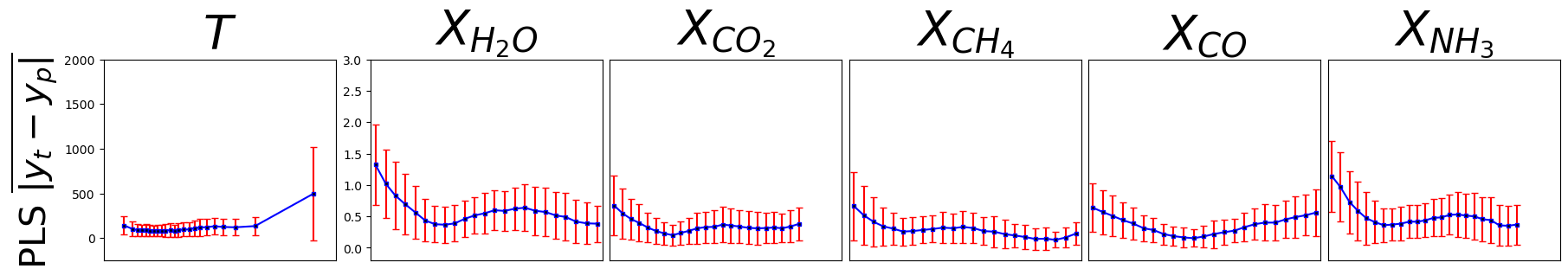}
\includegraphics[width=0.85\columnwidth]{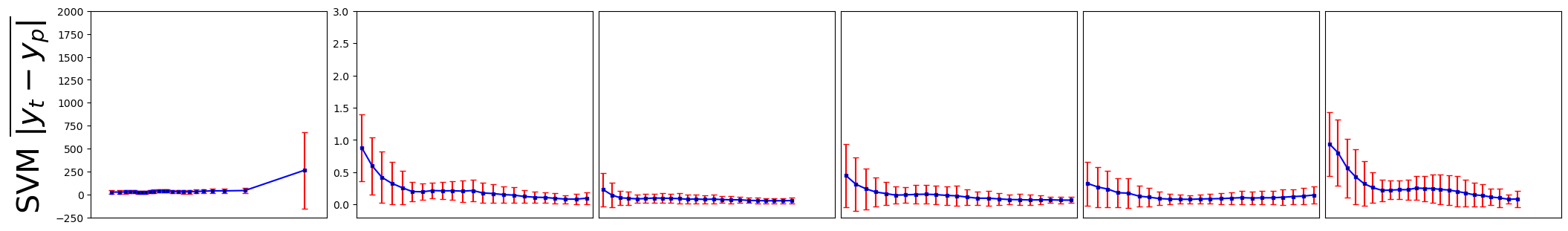}
\includegraphics[width=0.85\columnwidth]{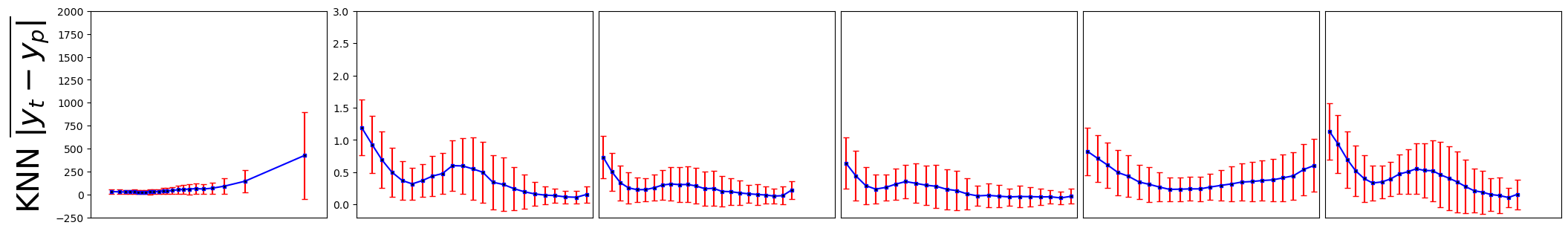}
\includegraphics[width=0.85\columnwidth]{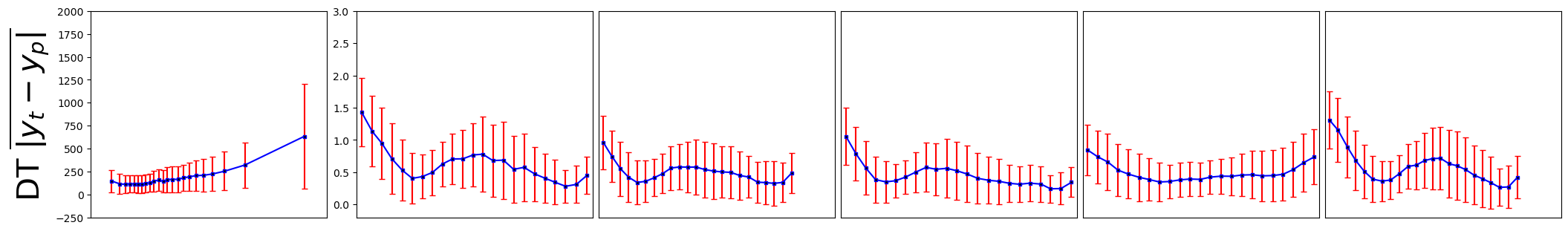}
\includegraphics[width=0.85\columnwidth]{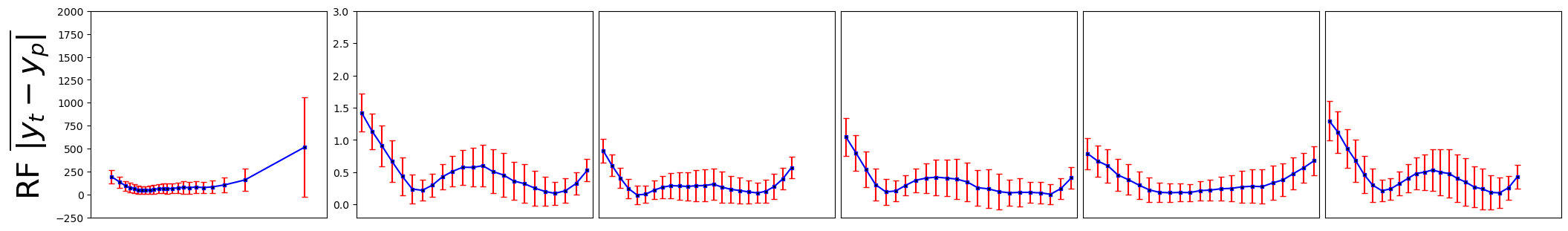}
\includegraphics[width=0.85\columnwidth]{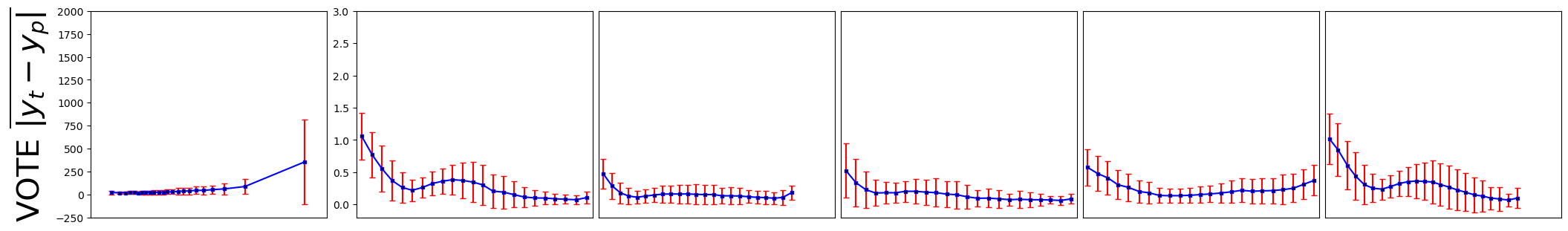}
\includegraphics[width=0.85\columnwidth]{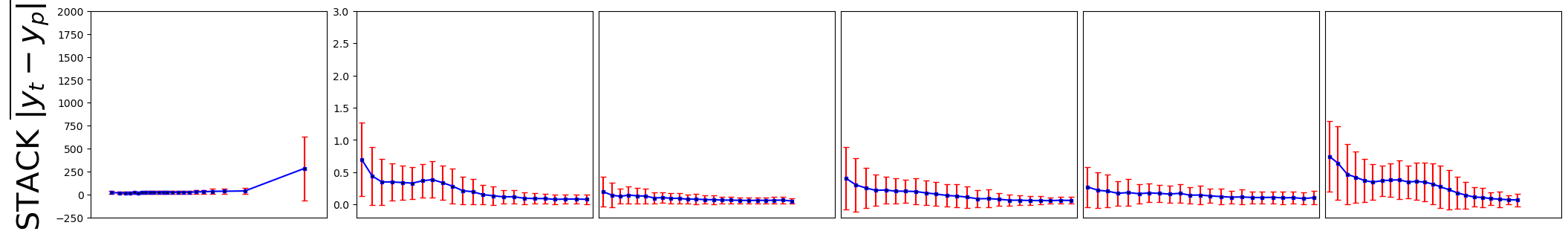}
\includegraphics[width=0.85\columnwidth]{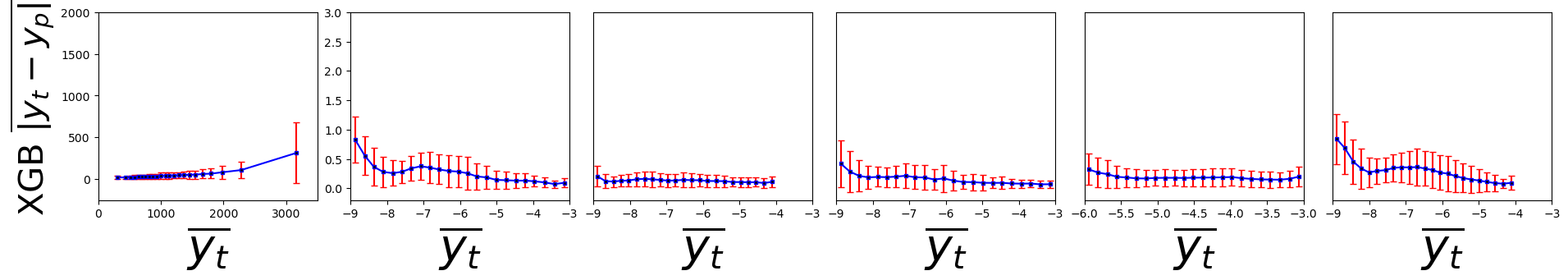}
\end{center}
    \caption{($\mathcal{NL}$) The same as Figure~\ref{fig:n_true}, but using log concentrations, $\text{log}(X)$, as target variables during training and testing.
    }
    \label{fig:nl_true}
\end{figure}

\begin{figure}[t]
\begin{center}
\includegraphics[width=0.85\columnwidth]{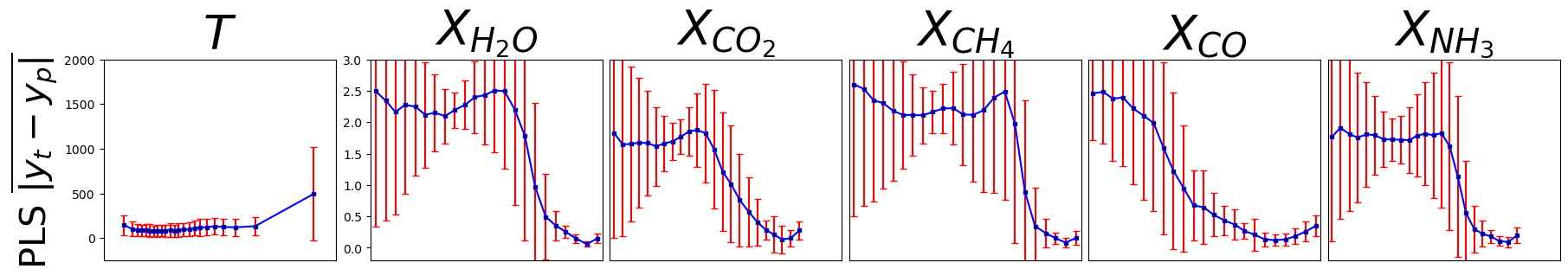}
\includegraphics[width=0.85\columnwidth]{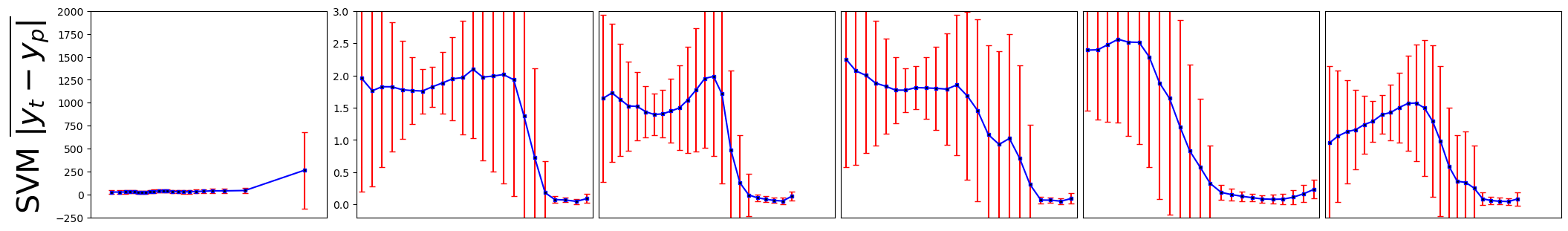}
\includegraphics[width=0.85\columnwidth]{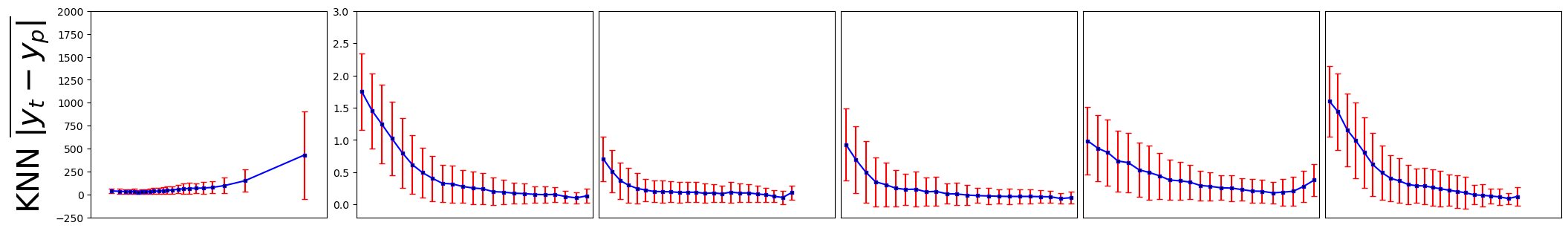}
\includegraphics[width=0.85\columnwidth]{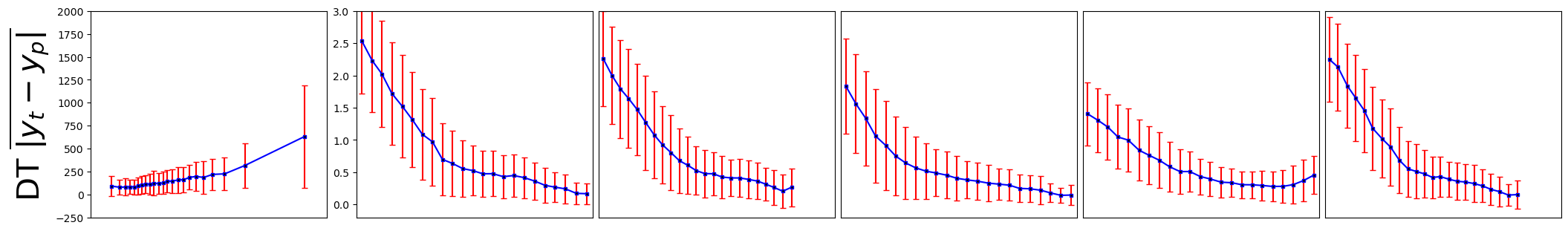}
\includegraphics[width=0.85\columnwidth]{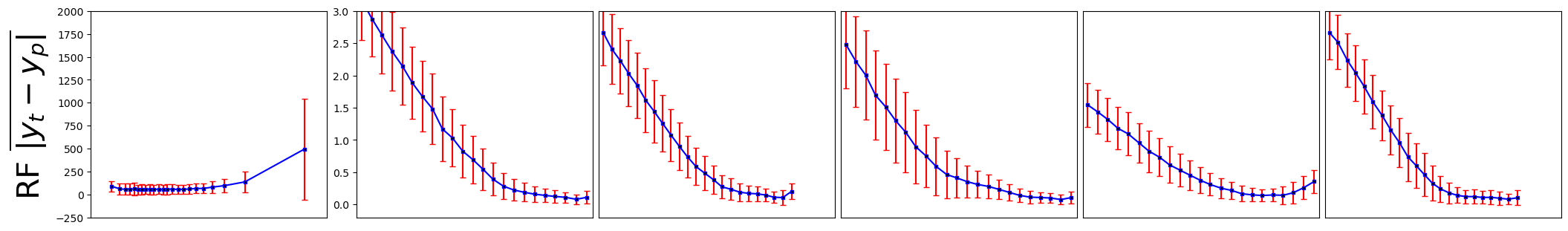}
\includegraphics[width=0.85\columnwidth]{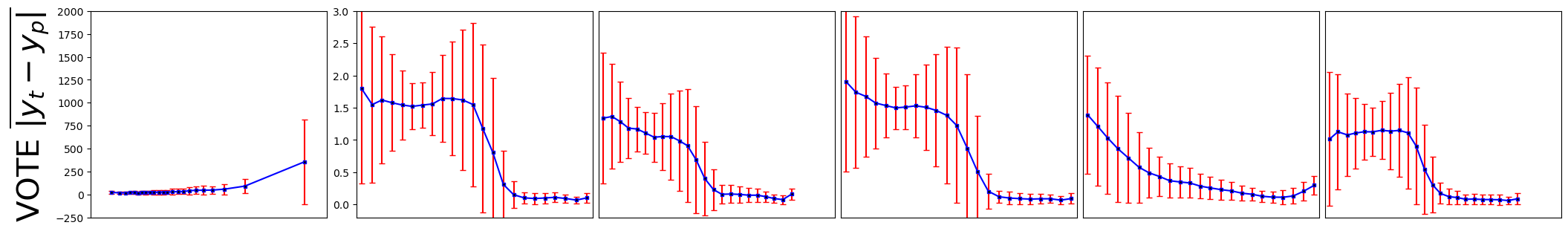}
\includegraphics[width=0.85\columnwidth]{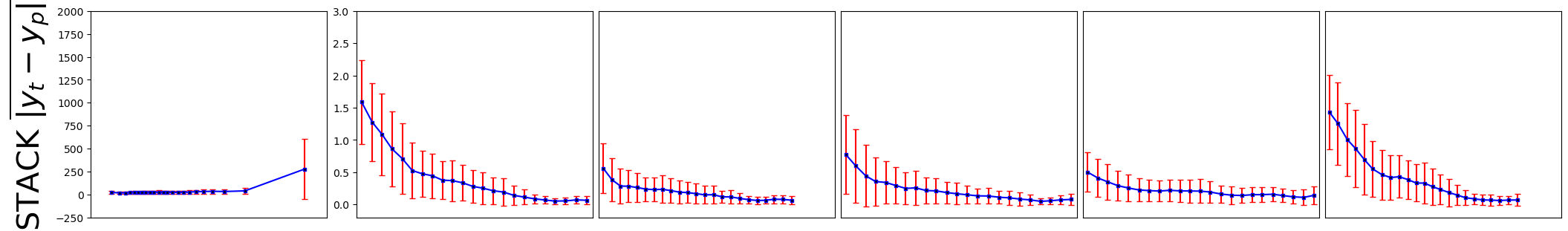}
\includegraphics[width=0.85\columnwidth]{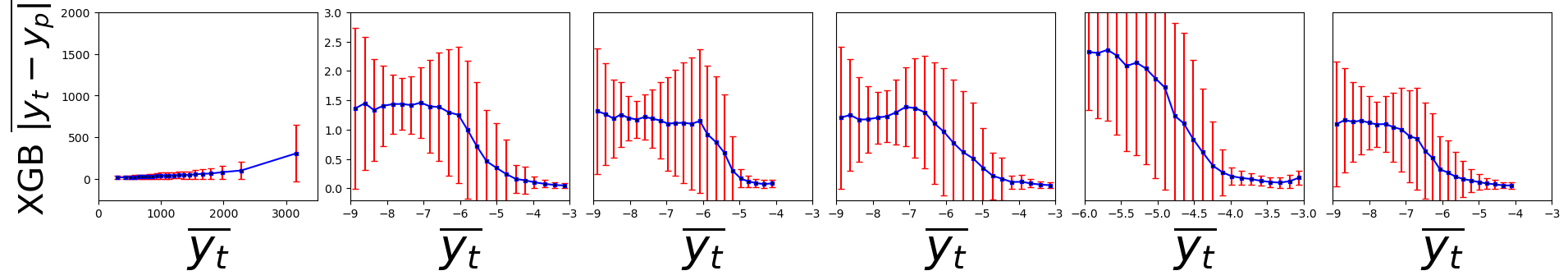}
\end{center}
    \caption{($\mathcal{NM}$) The same as Figure~\ref{fig:n_true}, but using in addition the spectral mean and standard deviation as feature variables.
    }
    \label{fig:nm_true}
\end{figure}

\begin{figure}[t]
\begin{center}
\includegraphics[width=0.82\columnwidth]{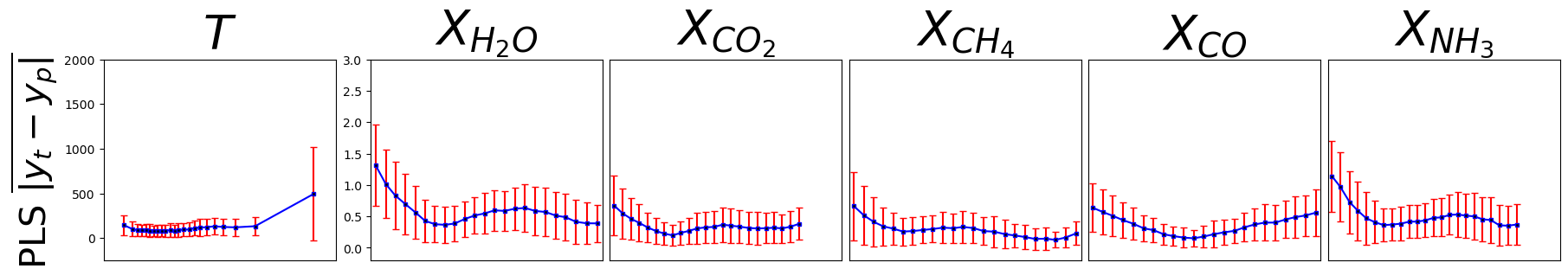}
\includegraphics[width=0.82\columnwidth]{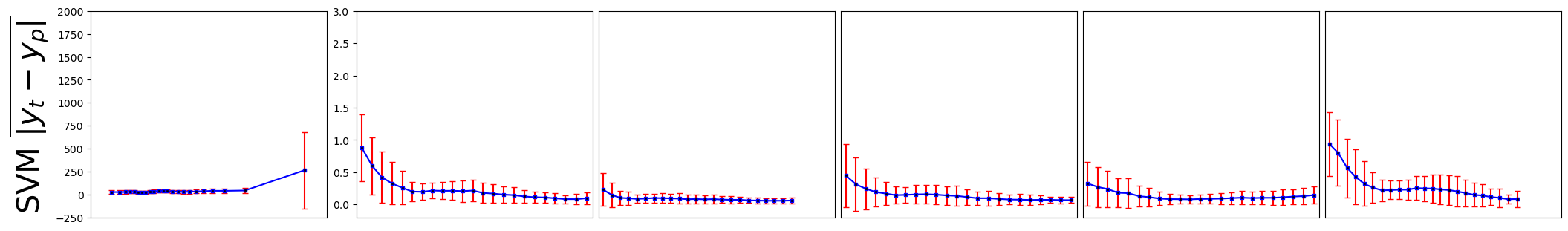}
\includegraphics[width=0.82\columnwidth]{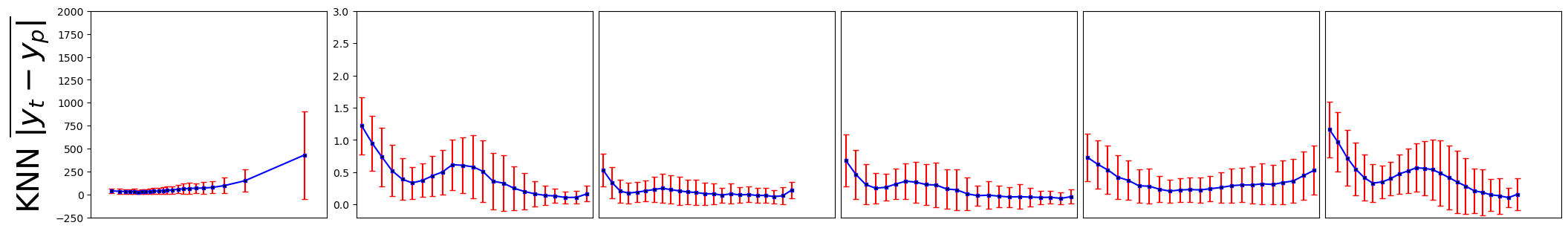}
\includegraphics[width=0.82\columnwidth]{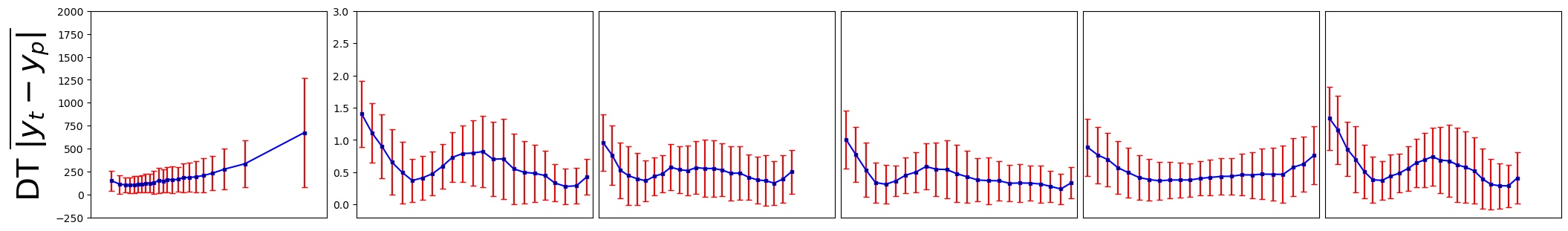}
\includegraphics[width=0.82\columnwidth]{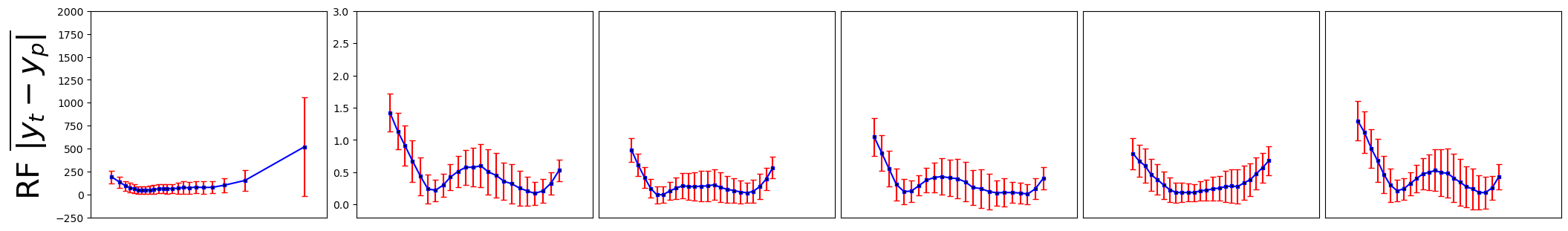}
\includegraphics[width=0.82\columnwidth]{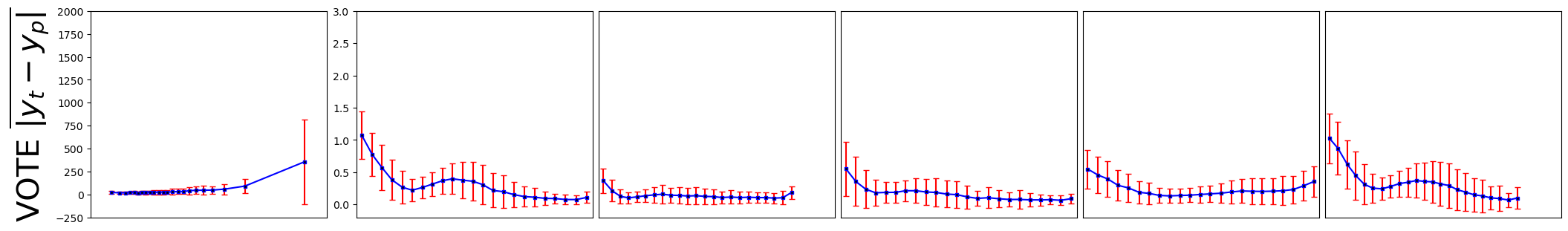}
\includegraphics[width=0.82\columnwidth]{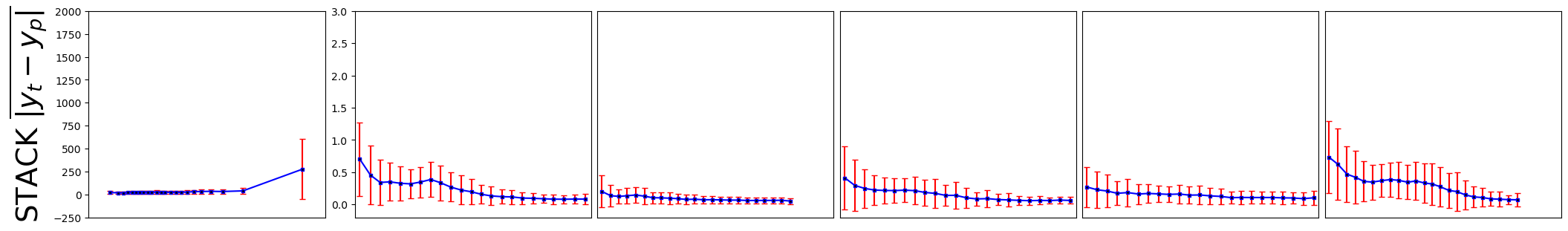}
\includegraphics[width=0.82\columnwidth]{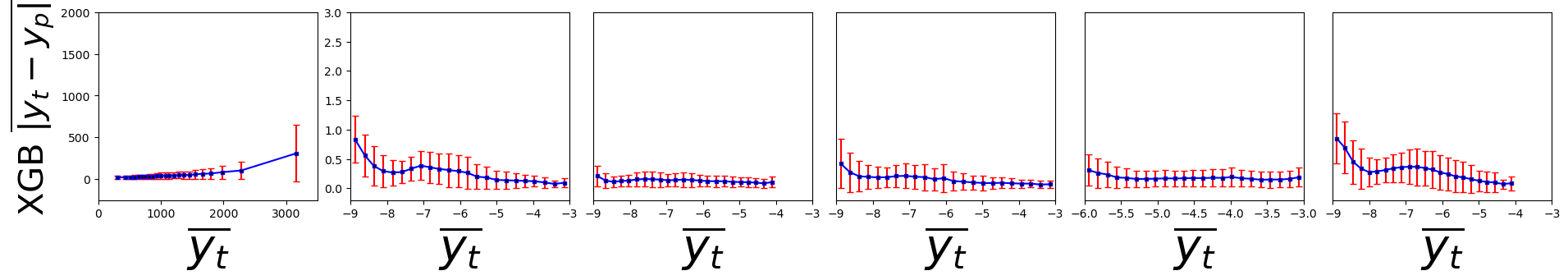}
\end{center}
    \caption{($\mathcal{NML}$) The same as Figure~\ref{fig:nm_true}, but using log concentrations, $\text{log}(X)$, as target variables during training and testing.
    }
    \label{fig:nml_true}
\end{figure}

\begin{figure}[t]
\begin{center}
\includegraphics[width=0.85\columnwidth]{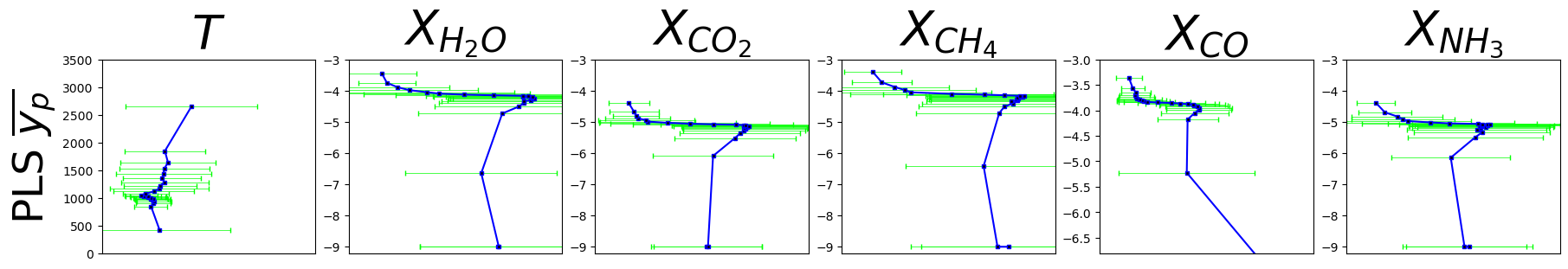}
\includegraphics[width=0.85\columnwidth]{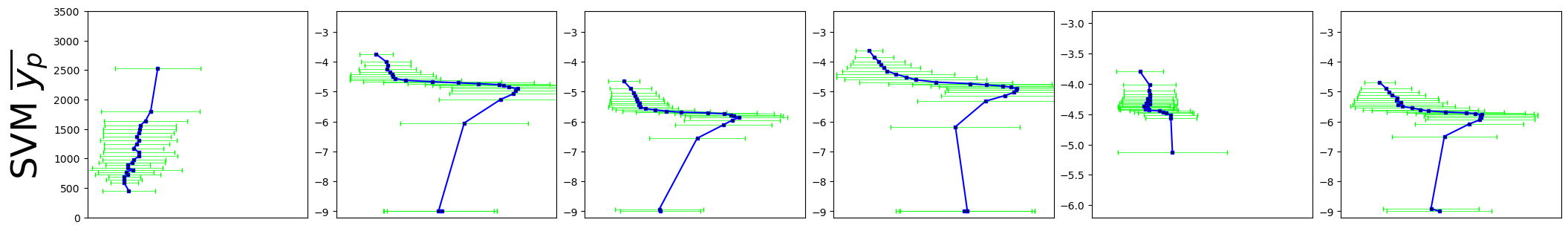}
\includegraphics[width=0.85\columnwidth]{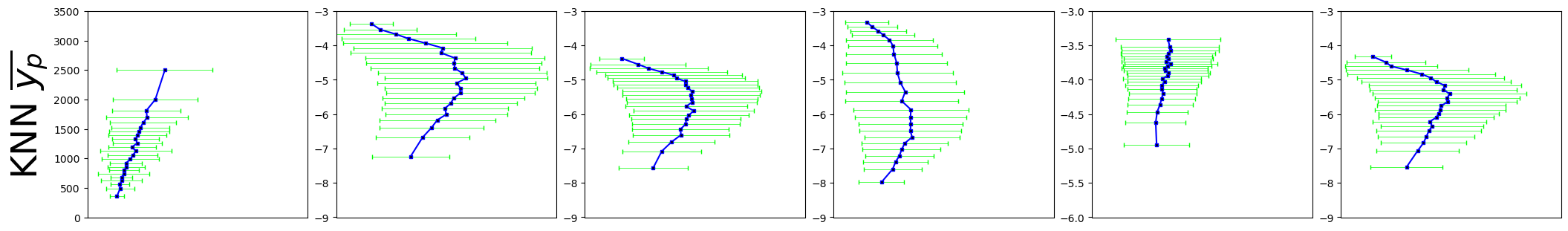}
\includegraphics[width=0.85\columnwidth]{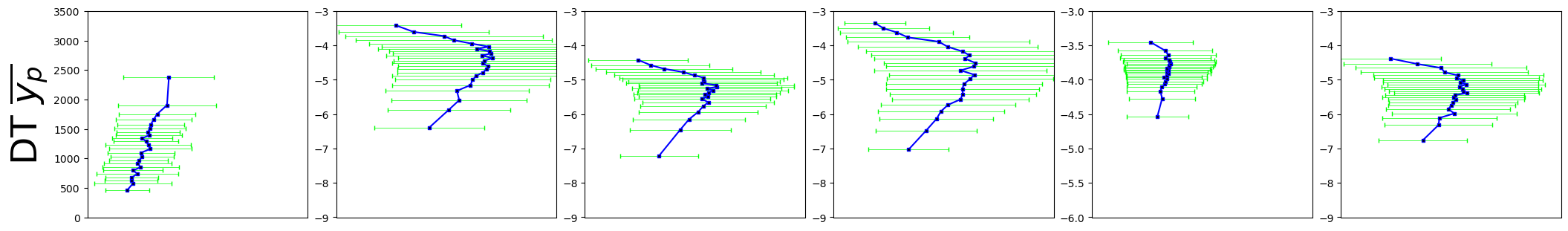}
\includegraphics[width=0.85\columnwidth]{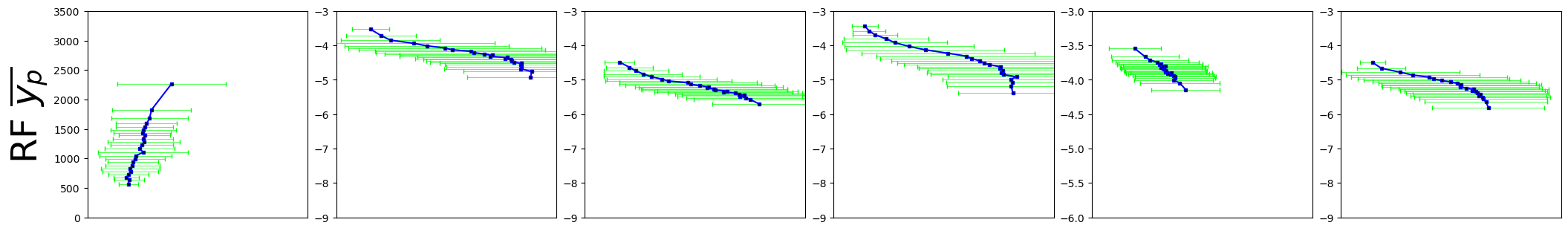}
\includegraphics[width=0.85\columnwidth]{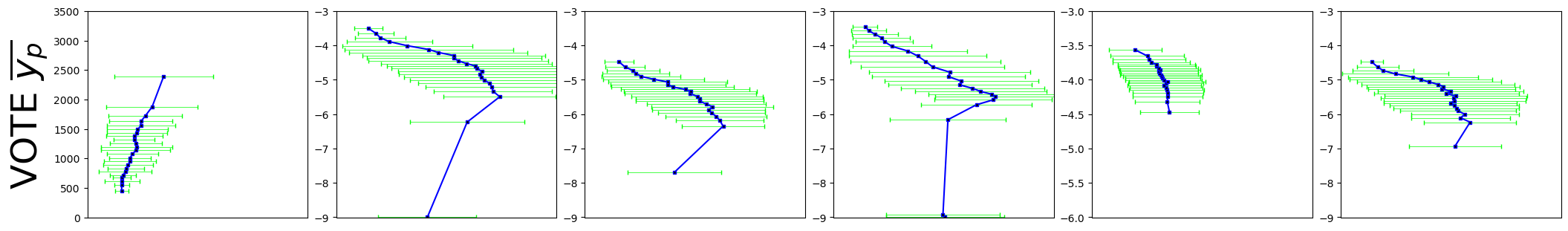}
\includegraphics[width=0.85\columnwidth]{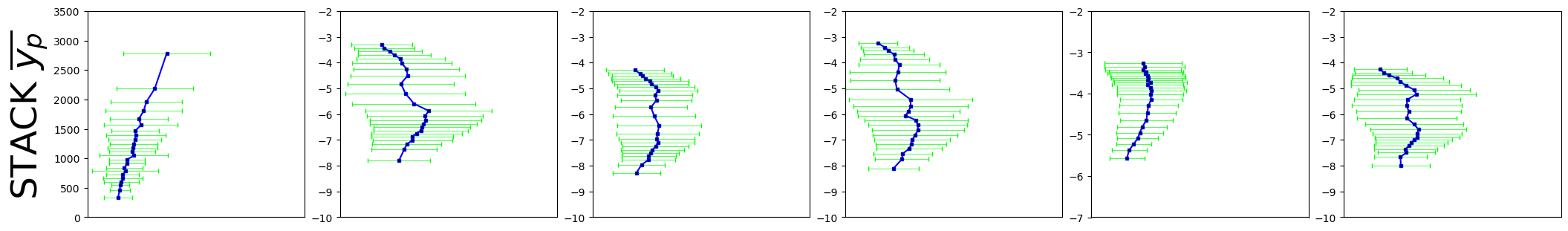}
\includegraphics[width=0.85\columnwidth]{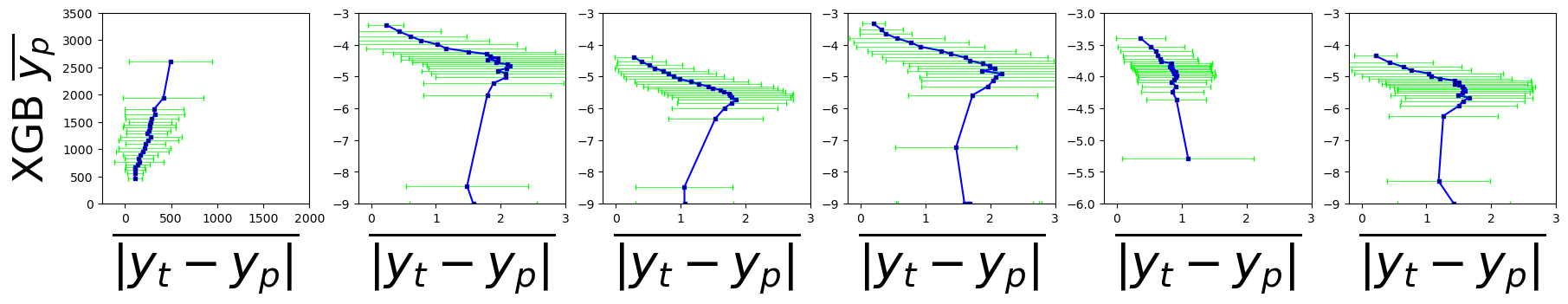}
\end{center}
    \caption{($\mathcal{S}$) Plots of the average predicted value $\bar{y}_p$ versus the average absolute deviation $|y_t-y_p|$ of the target prediction $y_p$ from the true value $y_t$. The model is trained and tested on the standardized spectral data $\{S[M],S[y]\}$ (see Section~\ref{sec:training_data}). The quantile bins were formed based on the sorted predicted target values $y_p$.  The error bars indicate the standard deviations of the quantity $|y_t-y_p|$ within each bin.     
    }
    \label{fig:s_pred}
\end{figure}

\begin{figure}[p]
\begin{center}
\includegraphics[width=0.85\columnwidth]{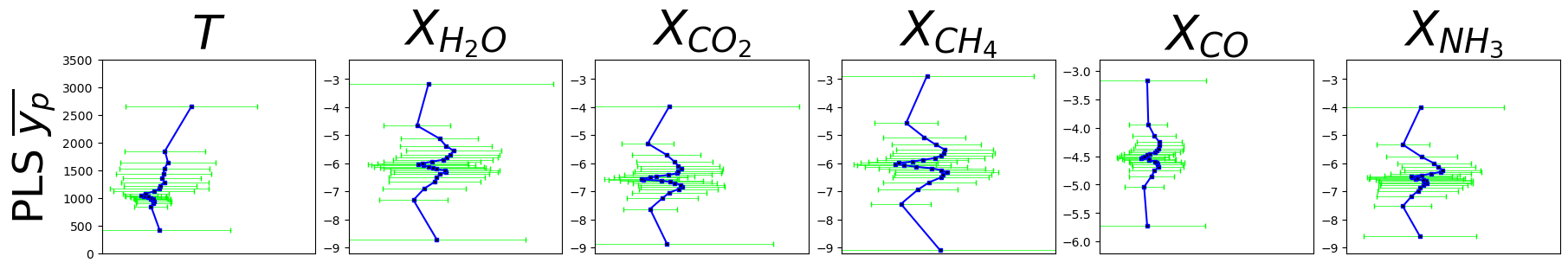}
\includegraphics[width=0.85\columnwidth]{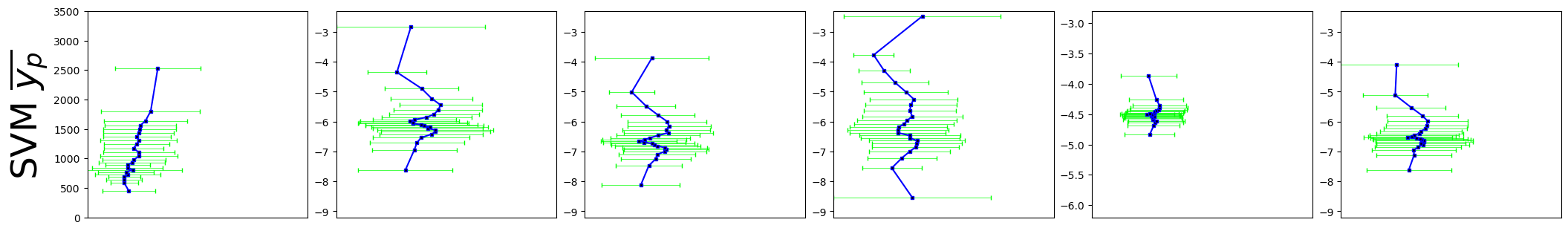}
\includegraphics[width=0.85\columnwidth]{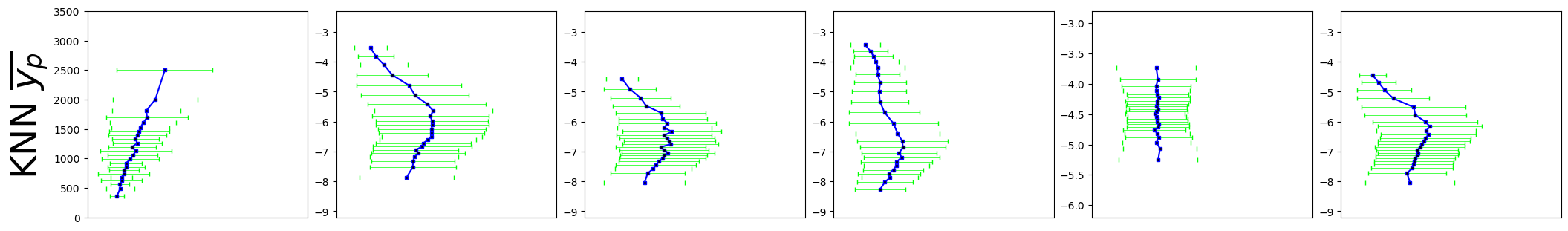}
\includegraphics[width=0.85\columnwidth]{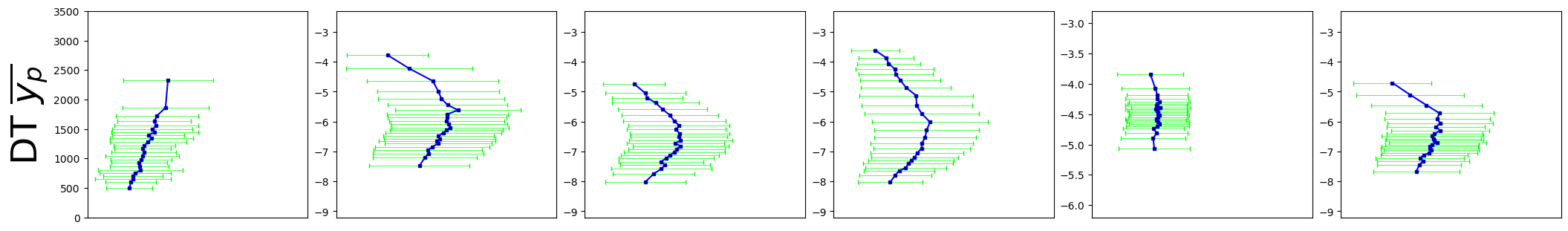}
\includegraphics[width=0.85\columnwidth]{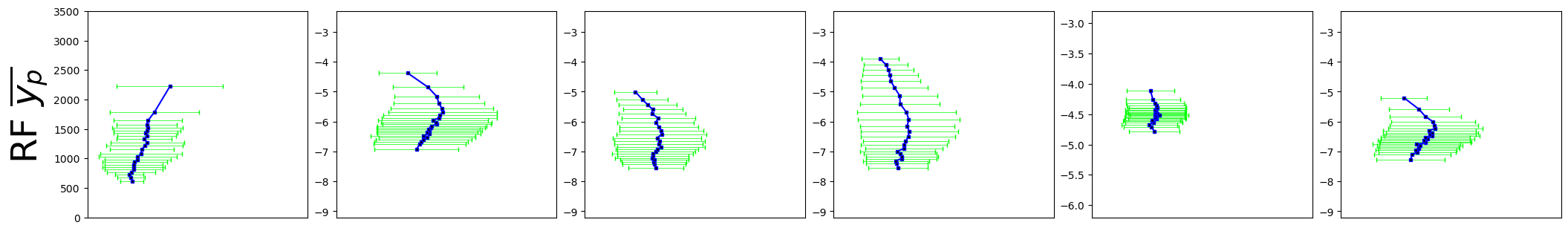}
\includegraphics[width=0.85\columnwidth]{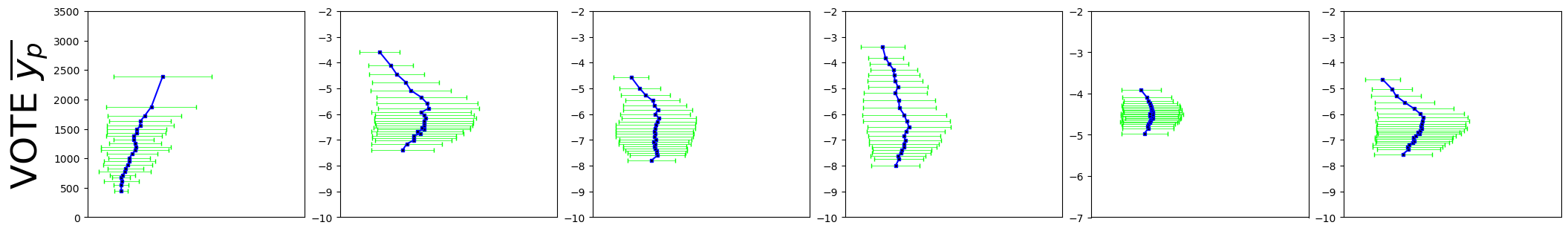}
\includegraphics[width=0.85\columnwidth]{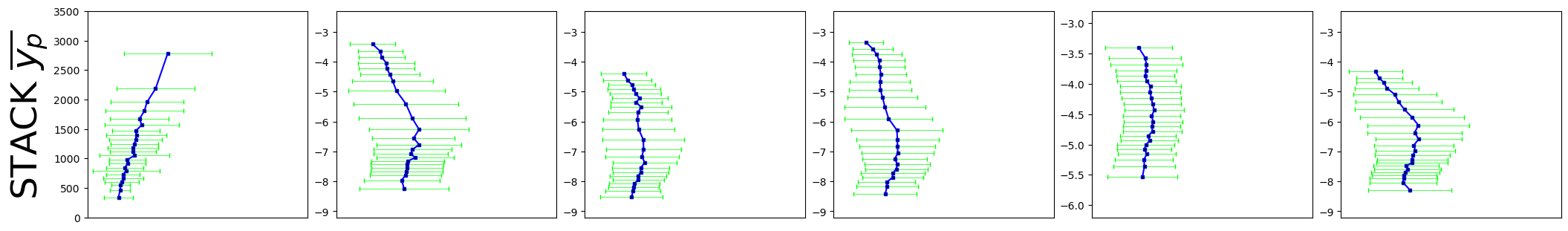}
\includegraphics[width=0.85\columnwidth]{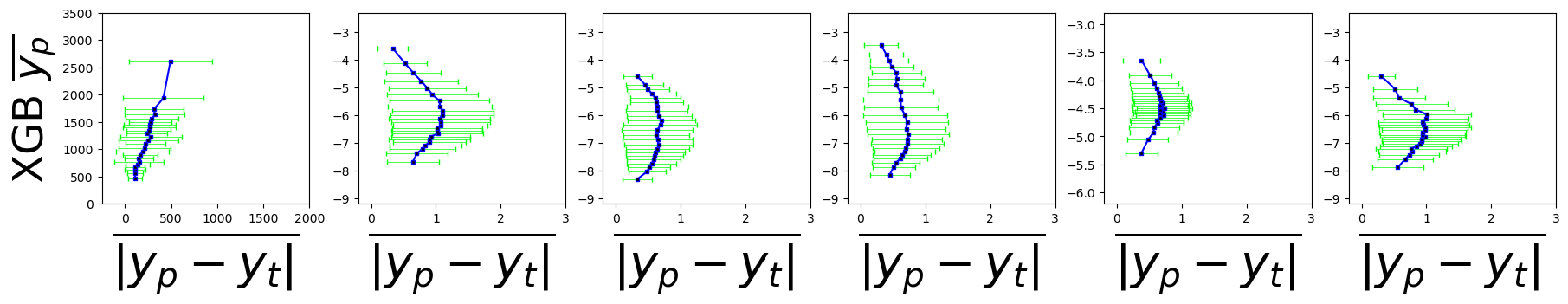}
\end{center}
    \caption{($\mathcal{SL}$) 
    The same as Figure~\ref{fig:s_pred}, but using log concentrations, $\text{log}(X)$, as target variables during training and testing.
    }
    \label{fig:sl_pred}
\end{figure}


\begin{figure}[t]
\begin{center}
\includegraphics[width=0.85\columnwidth]{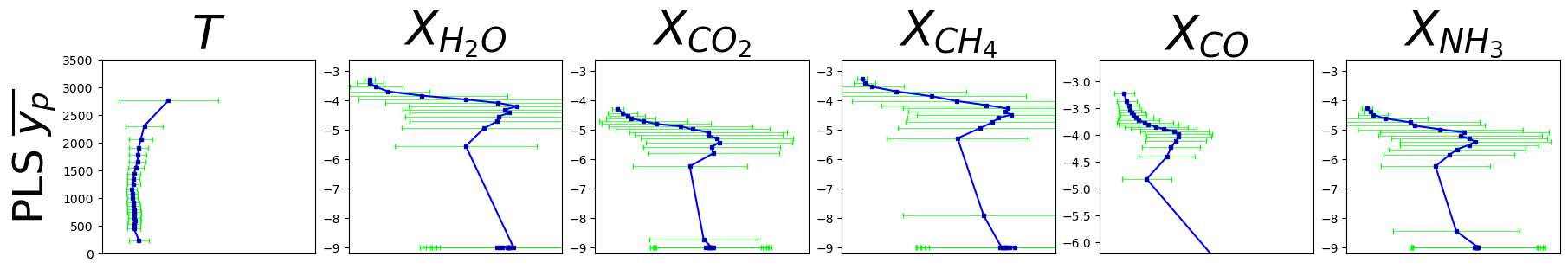}
\includegraphics[width=0.85\columnwidth]{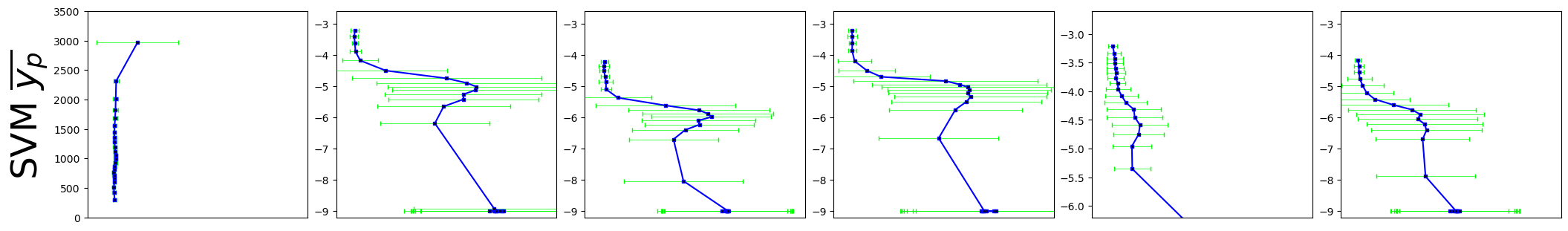}
\includegraphics[width=0.85\columnwidth]{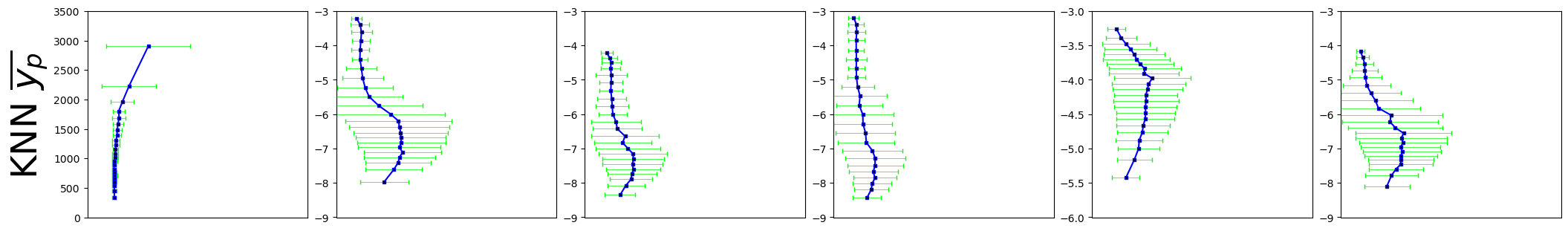}
\includegraphics[width=0.85\columnwidth]{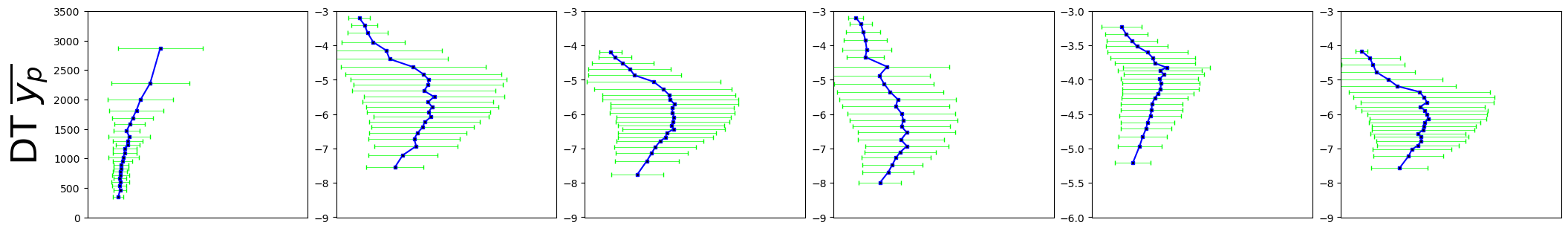}
\includegraphics[width=0.85\columnwidth]{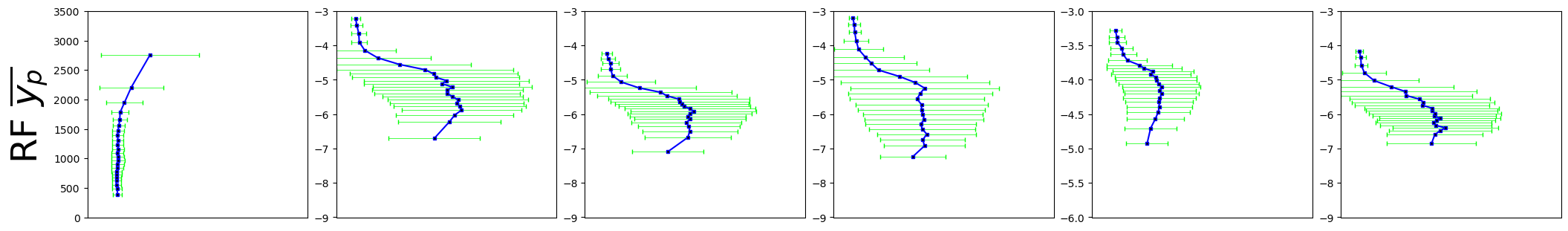}
\includegraphics[width=0.85\columnwidth]{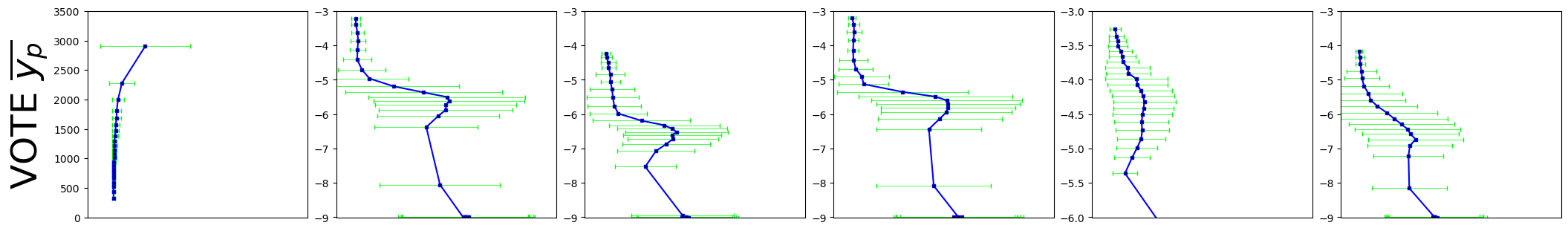}
\includegraphics[width=0.85\columnwidth]{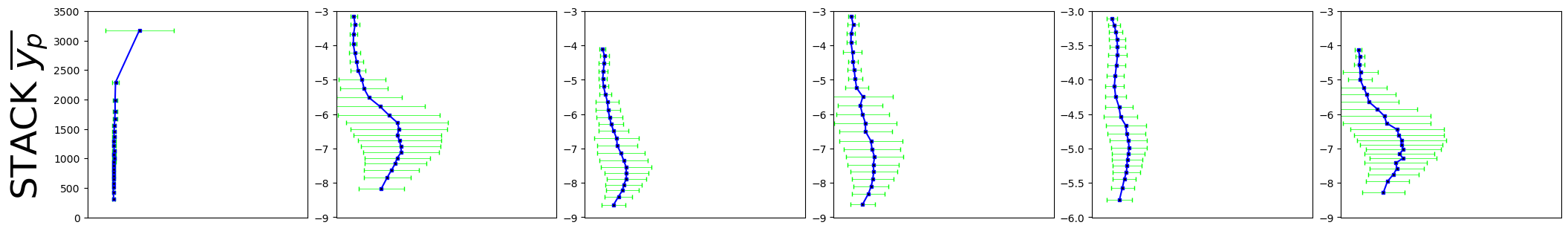}
\includegraphics[width=0.85\columnwidth]{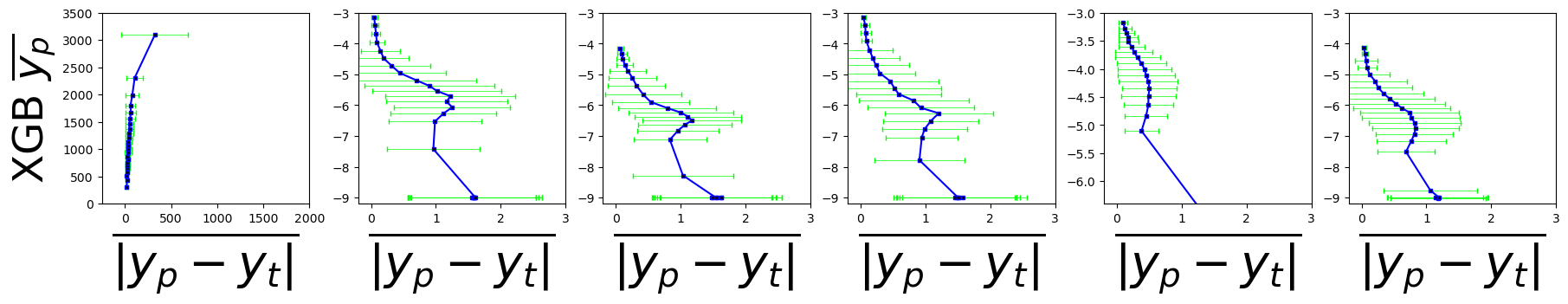}
\end{center}
    \caption{($\mathcal{N}$) The same as Figure~\ref{fig:s_pred}, but using the normalized spectral data $N[M]$ for training and testing.
    }
    \label{fig:n_pred}
\end{figure}

\begin{figure}[t]
\begin{center}
\includegraphics[width=0.85\columnwidth]{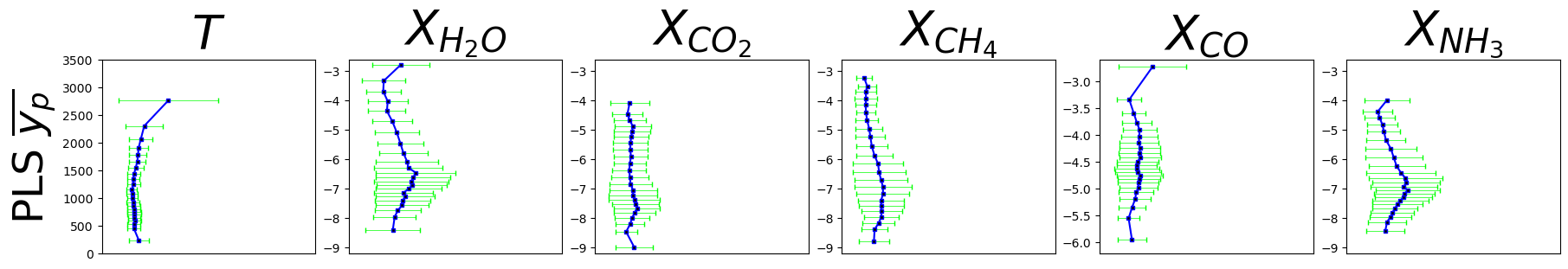}
\includegraphics[width=0.85\columnwidth]{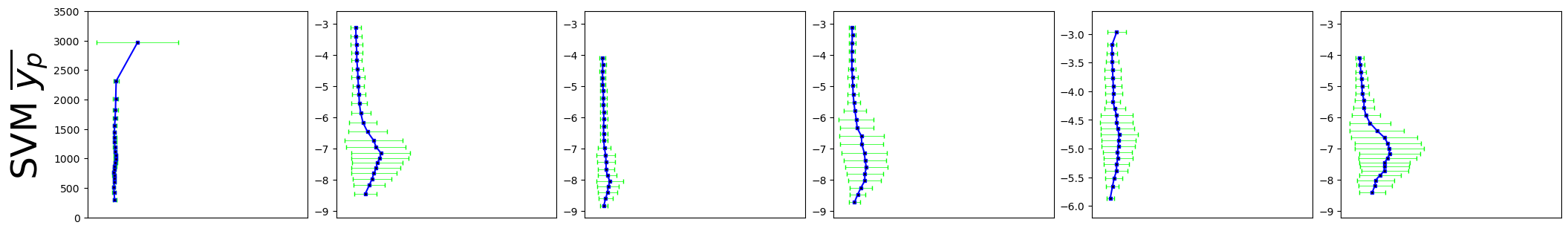}
\includegraphics[width=0.85\columnwidth]{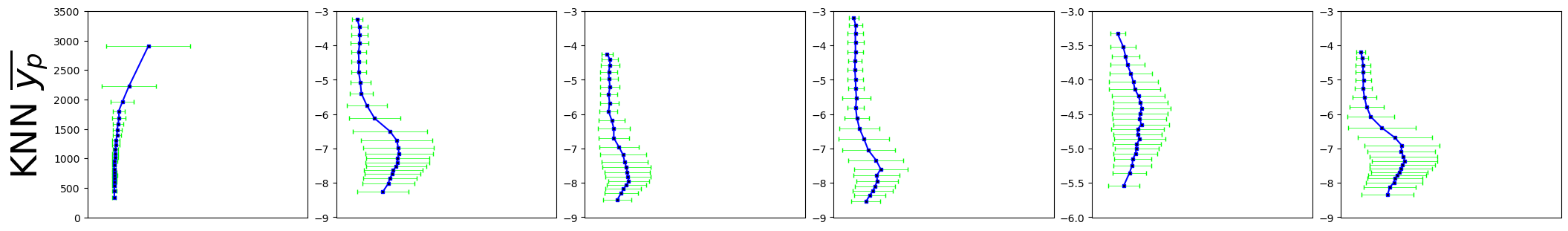}
\includegraphics[width=0.85\columnwidth]{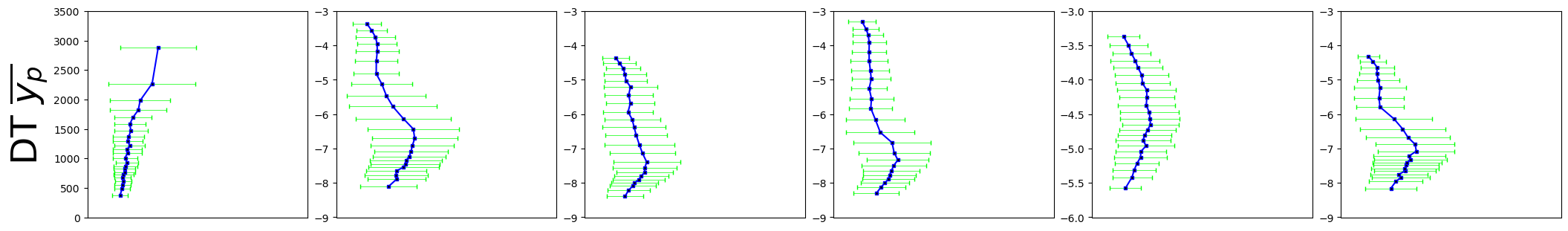}
\includegraphics[width=0.85\columnwidth]{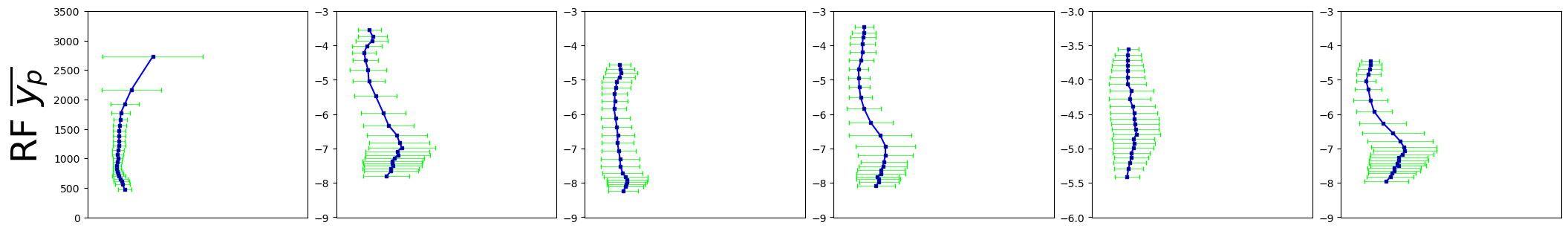}
\includegraphics[width=0.85\columnwidth]{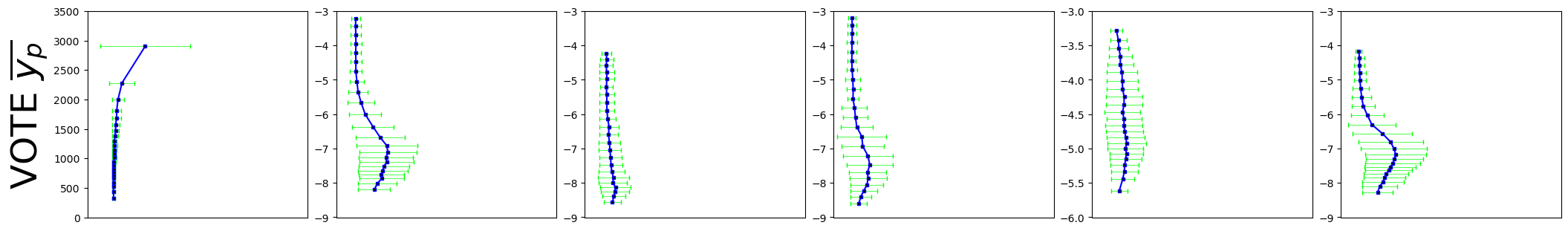}
\includegraphics[width=0.85\columnwidth]{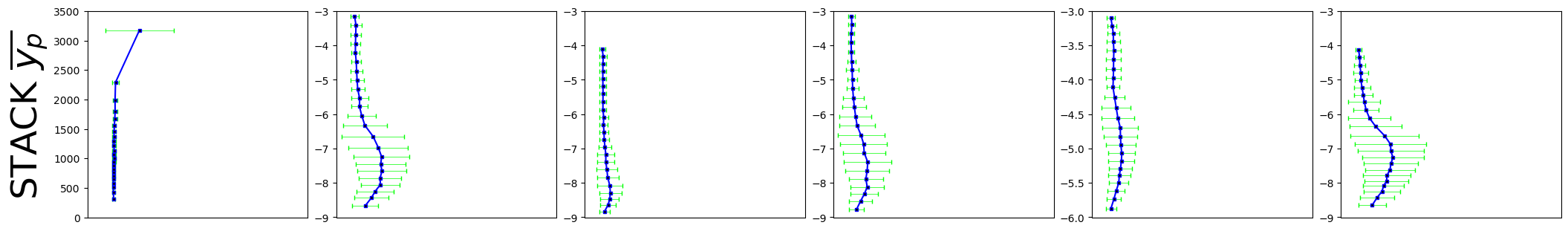}
\includegraphics[width=0.85\columnwidth]{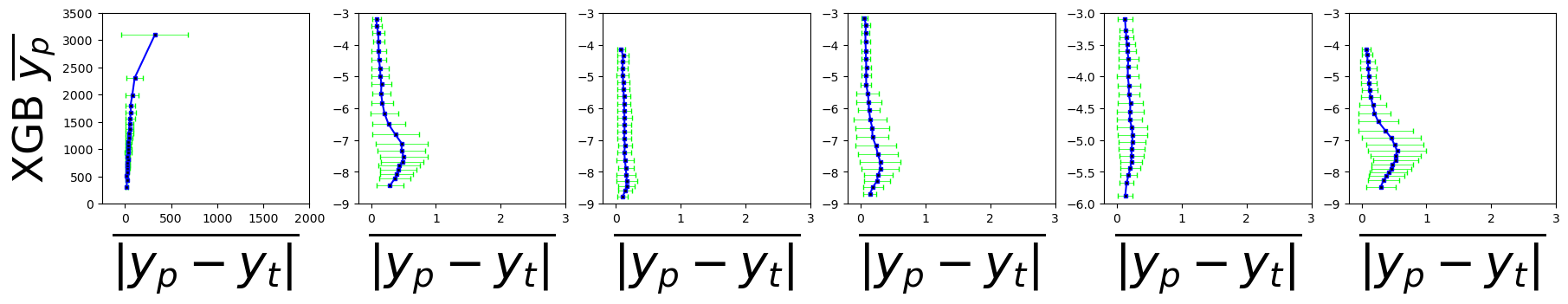}
\end{center}
    \caption{($\mathcal{NL}$) The same as Figure~\ref{fig:n_pred}, but using log concentrations, $\text{log}(X)$, as target variables during training and testing.
    }
    \label{fig:nl_pred}
\end{figure}

\begin{figure}[t]
\begin{center}
\includegraphics[width=0.85\columnwidth]{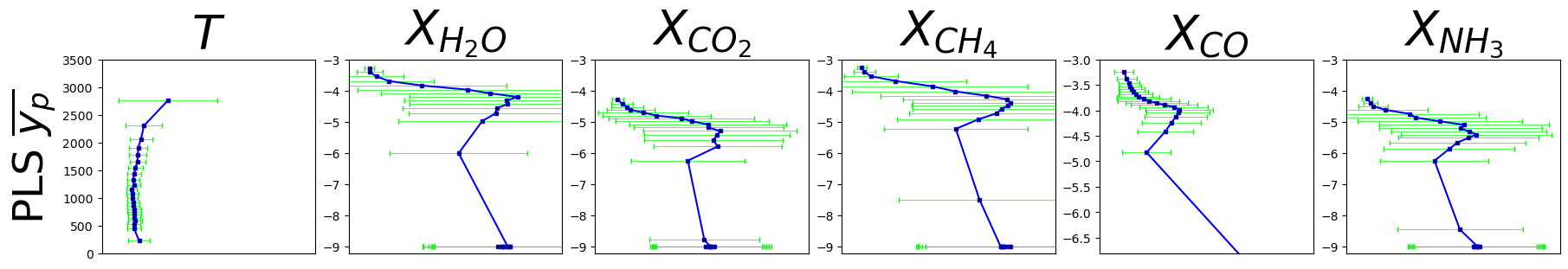}
\includegraphics[width=0.85\columnwidth]{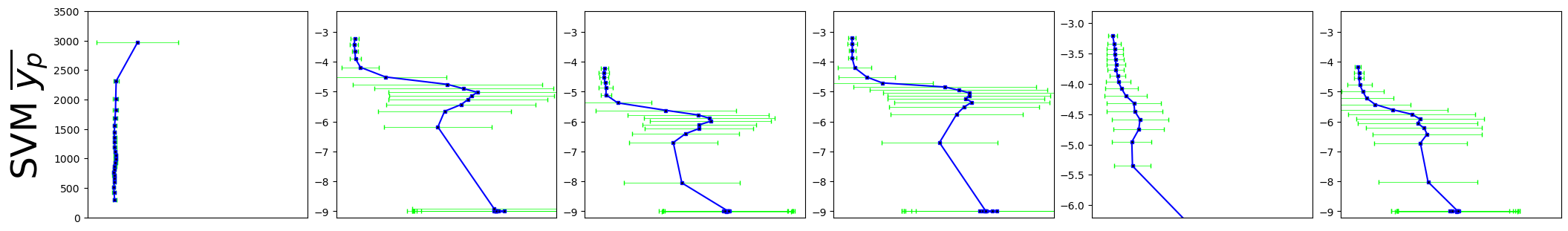}
\includegraphics[width=0.85\columnwidth]{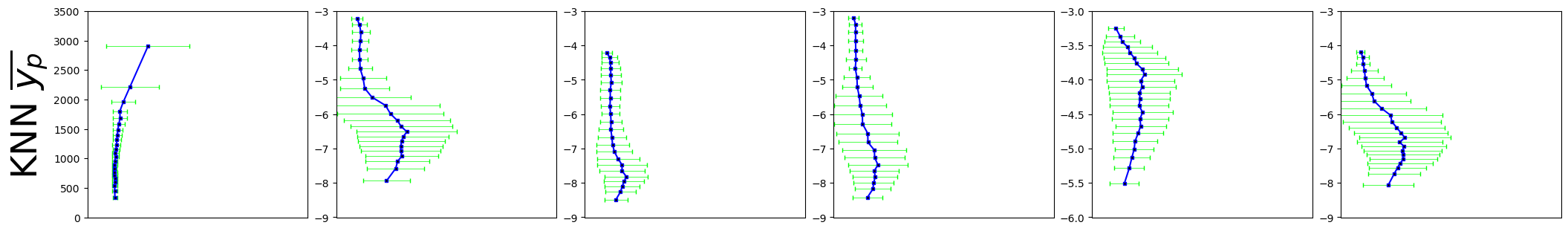}
\includegraphics[width=0.85\columnwidth]{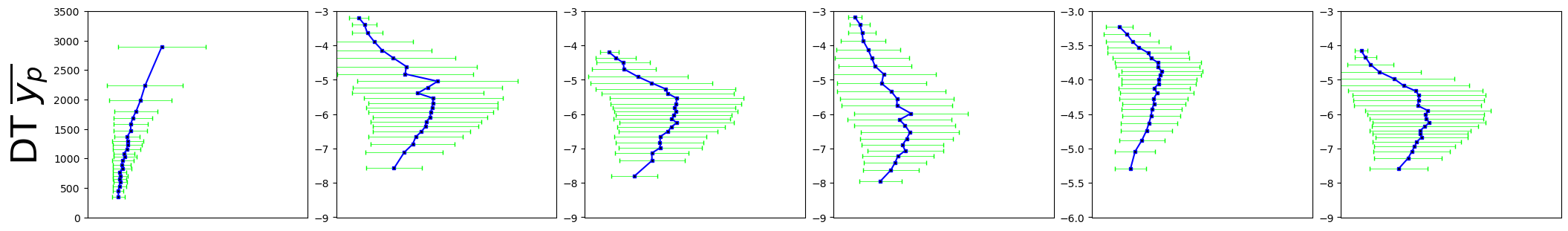}
\includegraphics[width=0.85\columnwidth]{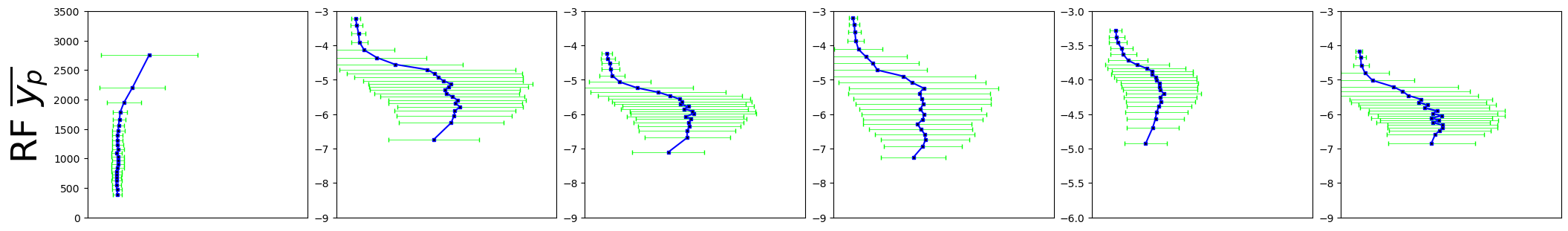}
\includegraphics[width=0.85\columnwidth]{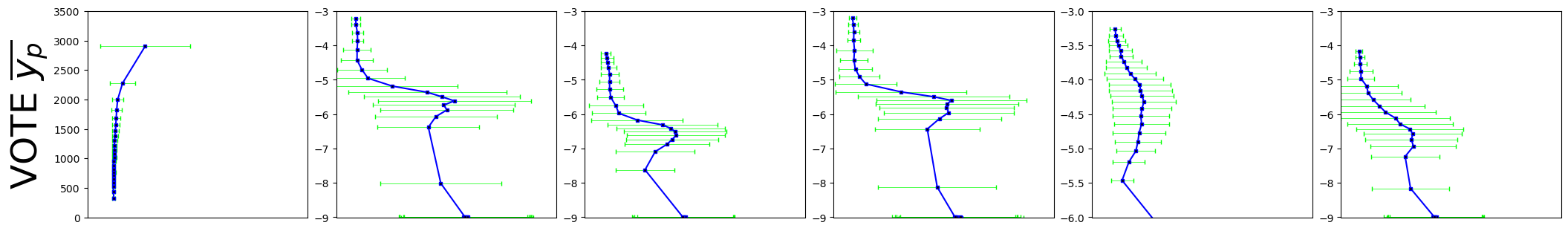}
\includegraphics[width=0.85\columnwidth]{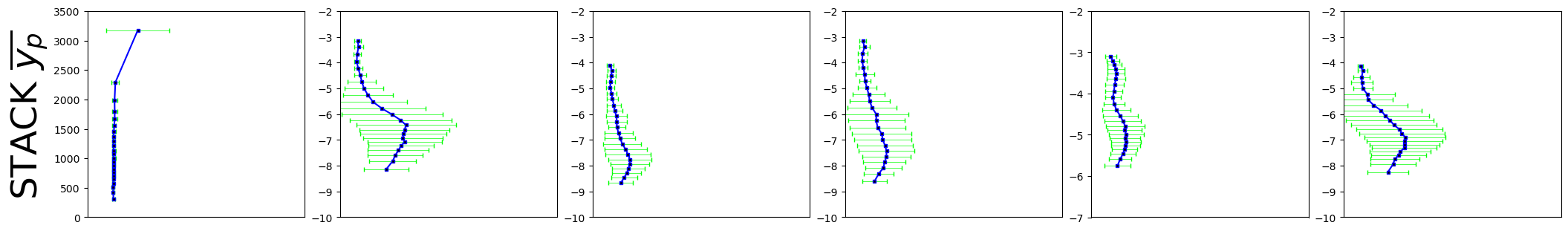}
\includegraphics[width=0.85\columnwidth]{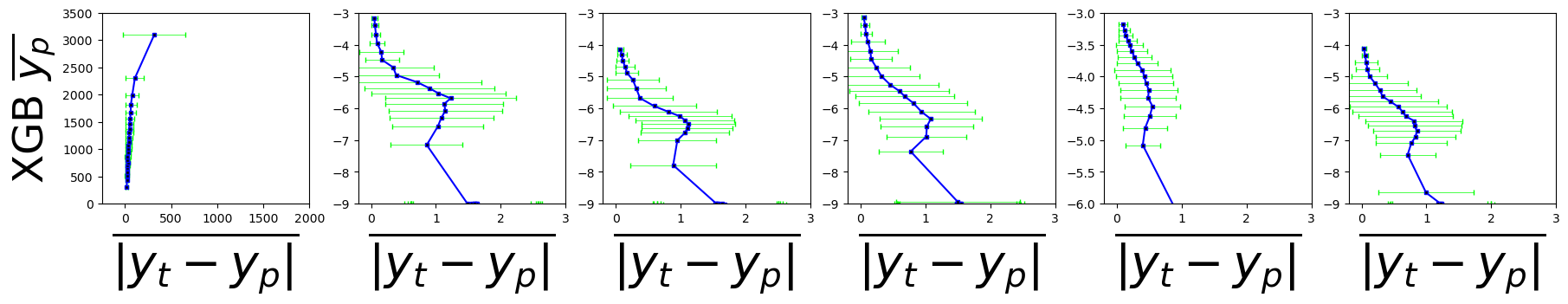}
\end{center}
    \caption{($\mathcal{NM}$) 
    The same as Figure~\ref{fig:n_pred}, but using in addition the spectral mean and standard deviation as feature variables.
    }
    \label{fig:nm_pred}
\end{figure}

\begin{figure}[t]
\begin{center}
\includegraphics[width=0.85\columnwidth]{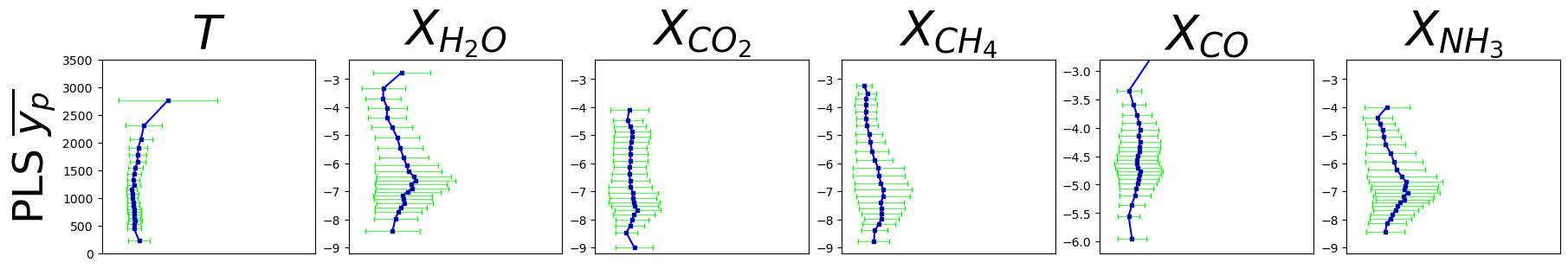}
\includegraphics[width=0.85\columnwidth]{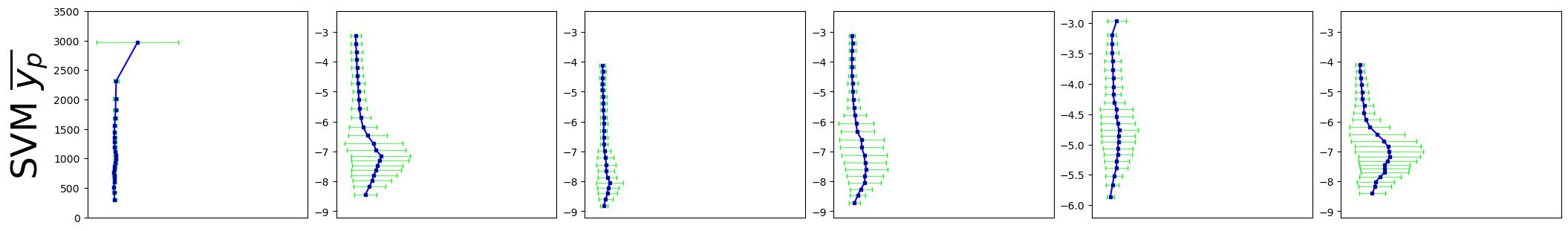}
\includegraphics[width=0.85\columnwidth]{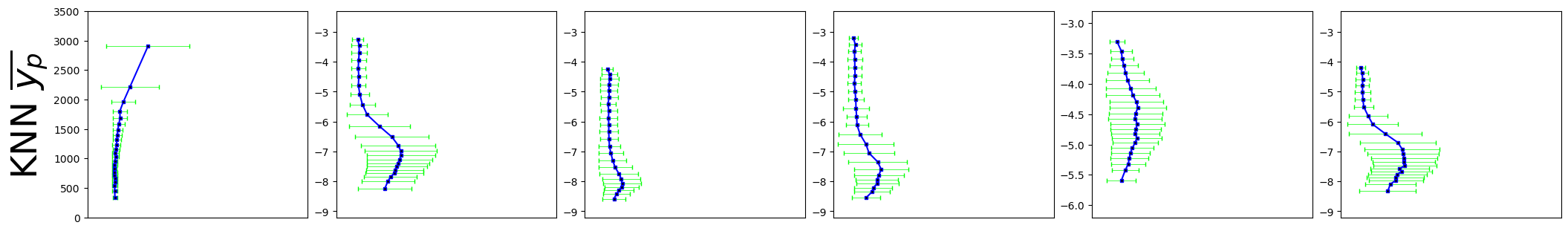}
\includegraphics[width=0.85\columnwidth]{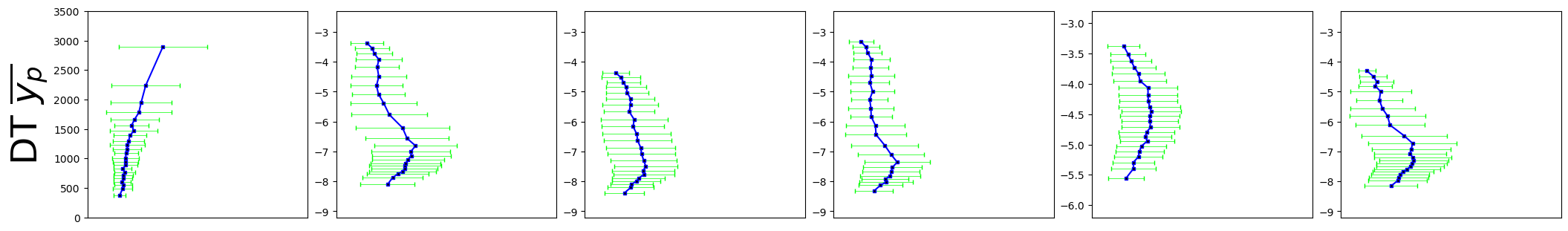}
\includegraphics[width=0.85\columnwidth]{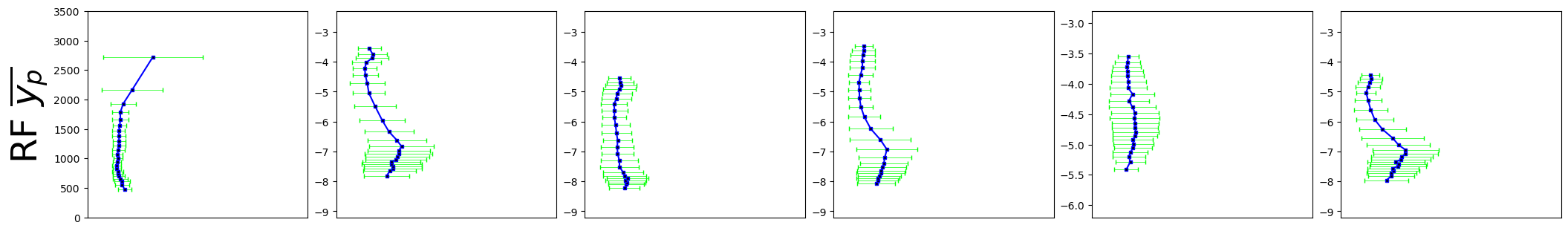}
\includegraphics[width=0.85\columnwidth]{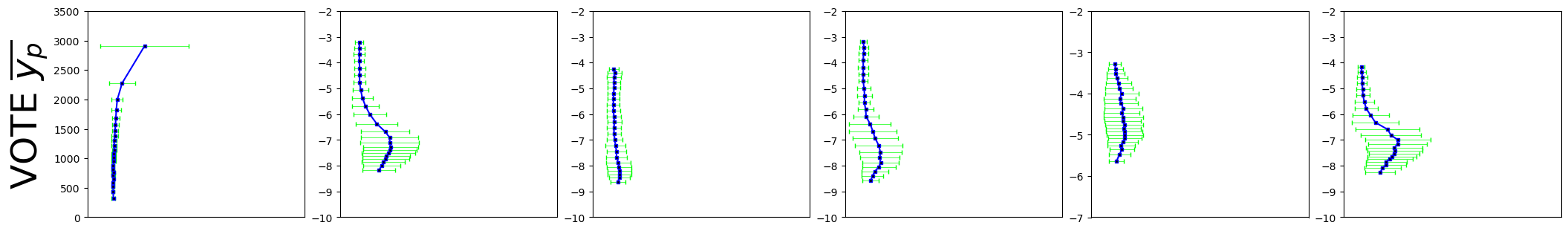}
\includegraphics[width=0.85\columnwidth]{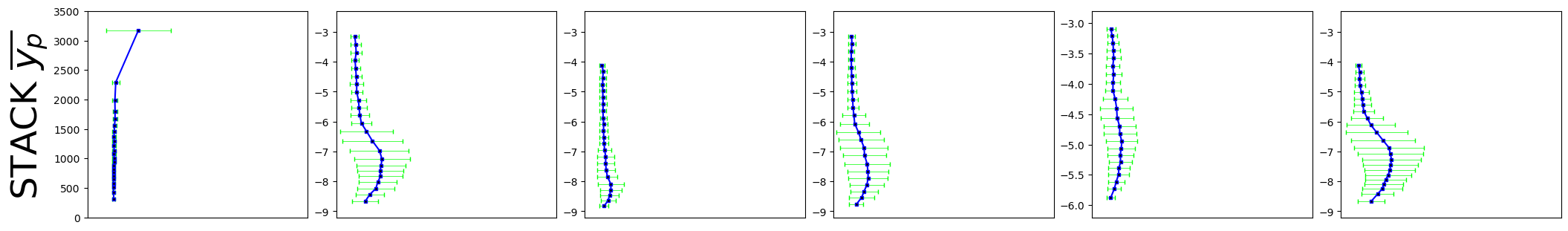}
\includegraphics[width=0.85\columnwidth]{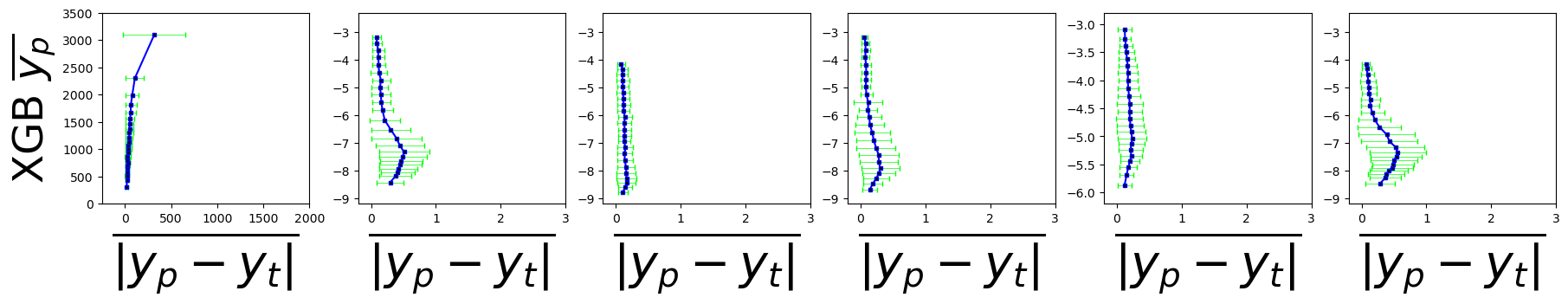}
\end{center}
    \caption{($\mathcal{NML}$) 
    The same as Figure~\ref{fig:nm_pred}, but using log concentrations, $\text{log}(X)$, as target variables during training and testing.
    }
    \label{fig:nml_pred}
\end{figure}

\end{document}